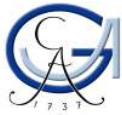
GEORG-AUGUST-UNIVERSITÄT
GÖTTINGEN

Fakultät für
Physik
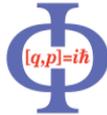

Master's Thesis

# Simultane Multischichtrekonstruktion durch regularisierte nichtlineare Inversion

# Simultaneous multi-slice reconstruction by Regularized Nonlinear Inversion

prepared by

**Sebastian Rosenzweig**

from Würzburg

at the Institut für Diagnostische und Interventionelle Radiologie Göttingen

| | |
|---|---|
| **Date of Submission:** | 14th September 2016 |
| **First Referee:** | Prof. Dr. Martin Uecker |
| **Second Referee:** | Prof. Dr. Christoph Schmidt |

# Contents

# 1. Introduction

Magnetic resonance imaging (MRI) is one of the most important imaging modalities in hospitals and clinical research. This non-invasive technique provides anatomical images from human or biological systems and offers a superior soft-tissue contrast compared to other imaging methods such as Computed Tomography (CT). Amongst other things, MRI can provide flow, diffusion and structural information in two and three dimensions. However, 3D measurements are not always feasible due to their long acquisition times. One possible way to bridge the gap between two- and three-dimensional imaging is the use of simultaneous multi-slice (SMS) MRI, where multiple slices of an object of study are acquired. To disentangle the multiple slices the spatial encoding information inherent in receiver coil arrays and/or special encoding schemes can be used.

The idea of SMS is quite old. In the year 1980, Maudsley proposed to use SMS for improving the efficiency of line-scan imaging techniques [1]. At the end of the same decade the work of Müller laid the basis for modern SMS radiofrequency (RF) excitation pulse design [2]. By the introduction of parallel imaging at the turn of the millennium, further improvements in SMS could be achieved [3]. Despite significant advances in theory and practice, for a long time simultaneous multi-slice remained a topic followed by just a few research groups. One reason for this lack of interest was the late availability of receiver coils with coil distribution in z-direction - a mandatory property for accelerating axial acquisitions. Then again, there had not been an obvious application for this method. As a consequence, the implementation of SMS sequences and reconstructions on vendor platforms is far from complete and often not available out of the box.

It was not until the beginning of the present decade that MRI has attracted increased interest among researchers. In particular, publications about Echo Planar Imaging (EPI) in combination with SMS by Moeller et al. [4] and Feinberg et al. [5], which demonstrated a significant scan time reduction and image quality improvements, drew widespread attention to this method. In general, the field of application for SMS are time critical acquisitions such as functional MRI or Diffusion Tensor Imaging, as well as abdominal or cardiac imaging. For all these experiments, common reconstruction approaches solve a linear equation by making use of previously determined coil sensitivities. Here we propose a new method for SMS MRI based on Regularized Nonlinear Inversion (NLINV) [6]. For this technique no prior knowledge about the coil profiles is required. Especially real-time applications benefit from the joint estimation of image content and coil sensitivities, since the latter may change due to motion or interactive changes to the slice position. For single-slice imaging, Uecker et al. achieved a time-resolution of 20 ms using NLINV and demonstrated significant image quality improvements compared to conventional autocalibrating parallel MRI - particularly for high reduction factors. The aim of the present thesis is to extend this nonlinear algorithm to multiple slices for Cartesian acquisitions.

In chapter 2 we present the basic quantum-mechanical theory of MRI and in chapter 3 we give a basic introduction into multi-slice MRI. The fundamentals of RF pulse design are covered in chapter 4 before we deduce the simultaneous multi-slice Regularized Nonlinear Inversion (SMS-NLINV) algorithm in chapter 5. Finally, in chapter 6 we present the utilized hardware and in chapter 7 we describe the experiments performed to validate the SMS sequence and to analyze the SMS-NLINV algorithm. We end with a conclusion and an outlook to non-Cartesian acquisitions and clinical applications in chapter 8.

# 2. Basics of magnetic resonance imaging

In this chapter we introduce the basics of magnetic resonance imaging. We will cover the physical principles of nuclear magnetic resonance (NMR), describe the actual NMR experiment and explain the fundamentals of image reconstruction.

## 2.1. Quantum mechanical description of nuclear magnetic resonance

NMR is based on the interaction of nuclear spins in an external magnetic field. Many phenomena can be understood using the classical or semi classical vector model. However, for a detailed description of NMR we use the quantum-mechanical formalism. Still, an in-depth study of the physical principles of NMR is beyond the scope of this thesis and is extensively described in the literature. We will only focus on main aspects that are relevant for this work. For further information, the interested reader is referred to the introductory textbooks [7–12].

### 2.1.1. Nuclear spin

NMR in bulk material was discovered independently by Purcell, Torrey and Pound at Harvard and by Bloch, Hansen and Packard at Stanford by the end of 1945 [13]. The effect of NMR can be observed in all atoms that possess a non-zero nuclear spin quantum number $s$. According to the Pauli principle only atoms with an uneven nucleon number or with both uneven proton and neutron numbers can possess such an angular momentum or spin, e.g. $^1$H ($s = 1/2$), $^{13}$C ($s = 1/2$), $^{14}$N ($s = 1$) or $^{17}$O ($s = 5/2$) [14]. The dominant nucleus in MRI studies is the proton in hydrogen. On the one hand, $^1$H is the most commonly occurring H-isotope and on the other hand, the body consists of tissue that is mainly composed of water and fat - both contain hydrogen [10]. Thus, the $^1$H hydrogen isotope is the natural choice to study the body.

## 2.1.2. Spin formalism[1]

In quantum mechanics, particles are described as complex wave functions $\Phi(\boldsymbol{r})$ and each measurable variable is associated with an operator that acts on these functions. $|\Phi(\boldsymbol{r})|^2$ is a measure for the probability of finding a particle at position $\boldsymbol{r}$. However, if a particle possesses a spin we have to ask for the probability of finding the particle at position $\boldsymbol{r}$ with a specific spin orientation. Thus, the Hilbert space of wave functions has to be extended by a spin state space. We therefore define the spin operator[2]

$$\hat{\boldsymbol{S}} := (\hat{S}_x, \hat{S}_y, \hat{S}_z). \tag{2.1}$$

As the spin has all the characteristics of an angular momentum, this operator has to satisfy the commutator relation[3]

$$[\hat{S}_k, \hat{S}_l] = i\hbar\epsilon_{klm}\hat{S}_m, \tag{2.2}$$

where $i$ is the imaginary unit, $\hbar$ the reduced Planck constant and $\epsilon_{klm}$ the Levi-Civita tensor.

Using (2.2) it can be shown that $\hat{S}^2$ and $\hat{S}_z$ have common eigenstates $|\mathcal{X}\rangle$ that span the spin state space. The respective eigenvalues are given by

$$\hat{S}^2 |\mathcal{X}\rangle = \hbar^2 s(s+1) |\mathcal{X}\rangle, \tag{2.3}$$

$$\hat{S}_z |\mathcal{X}\rangle = \hbar s_z |\mathcal{X}\rangle, \quad s_z = -s, -s+1, ..., s, \tag{2.4}$$

with $s$ being the integer or half-integer spin.

$^1$H nuclei are spin $s = 1/2$ particles, thus we get two possible eigenvalues $\hbar/2$ and $-\hbar/2$ for $\hat{S}_z$. The respective eigenstates are a basis of the spin-1/2 state space and defined as $|\uparrow\rangle$

---

[1][11, 12, 15] serve as general references for this section.
[2]For the general theory of spin we do not distinguish between electron spin and nuclear spin.
[3]Here, Einstein's sum convention is used.

and $|\downarrow\rangle$. They obey the relations

$$\hat{S}_z |\uparrow\rangle = \frac{\hbar}{2} |\uparrow\rangle , \tag{2.5a}$$

$$\hat{S}_z |\downarrow\rangle = -\frac{\hbar}{2} |\downarrow\rangle , \tag{2.5b}$$

$$\langle\uparrow|\uparrow\rangle = \langle\downarrow|\downarrow\rangle = 1, \quad \langle\uparrow|\downarrow\rangle = \langle\downarrow|\uparrow\rangle = 0. \tag{2.5c}$$

A general spin state is given by

$$|\mathcal{X}_{\text{gen}}\rangle = c_\uparrow |\uparrow\rangle + c_\downarrow |\downarrow\rangle \tag{2.6}$$

with $c_\uparrow$ and $c_\downarrow$ being complex numbers. The basis $\{|\uparrow\rangle,|\downarrow\rangle\}$ of the spin state space, in combination with the position-space wave functions $\Phi(\boldsymbol{r})$, make up elements of the extended state space

$$|\Psi(\boldsymbol{r})\rangle := \Phi_\uparrow(\boldsymbol{r}) |\uparrow\rangle + \Phi_\downarrow(\boldsymbol{r}) |\downarrow\rangle . \tag{2.7}$$

If the spin and the position states are uncorrelated, which we will assume in all further considerations, we can write

$$|\Psi(\boldsymbol{r})\rangle := \Phi(\boldsymbol{r})(c_\uparrow |\uparrow\rangle + c_\downarrow |\downarrow\rangle). \tag{2.8}$$

As we are only interested in the spin properties, the absence of any $\boldsymbol{r}$ dependence originates from neglecting the particle's orbital and translational motion,

$$|\Psi\rangle := c_\uparrow \Phi |\uparrow\rangle + c_\downarrow \Phi |\downarrow\rangle := c_\uparrow |\Psi_\uparrow\rangle + c_\downarrow |\Psi_\downarrow\rangle . \tag{2.9}$$

The elements $|\Psi\rangle$, $|\Psi_\uparrow\rangle$ and $|\Psi_\downarrow\rangle$ are generally normalized.

### 2.1.3. Spin in a constant homogeneous magnetic field

In this section we analyze the behavior of a spin in a constant homogeneous external magnetic field.

**Energy eigenstates.** According to the correspondence principle[4] the nuclear spin, which has the characteristics of an angular momentum, possesses an associated magnetic moment

---

[4] The correspondence principle states, that the quantum mechanical relations between variables are the same as the classical ones.

$$\hat{\boldsymbol{\mu}} = \gamma \hat{S}. \tag{2.10}$$

Here $\gamma$ is the gyromagnetic ratio, which is characteristic for every atom. In literature usually the value for the reduced gyromagnetic ratio is found, which for $^1$H is given by $\gamma/2\pi = 42.5774\,\mathrm{MHz\,T^{-1}}$ [10].

The coupling of the magnetic moment to an external magnetic field

$$\boldsymbol{B}_0 = B_0 \boldsymbol{e}_z, \tag{2.11}$$

which conventionally points in z-direction, leads to the Hamiltonian[5]

$$\hat{H} = -\hat{\boldsymbol{\mu}} \cdot \boldsymbol{B}_0 \stackrel{(2.11)}{=} -\hat{\mu}_z B_0 \stackrel{(2.10)}{=} -\gamma \hat{S}_z B_0. \tag{2.12}$$

$\hat{H}$ describes the energy of a quantum mechanical system. By solving the time-independent Schrödinger equation

$$\hat{H} \ket{\Psi} = E \ket{\Psi}, \tag{2.13}$$

the actual energies of the eigenstates $\ket{\uparrow}$ and $\ket{\downarrow}$ for $^1$H nuclei can be obtained.

$$E_\uparrow \ket{\uparrow} = \hat{H} \ket{\uparrow} \stackrel{(2.5a, 2.12)}{=} -B_0 \gamma \frac{\hbar}{2} \ket{\uparrow}, \tag{2.14a}$$

$$E_\downarrow \ket{\downarrow} = \hat{H} \ket{\downarrow} \stackrel{(2.5b, 2.12)}{=} B_0 \gamma \frac{\hbar}{2} \ket{\downarrow}, \tag{2.14b}$$

$$\Rightarrow E_\uparrow = -B_0 \gamma \frac{\hbar}{2}, \quad E_\downarrow = B_0 \gamma \frac{\hbar}{2}. \tag{2.15}$$

The corresponding eigenstates $\ket{\uparrow}$ and $\ket{\downarrow}$ are frequently referred to as the Zeeman states.

**Precession.**[6]    The absorbed or released energy by the proton spin system upon a transition between the two states according to (2.15) is given by

$$\Delta E = \gamma \hbar B_0 = \gamma \omega_0. \tag{2.16}$$

Here we have defined the so-called Larmor frequency

$$\omega_0 := \gamma B_0. \tag{2.17}$$

---

[5]Here we neglect kinetic energy terms.
[6][12, 16, 17] serve as general references for this paragraph.

In the classical vector model, the Larmor frequency corresponds to the precession of the magnetic moment vector around $\boldsymbol{B}_0$ (see appendix A.2). Note, however, that in the quantum mechanical description $\omega_0$ is related to a state transition. Still, the time-dependent Schrödinger equation

$$\hat{H}\ket{\Psi(t)} = i\hbar \frac{d}{dt}\ket{\Psi(t)}, \tag{2.18}$$

can be used to derive a quantum mechanical precession. The general solution $\ket{\Psi(t)}$ of (2.18) with $\hat{H}$ from (2.12) is a linear combination of the orthonormal solutions $\ket{\Psi_\uparrow}$ and $\ket{\Psi_\downarrow}$ of the time-independent Schrödinger equation,

$$\ket{\Psi(t)} = \alpha_\uparrow \ket{\Psi_\uparrow} + \alpha_\downarrow \ket{\Psi_\downarrow}, \tag{2.19}$$

$$\alpha_\uparrow := a_\uparrow e^{-iE_\uparrow t/\hbar}, \quad \alpha_\downarrow := a_\downarrow e^{-iE_\downarrow t/\hbar}. \tag{2.20}$$

$\ket{\Psi(t)}$ must be normalized, i.e. $|\alpha_\uparrow|^2 + |\alpha_\downarrow|^2 = 1$. With the relations

$$\hat{\mu}_z \ket{\Psi_\uparrow} \overset{(2.5a)(2.9)}{\underset{(2.10)}{=}} \gamma \frac{\hbar}{2} \ket{\Psi_\uparrow}, \tag{2.21a}$$

$$\hat{\mu}_z \ket{\Psi_\downarrow} \overset{(2.5b)(2.9)}{\underset{(2.10)}{=}} -\gamma \frac{\hbar}{2} \ket{\Psi_\downarrow}, \tag{2.21b}$$

the time evolution of the magnetic moment's z-component is given by the expectation value

$$\begin{aligned}
\langle \hat{\mu}_z \rangle &= \bra{\Psi(t)} \hat{\mu}_z \ket{\Psi(t)} \\
&= \bra{\alpha_1 \Psi_\uparrow + \alpha_2 \Psi_\downarrow} \hat{\mu}_z \ket{\alpha_1 \Psi_\uparrow + \alpha_2 \Psi_\downarrow} \\
&= \gamma \frac{\hbar}{2}(|\alpha_\uparrow|^2 - |\alpha_\downarrow|^2) \\
&\overset{(2.20)}{=} \gamma \frac{\hbar}{2}(|a_\uparrow|^2 - |a_\downarrow|^2)
\end{aligned} \tag{2.22}$$

Similar considerations (also see appendix A.3) lead to

$$\hat{\mu}_x \ket{\Psi_\uparrow} = \gamma \frac{\hbar}{2} \ket{\Psi_\downarrow}, \quad \hat{\mu}_x \ket{\Psi_\downarrow} = \gamma \frac{\hbar}{2} \ket{\Psi_\uparrow}, \tag{2.23}$$

$$\langle \hat{\mu}_x \rangle = \gamma \hbar |a_\uparrow||a_\downarrow| \cos(\omega_0 t + \varphi), \tag{2.24}$$

$$\hat{\mu}_y \ket{\Psi_\uparrow} = i\gamma \frac{\hbar}{2} \ket{\Psi_\downarrow}, \quad \hat{\mu}_y \ket{\Psi_\downarrow} = -i\gamma \frac{\hbar}{2} \ket{\Psi_\uparrow}, \tag{2.25}$$

$$\langle \hat{\mu}_y \rangle = -\gamma \hbar |a_\uparrow||a_\downarrow| \sin(\omega_0 t + \varphi), \tag{2.26}$$

with the phase factor $\varphi$ determined by initial conditions.

We conclude that the expectation value of the magnetic moment's z-component is time-independent, whereas the x- and y-component perform a clockwise precessional movement in the xy-plane with frequency $\omega_0$.

**Magnetization.**   In MRI we are not interested in the behavior of a single spin but in the properties of the spin ensemble inside a voxel.[7] Therefore, it is convenient to introduce the magnetization operator

$$\hat{\boldsymbol{m}} = \frac{\int_{\mathcal{V}} \rho(\boldsymbol{r})\hat{\boldsymbol{\mu}} \, d^3\boldsymbol{r}}{\mathcal{V}}, \tag{2.27}$$

with $\rho(\boldsymbol{r})$ being the proton density in a macroscopic voxel volume $\mathcal{V}$. In general, there is no fixed phase relationship between the precessing magnetic moments of spins, thus there is no net magnetization in x- and y-direction. Hence, the expectation value for the bulk magnetization in thermal equilibrium is constant and points along the positive z-direction

$$\langle \hat{\boldsymbol{m}} \rangle = \langle \hat{m}_z \rangle \boldsymbol{e}_z := m_0 \boldsymbol{e}_z. \tag{2.28}$$

Moreover, in thermal equilibrium, the probability of finding a particle in one of the eigenstates states $|\uparrow\rangle$ or $|\downarrow\rangle$ is governed by the Boltzmann distribution[8]

$$p_\uparrow = \frac{e^{-E_\uparrow/k_B T}}{e^{-E_\uparrow/k_B T} + e^{-E_\downarrow/k_B T}}, \quad p_\downarrow = \frac{e^{-E_\downarrow/k_B T}}{e^{-E_\uparrow/k_B T} + e^{-E_\downarrow/k_B T}}, \tag{2.29}$$

with $k_B$ the Boltzmann constant and $T$ the temperature. For $N$ nuclei of type $^1$H, $T = 293\,\text{K}$ and an external magnetic field of $B_0 = 3\,\text{T}$, the relative population excess of the lower energy state $|\uparrow\rangle$ is given by

$$\Delta N/N = p_\uparrow - p_\downarrow \underset{(2.29)}{\overset{(2.15)}{\approx}} 10^{-5}. \tag{2.30}$$

Hence, at room temperature the bulk magnetization $m_0$ is reduced by a factor of approximately $10^{-5}$ compared to a system where all spins populate the lower energy state.

In MRI we want to detect the magnetization because it contains information about the investigated tissue in a voxel (see (2.27)). It can be measured using a technique called RF spin tipping, which we introduce in the upcoming section.

---

[7] A voxel is a volume element and the 3D analogue to the pixel in 2D.

[8] For a rigorous derivation of the distribution of spins over energy states one has to use the Fermi-Dirac statistics (Fermions, i.e. half-integer spin particles) or the Bose-Einstein statistics (bosons, i.e. integer spin particles). However, at room temperature, i.e. in the high-temperature approximation, they both approach the Boltzmann distribution [18].

## 2.1.4. Radiofrequency spin tipping[9]

We introduce the theory of spin tipping for a single spin before we generalize it to a spin ensemble. In contrast to the previous section, we now consider time-dependent magnetic fields

$$\boldsymbol{B}(t) := \boldsymbol{B}_0 + \boldsymbol{B}_{\mathrm{ad}}(t). \tag{2.31}$$

Here $\boldsymbol{B}_0 := B_0\boldsymbol{e}_z$ is constant and $\boldsymbol{B}_{\mathrm{ad}}(t) := B_x(t)\boldsymbol{e}_x + B_y(t)\boldsymbol{e}_y$ is an additional field that varies in the xy-plane. To solve the time-dependent Schrödinger equation (2.18) we have to introduce an additional time dependency in ansatz (2.19),

$$|\Psi(t)\rangle = \beta_\uparrow |\Psi_\uparrow\rangle + \beta_\downarrow |\Psi_\downarrow\rangle, \tag{2.32}$$

$$\beta_\uparrow := b_\uparrow(t)e^{-iE_\uparrow t/\hbar}, \quad \beta_\downarrow := b_\downarrow(t)e^{-iE_\downarrow t/\hbar}. \tag{2.33}$$

This yields two independent differential equations

$$(2/\gamma)i\dot{\beta}_\uparrow = -\beta_\uparrow B_z - \beta_\downarrow(B_x - iB_y), \tag{2.34a}$$

$$(2/\gamma)i\dot{\beta}_\downarrow = \beta_\downarrow B_z - \beta_\uparrow(B_x + iB_y). \tag{2.34b}$$

We choose the magnetic field components to be $B_x \sim \cos(\omega t)$ and $B_y \sim -\sin(\omega t)$, i.e. a clockwise rotating magnetic field, and define $B_x \pm iB_y := Fe^{\mp i\omega t}$. Transforming (2.34) with (2.15), (2.17) and (2.33) yields

$$\dot{b}_\uparrow = i\frac{\gamma}{2}b_\downarrow Fe^{-i(\omega_0-\omega)t}, \tag{2.35a}$$

$$\dot{b}_\downarrow = i\frac{\gamma}{2}b_\uparrow Fe^{i(\omega_0-\omega)t}. \tag{2.35b}$$

We can further simplify these equations by setting $\omega \overset{!}{=} \omega_0$, i.e. the magnetic field is chosen to oscillate with the Larmor frequency. By defining $\Omega := (\gamma/2)F$ we get

$$\dot{b}_\uparrow = i\Omega b_\downarrow, \tag{2.36a}$$

$$\dot{b}_\downarrow = i\Omega b_\uparrow, \tag{2.36b}$$

and find the solutions

$$b_\uparrow = c \cdot \sin(\Omega t + \varphi), \quad b_\downarrow = -c \cdot i\cos(\Omega t + \varphi), \tag{2.37}$$

---

[9][7, 19] serve as general references for this section.

with a constant $c$ and a phase factor $\varphi$ defined by initial conditions. With (2.37), (2.32) and (2.33) we obtain a solution for the time-dependent Schrödinger equation

$$|\Psi(t)\rangle = \sin(\Omega t + \varphi)e^{i\omega_0 t/2}|\Psi_\uparrow\rangle - i\cos(\Omega t + \varphi)e^{-i\omega_0 t/2}|\Psi_\downarrow\rangle. \tag{2.38}$$

In analogy to section 2.1.3 we derive

$$\langle\hat{\mu}_x\rangle = -\gamma\frac{\hbar}{2}\sin(2\Omega t + \varphi')\sin(\omega_0 t), \tag{2.39}$$

$$\langle\hat{\mu}_y\rangle = -\gamma\frac{\hbar}{2}\sin(2\Omega t + \varphi')\cos(\omega_0 t), \tag{2.40}$$

$$\langle\hat{\mu}_z\rangle = -\gamma\frac{\hbar}{2}\cos(2\Omega t + \varphi'), \tag{2.41}$$

with $\varphi'$ a phase factor determined by the initial conditions. We find the expectation value of the magnetic moment's z-component $\langle\hat{\mu}_z\rangle$ to be no longer constant but to vary with time. Equations (2.39)-(2.41) describe a nutation, i.e. the magnetic moment's z-component is tipped away from the z-axis. Hence, the size of the flip angle depends on the strength and duration of $\boldsymbol{B}_{ad}(t)$. In NMR experiments, this additional field is provided by means of a RF wave and is known as RF excitation pulse.

It is important to note that all tipped magnetic moments will be in phase with one another. As soon as $\boldsymbol{B}_{ad}$ is switched off, they will precess as described in section 2.1.3, but due to the fixed phase relationship, the x- and y-component of the bulk magnetization will not cancel out but oscillate, $\langle\hat{m}_x + i\hat{m}_y\rangle \sim e^{-i\omega_0 t}$. Basic electrodynamics tells us that an oscillation magnetization produces an electromagnetic wave which we can measure.

## 2.1.5. Bloch equation

Using the definition (2.31) for $\boldsymbol{B}$ and considering the magnetization (2.27) in a voxel rather than the magnetic moment of single spins, we can derive an alternative representation for equations (2.39)-(2.41),

$$\langle\dot{\hat{m}}_x\rangle = \gamma(\langle\hat{m}_y\rangle B_z - \langle\hat{m}_z\rangle B_y), \tag{2.42}$$

$$\langle\dot{\hat{m}}_y\rangle = \gamma(\langle\hat{m}_z\rangle B_x - \langle\hat{m}_x\rangle B_z), \tag{2.43}$$

$$\langle\dot{\hat{m}}_z\rangle = \gamma(\langle\hat{m}_x\rangle B_y - \langle\hat{m}_y\rangle B_x), \tag{2.44}$$

or in vector notation

$$\langle \dot{\hat{\boldsymbol{m}}} \rangle = \gamma \langle \hat{\boldsymbol{m}} \rangle \times \boldsymbol{B}. \tag{2.45}$$

In 1946, Felix Bloch developed in a more general form the classical analogue of this equation for a macroscopic magnetization using the vector model [20]. This Bloch equation is given by

$$\frac{d}{dt} \begin{pmatrix} \langle \hat{m}_x \rangle \\ \langle \hat{m}_y \rangle \\ \langle \hat{m}_z \rangle \end{pmatrix} = \gamma \begin{pmatrix} \langle \hat{m}_x \rangle \\ \langle \hat{m}_y \rangle \\ \langle \hat{m}_z \rangle \end{pmatrix} \times \begin{pmatrix} B_x \\ B_y \\ B_z \end{pmatrix} + \begin{pmatrix} -\frac{1}{T_2}\langle \hat{m}_x \rangle \\ -\frac{1}{T_2}\langle \hat{m}_y \rangle \\ \frac{m_0 - \langle \hat{m}_z \rangle}{T_1} \end{pmatrix}, \tag{2.46}$$

with $\langle \hat{\boldsymbol{m}} \rangle = \left( \langle \hat{m}_x \rangle, \langle \hat{m}_y \rangle, \langle \hat{m}_z \rangle \right)^T$ the magnetization's expectation value and $m_0$ the thermal equilibrium value of $\langle \hat{m}_z \rangle$ in the presence of $\boldsymbol{B}_0 = B_0 \boldsymbol{e}_z$ only. This equation considers relaxation effects governed by the factors $T_1$ and $T_2$ [8, 10]. The relaxation time $T_1$ describes the spin-lattice or longitudinal relaxation. It characterizes the time for $\langle \hat{m}_z \rangle$ to recover its equilibrium value $m_0$ after excitation,

$$\langle \hat{m}_z(t) \rangle = m_0(1 - e^{t/T_1}). \tag{2.47}$$

The relaxation time $T_2$ describes the spin-spin or transversal relaxation.[10] It is a measure for the decay of the magnetization's transverse component $\langle \hat{m}_{xy} \rangle = \sqrt{\langle \hat{m}_x \rangle^2 + \langle \hat{m}_y \rangle^2}$. The reason for this decline is the loss of coherence between the single precessing magnetic moments. This phenomena is also called free induction decay and follows approximately the exponential law

$$\langle \hat{m}_{xy}(t) \rangle = \langle \hat{m}_{xy}^{\max} \rangle e^{-t/T_2}. \tag{2.48}$$

## 2.1.6. Conclusion for MRI

Although nuclear magnetic resonance is a complex quantum mechanical phenomenon, magnetic resonance imaging can be understood with only a few statements that summarize the previous sections. (i) In an external magnetic field, the equilibrated magnetization of protons is pointing towards the field's axis. (ii) By applying a transversal RF excitation pulse in resonance, the magnetization is tipped to the xy-plane where it starts to precess and thus emits a RF wave. The flip angle is determined by the pulse duration and amplitude. (iii) The emitted wave can be detected and analyzed. It provides information about the magnetization and therefore about the investigated object.

---

[10]Sometimes $T_2^*$ instead of $T_2$ is used if dephasing due to magnetic field inhomogeneities is explicitly considered.

## 2.2. The pulsed NMR experiment

The aim of MRI is to obtain spatially resolved information about the tissue of a patient or phantom. In this section we introduce the basic setup of a pulsed NMR experiment, describe how we achieve spatial resolution and introduce the fundamentals of image reconstruction.

### 2.2.1. Experimental setup

The MR scanner is composed of several types of coils (see Fig. 2.1). The main magnetic coil generates a strong and homogeneous external magnetic field $B_0$. Shim coils are used to eliminate field inhomogeneities. The RF excitation coils are used to apply RF pulses to tip the magnetization. RF receiver coils are utilized to detect the RF waves emitted by the precessing magnetization. Gradient coils for each axis x, y and z are used for slice selection and for spatial encoding, as described in the following sections.

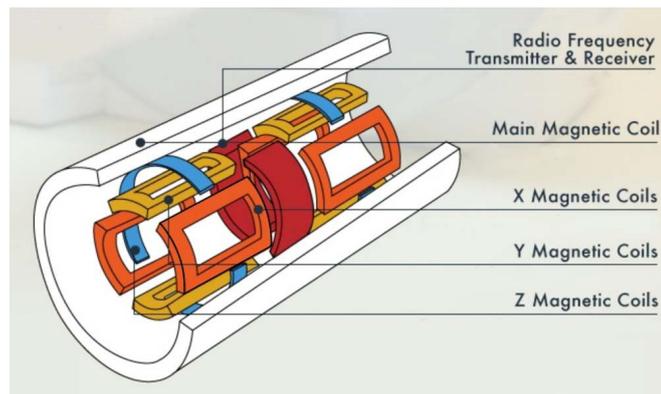

Fig. 2.1: Schematic of a MRI coil system. Adapted from [21].

### 2.2.2. Slice selection

In MRI we have the possibility to either acquire and reconstruct a full 3D data set or to choose one or more specific slices of interest. The latter will be the topic of this thesis. The basic idea of slice selection is to switch on an additional linear field gradient (slice selection gradient) that causes a variation in the Larmor frequency of the spins in gradient direction. We then use a bandwidth limited RF pulse to excite a slice of certain thickness.

**Slice selection gradient.** We denote the additional linear field gradient

$$G_{ss} := G_{ss}e_{ss},\tag{2.49}$$

with gradient direction $e_{ss}$. The corresponding additional magnetic field is

$$B_{ss}(r) = (G_{ss} \cdot r)e_z,\tag{2.50}$$

where $r$ denotes the spatial position. The total magnetic field can be written as[11]

$$B(r) = (B_0 + B_{ss}(r)) \overset{(2.11)(2.50)}{=} (B_0 + G_{ss} \cdot r)e_z.\tag{2.51}$$

The corresponding Larmor frequency also becomes spatially dependent,

$$\omega_L(r) \overset{(2.17)}{=} \gamma|B(r)| \overset{(2.51)}{=} \gamma(B_0 + G_{ss} \cdot r) = \omega_0 + \gamma\, G_{ss} \cdot r.\tag{2.52}$$

Without loss of generality we assume that we want to select a slice with thickness $\Delta z$ perpendicular to the z-axis. We therefore have to vary the field strength along the z-direction and the gradient is given by $G_{ss} = (0, 0, G_z)^T = G_z e_z$. Hence, the magnetic field strength only depends on $z$,

$$B(r) \overset{(2.51)}{=} B(z)e_z = (B_0 + B_{ss}(z))e_z \overset{(2.50)}{=} (B_0 + G_z z)e_z.\tag{2.53}$$

Likewise, the Larmor frequency (2.52) varies along the z-axis. Let $z_c$ denote the slice center. Then the corresponding Larmor frequency at the center of the slice is

$$\omega_c = \gamma(B_0 + G_z z_c).\tag{2.54}$$

Inside the slice we obtain a total frequency deviation in z-direction given by

$$\Delta\omega = \gamma G_z \Delta z,\tag{2.55}$$

---

[11]Strictly speaking, equation (2.51) is an approximation for high magnetic field strength. The Maxwell equations $\nabla \cdot B(r) = 0$ and $\nabla \times B(r) = 0$ for a static magnetic field imply that an inhomogeneous field cannot have a single non-zero component. Hence, if we desire an inhomogeneous magnetic field which points in, e.g. $e_z$-direction (see (2.53)), the actual magnetic field which prevails in the scanner will also include concomitant components that point in $e_x$- and $e_y$-direction. However, these fields are inversely proportional to the external field $B_0$ and as such we can neglect them for the field strength $B_0 = 3\,\text{T}$ which is used in the scope of this thesis [9]. For very low magnetic field MRI, the concomitant fields must be explicitly taken into account as they cause artifacts [22].

which explicitly depends on the slice thickness and the gradient strength. A visualization of the relationship of $B_{ss}$, $\Delta z$ and $\Delta \omega$ is given in Fig. 2.2.

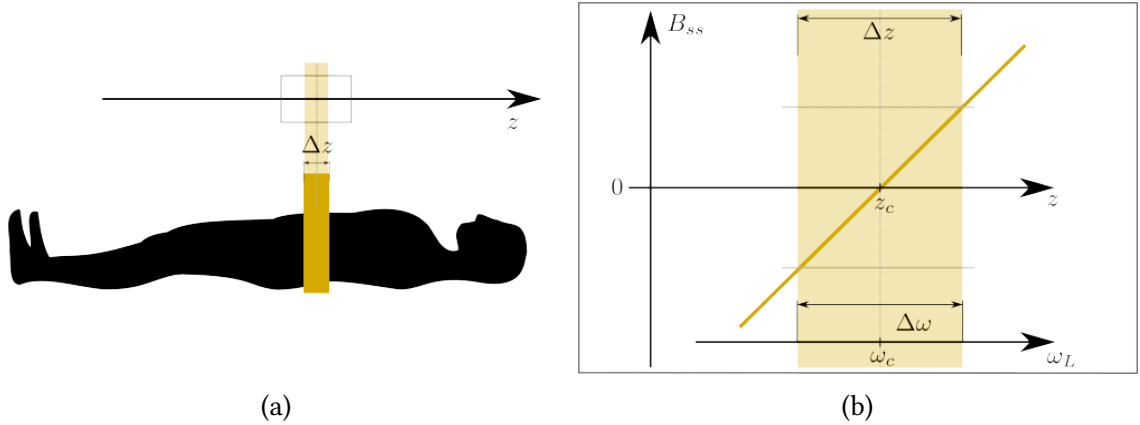

(a)

(b)

Fig. 2.2: Schematic of the quantities relevant for slice selection. $\Delta z$: Slice thickness. $z_c$: Slice center. $B_{ss}$: Additional magnetic gradient field strength. $\omega_L$: Larmor frequency. $\omega_c$: Larmor frequency at the slice center. $\Delta \omega$: Total frequency interval inside the slice. (b) shows a zoomed version of (a).

**Radiofrequency excitation pulse.**[12]    The Larmor frequency gradient inside the probe in combination with a frequency selective RF pulse

$$B_{\mathrm{rf}}(t) := B_{\mathrm{env}}(t) e^{-i\omega_{\mathrm{rf}} t} \tag{2.56}$$

can be used for slice selective excitation.

For a rigorous derivation of the actual pulse shape we have to consider the Bloch equation (2.46). The relaxation term can be neglected, as typical pulse durations are much smaller than the relaxation times $T_1$ and $T_2$. Calculations become more feasible if we transform the Bloch equation into a rotating frame system (see appendix A.4) with an angular frequency that matches the RF pulse frequency $\omega_{\mathrm{rf}}$.

$$\left( \frac{d}{dt} \langle \hat{\boldsymbol{m}} \rangle \right)_{\mathrm{r}} = \gamma \langle \hat{\boldsymbol{m}} \rangle \times \begin{pmatrix} B_{\mathrm{env}}(t) \\ 0 \\ B(z) - \omega_{\mathrm{rf}}/\gamma \end{pmatrix}, \tag{2.57}$$

with the position dependent magnetic field $B(z) := B_0 + G_z z$. Without loss of generality, we assume that $B_{\mathrm{env}}(t)$ is applied along the x-axis in the rotating frame. By introducing

---

[12] [8, 9] serve as general references for this paragraph.

the frequency offset

$$\Delta\Omega(z) := \gamma B(z) - \omega_{rf} \qquad (2.58)$$

we can write (2.57) explicitly as[13]

$$\langle \dot{\hat{m}}_x \rangle = \Delta\Omega \langle \hat{m}_y \rangle, \qquad (2.59a)$$

$$\langle \dot{\hat{m}}_y \rangle = \gamma B_{env}(t) \langle \hat{m}_z \rangle - \Delta\Omega \langle \hat{m}_x \rangle, \qquad (2.59b)$$

$$\langle \dot{\hat{m}}_z \rangle = -\gamma B_{env}(t) \langle \hat{m}_y \rangle. \qquad (2.59c)$$

We define the complex transverse magnetization

$$\langle \hat{m}_\perp \rangle := \langle \hat{m}_x \rangle + i \langle \hat{m}_y \rangle \qquad (2.60)$$

and reexpress (2.59a) and (2.59b) as

$$\frac{d}{dt} \langle \hat{m}_\perp \rangle = -i\Delta\Omega \langle \hat{m}_\perp \rangle + i\gamma B_{env}(t) \langle \hat{m}_z \rangle. \qquad (2.61)$$

The solution to this equation with initial conditions $\langle \hat{m}_\perp(t) \rangle|_{t=0} = 0$ was given by P.M. Joseph [23]

$$\langle \hat{m}_\perp(t) \rangle = i\gamma e^{-i\Delta\Omega t} \int_0^t \langle \hat{m}_z(\tau) \rangle B_{env}(\tau) e^{i\Delta\Omega \tau} d\tau. \qquad (2.62)$$

Often the approximation $\langle \hat{m}_z(t) \rangle \approx m_0$ for small flip angles $\alpha$ is used.[14] We then obtain

$$|\langle \hat{m}_\perp \rangle| = \sqrt{\langle \hat{m}_x \rangle^2 + \langle \hat{m}_y \rangle^2} \approx \gamma m_0 \left| \int_0^t B_{env}(\tau) e^{i\Delta\Omega \tau} d\tau \right|. \qquad (2.63)$$

Basic geometry and the small flip angle approximation yields the excitation profile

$$\sin\alpha(\Delta\Omega) \approx \frac{|\langle \hat{m}_\perp \rangle|}{m_0} \approx \alpha(\Delta\Omega) \stackrel{(2.63)}{\approx} \gamma \left| \int_0^t B_{env}(\tau) e^{i\Delta\Omega \tau} d\tau \right|. \qquad (2.64)$$

The integral interval $[0, t]$ can be replaced by $[-\infty, \infty]$ because $B_{rf}(t) \equiv 0$ except during the pulse. Hence, the excitation profile is given by the absolute value of the pulse envelope's Fourier transform.

By setting $\omega_{rf} \stackrel{!}{=} \omega_c$ in (2.58) we get $\Delta\Omega(z) = \omega(z) - \omega_c$. Then, in a slice with thickness

---

[13]For convenience we drop the index r which denotes the rotating frame system.
[14]This approximation holds well for flip angles up to 30° [9].

$\Delta z$ we find all frequencies of the interval

$$I_\omega := \left[ \omega_c - \frac{\Delta\omega}{2}, \omega_c + \frac{\Delta\omega}{2} \right].$$ (2.65)

The excitation profile for ideal slice selection is therefore a boxcar function

$$\alpha(\Delta\Omega) \sim \Pi\left(\frac{\omega - \omega_c}{\Delta\omega}\right) := \left\{ \begin{array}{ll} 1 & \text{for } \omega \in I_\omega \\ 0 & \text{else} \end{array} \right. .$$ (2.66)

According to (2.64) we obtain the corresponding time-domain representation of the envelope function by Fourier transforming (2.66)

$$B_{\text{env}}(t) \sim \int_{-\infty}^{\infty} \Pi\left(\frac{\omega - \omega_c}{\Delta\omega}\right) e^{i\omega t} d\omega \sim \text{sinc}\left(\frac{\Delta\omega}{2}t\right).$$ (2.67)

To sum up, for slice selective excitation we apply a magnetic field gradient of a certain strength perpendicular to the desired slice and excite the spins with a transversal RF pulse of type

$$B_{\text{rf}}(t) \sim \text{sinc}\left(\frac{\Delta\omega}{2}t\right) e^{-i\omega_c t}.$$ (2.68)

However, a sinc-function is not a feasible choice for the envelope as it has infinite length. We therefore have to further modify the envelope, which is the topic of the next section.

### 2.2.3. Pulse envelope truncation effects

The ideal excitation pulse envelope as suggested by (2.68) is an infinite sinc. We define its discrete representation as

$$B_{\text{env}}^{\text{sinc}}(t_i) := \text{sinc}\left(\frac{\Delta\omega}{2}t_i\right),$$ (2.69)

where $t_i$ are discrete times. The excitation pulse must be constrained to a finite duration $T$, i.e. $t_i \in [-T/2, T/2]$. However, the frequency domain representation of a cropped sinc pulse shows a heaped appearance and amplification of side-lobes. Better results can be accomplished by using window-functions for truncation [24]. One popular choice is the Hann function

$$W(t_i) := \frac{1}{2}\left(1 + \cos\left(\frac{2\pi t_i}{T}\right)\right).$$ (2.70)

The modified envelope function of a RF excitation pulse is then given by

$$B_{\mathrm{env}}(t_i) = W(t_i) \cdot B_{\mathrm{env}}^{\mathrm{sinc}}(t_i)$$
$$= \frac{1}{2} \left( 1 + \cos\left( \frac{2\pi t_i}{T} \right) \right) \cdot \mathrm{sinc}\left( \frac{\Delta\omega}{2} t_i \right). \tag{2.71}$$

The benefit of using a window function over simply cropping the sinc to the interval $I_T := [-T/2, T/2]$ is illustrated in Fig. 2.3.

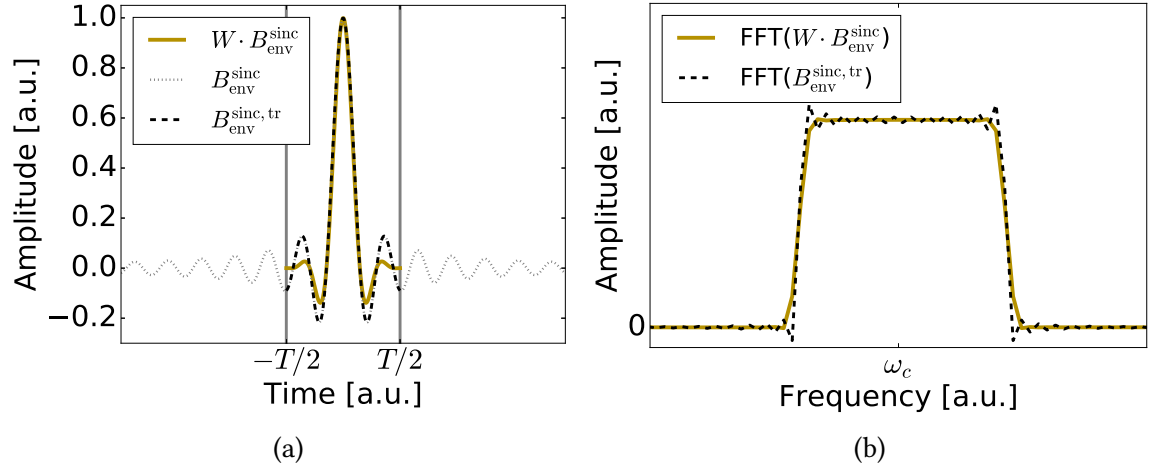

(a)        (b)

Fig. 2.3: The effect of windowing in time and frequency space. (a) shows an infinite sinc $B_{\mathrm{env}}^{\mathrm{sinc}}$, a cropped sinc $B_{\mathrm{env}}^{\mathrm{sinc,tr}}$ and Hann-windowed sinc $W \cdot B_{\mathrm{env}}^{\mathrm{sinc}}$. The vertical lines imply the finite pulse duration. (b) depicts the real part of the respective Fourier transforms.

In Fig. 2.3a an infinite sinc and its cropped (truncated) version is shown together with the Hann-windowed sinc. The actual pulse duration is limited to the interval $I_T$ which is indicated by the vertical lines. Apparently the windowing ensures a smooth decrease of amplitude towards the limits of the interval. Fig. 2.3b shows the discrete Fourier transforms of the Hann-windowed and the cropped sinc. The latter shows distinct side lobes, which is a typical effect of truncation. By contrast, the Hann-windowed sinc does not show any undesired oscillations and closely resembles a boxcar function, which is the Fourier transform of an infinite sinc.

## 2.2.4. Spatial encoding and signal equation

In MRI the objective is a spatially resolved image, i.e. to obtain information about the magnetization in every voxel. In this section we show how spatial information can be encoded in precessing transversal magnetization.

There are two possible encoding techniques dubbed frequency-encoding and phase-encoding. In both cases we need to apply an additional linear magnetic field gradient $B_{\mathrm{grad}}(\boldsymbol{r}, t) = (\boldsymbol{G}_{\mathrm{grad}}(t) \cdot \boldsymbol{r}) \boldsymbol{e}_z$. The total magnetic field composed of the static external field $B_0$ and the additional gradient field $B_{\mathrm{grad}}(\boldsymbol{r}, t)$ can then be written as

$$\boldsymbol{B}(\boldsymbol{r}, t) = (B_0 + \boldsymbol{G}_{\mathrm{grad}}(t) \cdot \boldsymbol{r}) \boldsymbol{e}_z. \tag{2.72}$$

With this position and time-dependent magnetic field, we can adjust phase and frequency of the precessing transversal magnetization at position $\boldsymbol{r}$. The RF wave emitted by the encoded precessing magnetization can be used to recover $m(\boldsymbol{r})$. In the following we describe the encoding process and neglect relaxation effects.

**Intuitive explanation.** Before giving a mathematical description, a more intuitive explanation is presented. For simplicity we consider a 2D slice of excited spins that we want to acquire with resolution $N_{\mathrm{RO}} \times N_{\mathrm{PE}}$. In the schematic Fig. 2.4 we choose $N_{\mathrm{RO}} = N_{\mathrm{PE}} = 3$. For each of the nine pixels we define a transversal magnetization vector $\langle \hat{m}_\perp(\boldsymbol{r}, t) \rangle$ represented by an arrow. If spins are excited by a RF pulse, the magnetizations for each pixel are in phase (Fig. 2.4, *top-left*). For phase-encoding, we switch on a gradient field in phase-encoding-direction (PE-direction) for a short period of time. Afterwards all spins possess the same frequency again but their phasing differs depending on their position in gradient direction (Fig. 2.4, *top-right*). In case of frequency-encoding, we use the gradient field in read-out-direction (RO-direction) to make the precession frequency spatially dependent (Fig. 2.4, *bottom-left*). In practice both techniques are combined (Fig. 2.4, *bottom-right*). First, in PE-direction phase-encoding is performed.[15] Then frequency-encoding is done during signal acquisition, i.e. we apply a gradient in RO-direction while the receiver coils record the emitted radiation. We can separate the signals coming from magnetizations with different frequencies using a Fourier transform. This yields spatial resolution in RO-direction. However, we do not have enough information to distinguish the different phases. Hence, to obtain spatial resolution in PE-direction we need to perform the same excitation and encoding procedure $N_{\mathrm{PE}}$ times, where each time we vary the phase-encoding gradient. With this set of data we are able to calculate the magnetization in each pixel. This idea can be expressed in a formal way.

---

[15] For 3D acquisition phase-encoding is performed in PE- and z-direction.

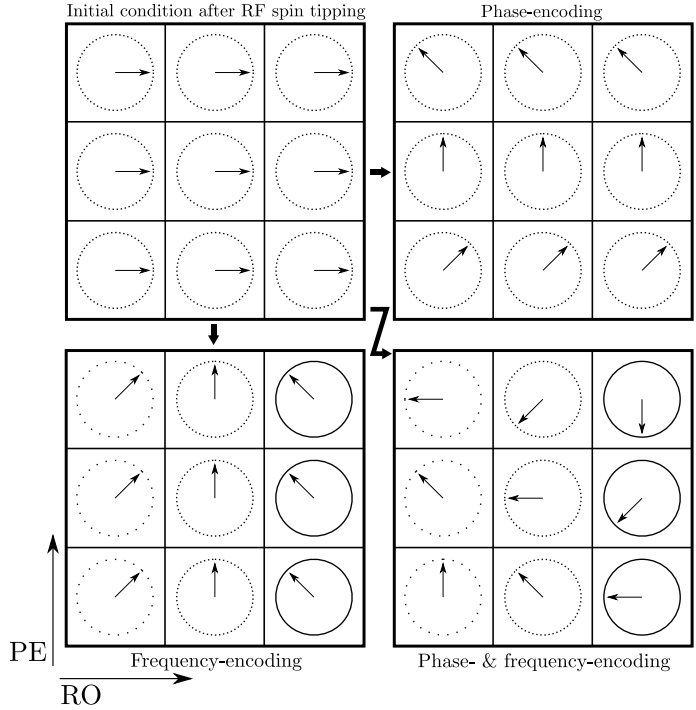

Fig. 2.4: Schematic of phase-encoding (PE-direction) and frequency-encoding (RO-direction) for a 2D slice containing nine precessing magnetization vectors represented by arrows. The phase is given by the arrow's orientation. The more dense the dots of the circle, the faster is the respective precessing frequency. *Top-left*: Initial condition: All vectors in phase. *Top-right*: After phase-encoding: The phases of the vectors differ in PE-direction. *Bottom-left*: During frequency-encoding: The precessing frequency differs in RO-direction. *Bottom-right*: Combined phase- and frequency-encoding: Each vector has unique phase-frequency values.

**Mathematical description.** For presentational purpose we assume an excitation flip angle of $\alpha = 90°$, so the entire equilibrium magnetization in z-direction $m_0(\boldsymbol{r})$ is tipped into the xy-plane. In the static laboratory-frame, the corresponding Bloch equation (2.46) without relaxation terms reads

$$\frac{d}{dt}\begin{pmatrix} \langle \hat{m}_x \rangle \\ \langle \hat{m}_y \rangle \\ 0 \end{pmatrix} = \gamma \begin{pmatrix} \langle \hat{m}_x \rangle \\ \langle \hat{m}_y \rangle \\ 0 \end{pmatrix} \times \begin{pmatrix} 0 \\ 0 \\ B_z \end{pmatrix}. \tag{2.73}$$

Using (2.60) we can write

$$\frac{d}{dt}\langle \hat{m}_\perp \rangle = -i\gamma B_z \langle \hat{m}_\perp \rangle, \tag{2.74}$$

with $B_z \stackrel{(2.72)}{=} B_0 + G_{\text{grad}}(t) \cdot r$. Equation (2.74) is a first-order linear ordinary differential equation with the well known solution

$$\langle \hat{m}_\perp(r, t) \rangle = m(r) e^{-i(\omega_0 t + \varphi(r, t))}, \tag{2.75}$$

with $\omega_0$ from (2.17). This equation describes the precession of the transverse magnetization. The phase factor $\varphi(r, t)$ is given by

$$\varphi(r, t) = \int_0^t \gamma G_{\text{grad}}(\tau) \cdot r \, d\tau. \tag{2.76}$$

With the definition

$$k(t) := \int_0^t \gamma G_{\text{grad}}(\tau) \, d\tau \tag{2.77}$$

we can write

$$\langle \hat{m}_\perp(r, t) \rangle = m(r) e^{-i(\omega_0 t + k(t) \cdot r)}. \tag{2.78}$$

The k-space trajectory $k(t)$ is determined by the time-dependent choice of the additional gradient fields $G_{\text{grad}}(\tau)$. With an appropriate choice of phase- and frequency-encoding, we can make the trajectory traverse large parts of k-space. For more details see section 2.2.6.

All voxels with excited spins contribute to the time-dependent complex signal $U(t)$ induced in the receiver coils of the scanner. We can express this signal by the integral

$$U(t) = \int m(r) e^{-i(\omega_0 t + k(t) \cdot r)} dr. \tag{2.79}$$

The Larmor frequency $\omega_0$ is known and can be demodulated by a quadrature detector. The actually measured signal is then given by

$$S(t) = \int m(r) e^{-ik(t) \cdot r} dr. \tag{2.80}$$

This so called signal equation shows a Fourier relationship between the recorded signal $S(t)$ and the underlying magnetization $m(r)$. Modern MRI scanner possess not only one but several receiver coils, each with a certain complex sensitivity $c^j(r)$. The respective signal equation for coil $j$ is

$$S^j(t) = \int c^j(r) m(r) e^{-ik(t) \cdot r} dr. \tag{2.81}$$

Given this signal, we have all information that we need to reconstruct the magnetization $m(\boldsymbol{r})$.

### 2.2.5. Fundamental idea of image reconstruction

To make the idea behind image reconstruction more clear, we rewrite the signal equation (2.80). Instead of interpreting the recorded signal as a function of time, we can denote

$$S(t) \stackrel{(2.80)}{=} S(\boldsymbol{k}(t)) = S(\boldsymbol{k}) = \mathcal{F}\left(m(\boldsymbol{r})\right). \tag{2.82}$$

Here, $\mathcal{F}$ stands for the Fourier transform

$$\mathcal{F}(m(\boldsymbol{r})) := \int m(\boldsymbol{r})e^{-i\boldsymbol{k}\cdot\boldsymbol{r}}d\boldsymbol{r} \tag{2.83}$$

For all future considerations it is more natural to no longer speak of a time signal $S(t)$ but of a k-space with complex intensity $S(\boldsymbol{k})$ at spatial frequency position $\boldsymbol{k}$. This k-space is related to the magnetization in image space by an inverse Fourier transform

$$m(\boldsymbol{r}) = \mathcal{F}^{-1}(S(\boldsymbol{k})) := \frac{1}{(2\pi)^D} \int S(\boldsymbol{k})e^{i\boldsymbol{k}\cdot\boldsymbol{r}}d\boldsymbol{k}, \tag{2.84}$$

where the power $D$ in the normalization is given by the dimension of the integral. Equation (2.84) is fundamental for all image reconstruction strategies in MRI.

### 2.2.6. k-Space sampling and sequence design[16]

In MRI we do not deal with continuous but discretized signals, i.e. a sampled k-space. This k-space is acquired along a trajectory determined by (2.77). We have to choose a sensible trajectory and collect sufficient samples along that path in k-space to make image reconstruction feasible. Commonly, a Cartesian sampling scheme is used where the k-space is sampled line by line. Thus, a Fast Fourier transform (FFT) for image reconstruction according to (2.84) can be applied directly, since the data lie on a Cartesian grid.

We realize the Cartesian trajectory using a 2D FLASH sequence [25] (Fig. 2.5a): We first perform slice selection using a RF excitation pulse (a) and a slice selection gradient ($b_1$). A rewinder gradient (b2) is added to compensate for the phase evolution caused by the slice selection gradient. At this point all spins are in phase, thus the trajectory starts in the center of k-space. Then, the phase-encoding gradient (d) is applied together with a dephase lobe ($c_1$) of the read gradient. Hence, the trajectory proceeds diagonally into the negative RO-direction. The position $k_{PE}$ is determined by the strength of the phase-encoding gradient. Now, the frequency-encoding or read-out gradient ($c_2$) is turned on and k-space is traversed with constant speed in positive RO-direction. The signal is acquired (f) during the flat top of the read-out gradient at a predetermined sampling distance $\Delta k_{RO}$. Finally, residual transversal magnetization is destroyed by spoiling gradients (e). This scheme is repeated for several phase-encoding gradient strengths which results in a certain sampling distance $\Delta k_{PE}$ in PE-direction (Fig. 2.5b).

---

[16][8, 9] serve as general references for this section.

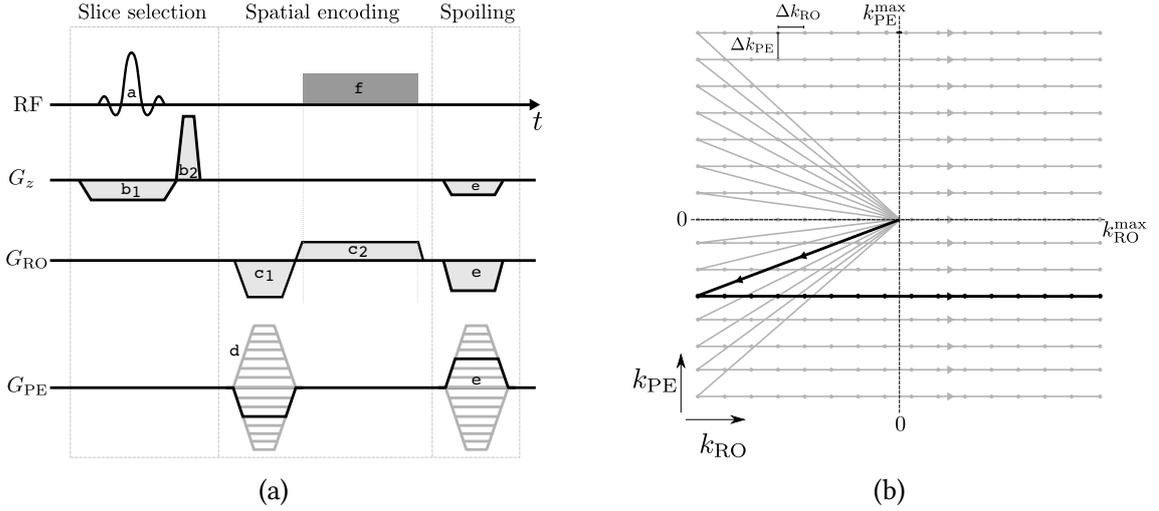

Fig. 2.5: Sequence diagram and corresponding (schematic) k-space trajectory. (a) 2D FLASH sequence diagram. $G_{SS}$: Slice selection gradient. $G_{RO}$: Read-out gradient. $G_{PE}$: Phase-encoding gradient. a: RF excitation pulse. $b_1$: Slice selection gradient. $b_2$: Rewinder gradient. $c_1$: Dephase gradient. $c_2$: Readout gradient. d: Phase-encoding gradient. The grey lines represent the different gradient strengths. The black one corresponds to the black trajectory line in (b). e: Spoiling gradients. f: Signal recording. (b) k-space trajectory. The dots represent samples.

Discrete signals in one domain lead to periodicity in the corresponding Fourier domain. In our case k-space is sampled, so we have to deal with replicates in the image domain which can cause aliasing artifacts. To prevent this from happening we have to consider the Nyquist criterion when sampling k-space, i.e. we have to choose the sampling distance sufficiently small,

$$\Delta k_{RO} \leqslant \frac{1}{\text{FOV}_{RO}}, \quad \Delta k_{PE} \leqslant \frac{1}{\text{FOV}_{PE}}, \tag{2.85}$$

where $\text{FOV}_{RO/PE}$ is the desired field of view in RO-/PE-direction.

### 2.2.7. Reduction factor and aliasing

One major drawback of MRI are the long measurement times. Especially in clinical practice fast imaging is crucial. In general, the acquisition of samples in RO-direction is fast, whereas phase-encoding is time consuming. Hence, the obvious way to get a speed-up is to skip phase-encoding steps, i.e. to undersample the k-space by acquiring fewer k-space lines. A schematic of a Cartesian undersampling pattern is depicted in Fig. 2.6. Typically, reconstruction algorithms need good knowledge about low spatial frequencies. Therefore, we fully sample $L_{\text{ref}}$ reference lines in the center of k-space. In the periphery we perform undersampling, i.e. we acquire $R$ times fewer samples compared to the full k-space. To quantify the undersampling we introduce the effective reduction factor

$$R_{\text{eff}} := \frac{n_K^{\text{full}}}{n_K^{\text{red}}}, \tag{2.86}$$

where $n_K^{\text{full/red}}$ are the number of samples in the full/reduced k-space. The reduced acquisition time is then given by

$$T_{\text{aq}}^{\text{red}} = \frac{T_{\text{aq}}^{\text{full}}}{R_{\text{eff}}}. \tag{2.87}$$

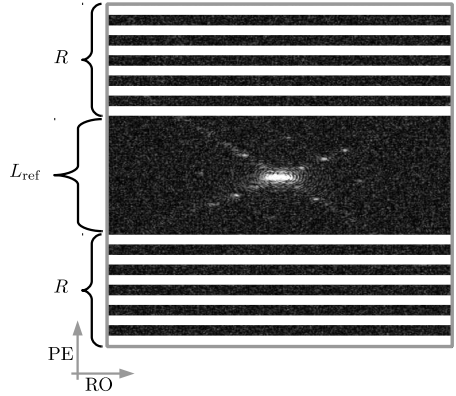

Fig. 2.6: Schematic of a Cartesian k-space with undersampling pattern. The white vertical lines depict not acquired k-space lines. $L_{\text{ref}}$: Number of reference lines in the k-space center. $R$: Reduction factor in the k-space periphery. RO and PE denote the read-out- and phase-encoding-direction.

By skipping phase-encoding steps we increase the sampling distance $\Delta k_{\text{PE}}$, which may result in a violation of the Nyquist criterion. This leads to wrap-around artifacts as demonstrated in Fig. 2.7. On the left we show the accurate reconstruction of an image that we get from a fully sampled (Nyquist) grid. On the right we omit every second k-space line and after reconstruction we observe a superposition of the original image with a
FOV/2 shifted replicate (aliasing). Reconstruction algorithms like SENSE (SENSitivity

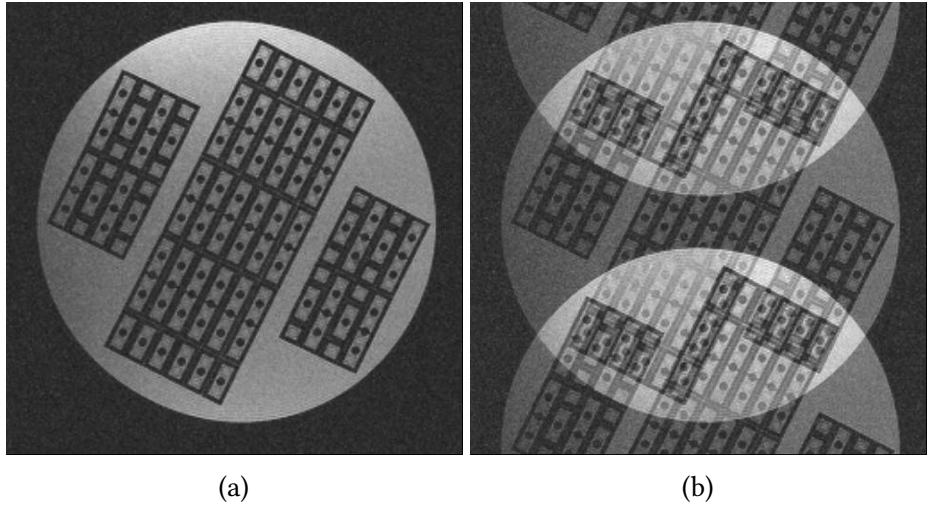

|                  (a)                  |                  (b)                  |

Fig. 2.7: Effects of k-space sampling. (a) Full k-space reconstruction. The Nyquist
criterion is satisfied, we get an accurate image. (b) Reduced k-space recon-
struction. Every second k-space line is omitted ($R_{\text{eff}} = R = 2$, $L_{\text{ref}} = 0$). We
observe wrap-around artifacts (aliasing).

Encoding) [26], GRAPPA (GeneRalized Autocalibrating Partially Parallel Acquisitions)
[27] or NLINV (Regularized NonLinear INVersion) [6] make use of information from coil
arrays[17] to eliminate these artifacts. In this way high acceleration factors are feasible.

## 2.2.8. Elimination of aliasing - SENSE

We introduce the basic idea of SENSE as an example for a linear reconstruction method.
Images suffering from wrap-around artifacts imply a superposition of magnetization
values from different spatial locations $\boldsymbol{r}_\rho$ in each coil $j$. Let $c_\rho^j = c^j(\boldsymbol{r}_\rho)$ be the coil sensitivity
of coil $j$ at a location indexed by $\rho$. Let furthermore $a_\rho$ be the actual magnetization at
position $\boldsymbol{r}_\rho$. The forward model for a measurement is then given by

$$v^j = c_\rho^j a_\rho. \tag{2.88}$$

From the measured data $v^j$ in each coil $j$ we can recover the original pixel values $a_\rho$ as
long as the matrix $c_\rho^j$ is invertible. In particular, this means that the number of superposed
pixels may not exceed the size of the coil array. We speak of autocalibrating MRI when coil sensitivities[18] are obtained from the same data used for reconstruction. The general theory of SENSE is beyond the scope of this thesis and can be found in [26, 28].

---

[17]See equation (2.81).

### 2.2.9. Noise

In practical measurements we do not only acquire signal from the object of study but also have to deal with noise. In NMR, this noise mainly arises through thermal motion (Brownian motion) of free electrons inside the receiver coils or the investigated tissue. Therefore, we commonly assume k-space data to be contaminated by additive Gaussian white noise $\eta$ of zero mean and variance $\sigma_\eta^2$ [8].

In image space the noise distribution for complex images is also Gaussian, as long as a linear and orthogonal transform - such as the complex Fourier transform - is used for reconstruction. However, magnitude images are more common in MRI because phase artifacts are avoided by discarding the phase information. Going from the complex to the magnitude image is a non-linear transformation, thus the noise distribution is no-longer Gaussian but Rician. Nevertheless, even in magnitude images we can assume to have Gaussian noise as long as ROIs with signal-to-noise ratios SNR := $S/\sigma > 3$ are considered [29]. Here $S$ is the image pixel intensity in the absence of noise and $\sigma$ the standard deviation of the noise.

When dealing with undersampled k-spaces we get aliasing artifacts, which have to be eliminated by the reconstruction algorithm. This process causes noise amplification in the image space. Pruessmann et al. proposed to use the geometry factor or g-factor (2.89) to quantify this noise amplification [26].[19] The g-factor is defined as

$$g(\boldsymbol{r}) := \frac{\sqrt{\sigma_{\text{red}}^2(\boldsymbol{r})}}{\sqrt{R_{\text{eff}}} \cdot \sqrt{\sigma_{\text{full}}^2(\boldsymbol{r})}}, \tag{2.89}$$

where $R_{\text{eff}}$ is the effective reduction factor of the undersampled k-space. The factors $\sigma_{\text{red}}^2(\boldsymbol{r})$ and $\sigma_{\text{full}}^2(\boldsymbol{r})$ are noise variances at pixel position $\boldsymbol{r}$ in the reconstructed image obtained from a reduced and a full k-space respectively. In general, the local noise amplification depends on the coil sensitivities and the undersampling pattern. A high noise amplification even in a small, localized region of the reconstruction can disguise essential information and possibly makes the whole image useless. Therefore, we use the maximum g-factor as a quantitative measure of reconstruction quality,

$$g_{\text{max}} = \max(g(\boldsymbol{r})). \tag{2.90}$$

---

[18]Or in the case of GRAPPA: convolution kernels.

[19]Originally, the g-factor was introduced for the linear SENSE reconstruction. However, we use the same formula to determine noise amplification for the reconstruction with simultaneous multi-slice Regularized Nonlinear Inversion (SMS-NLINV).

## 2.2.10. Signal intensity and flip angle

An important characteristic of NMR experiments is the resulting signal intensity $S$. For a FLASH sequence, Haase [30] showed the signal intensity to be given by

$$S = m_0 \frac{(1 - e^{\frac{-TR}{T_1}}) \cdot \sin \alpha}{1 - \cos \alpha \cdot e^{\frac{-TR}{T_1}}} e^{\frac{-TE}{T_2^*}}. \tag{2.91}$$

Here $TR$ is the repetition time, $TE$ is the echo time, $T_1$ and $T_2^*$ are relaxation constants, $m_0$ is the equilibrium magnetization and $\alpha$ is the flip angle by which the spins are tipped. In the special case of $TR \gg T_1$ and $TE \ll T_2^*$ we can approximate (2.91) to

$$S \approx m_0 \sin \alpha. \tag{2.92}$$

Hence, the signal intensity is mainly determined by the flip angle, which depends on the duration and strength of the RF excitation pulse (see section 2.1.4).

# 3. Multi-slice MRI

In multi-slice MRI experiments the aim is to get images of several slices. We can distinguish between conventional multi-slice and simultaneous multi-slice measurements.

## 3.1. Conventional multi-slice

In conventional multi-slice (MS) experiments, each slice is acquired separately which corresponds to successively performed single-slice measurements. For the acquisition of $M$ slices we perform $M$ measurements and each slice is excited only once (Fig. 3.1a). With this technique we gain information about the third dimension without having to do a time consuming 3D scan. However, 3D scans benefit from Fourier averaging which leads to an improved signal-to-noise ratio (SNR) compared to single-slice experiments.

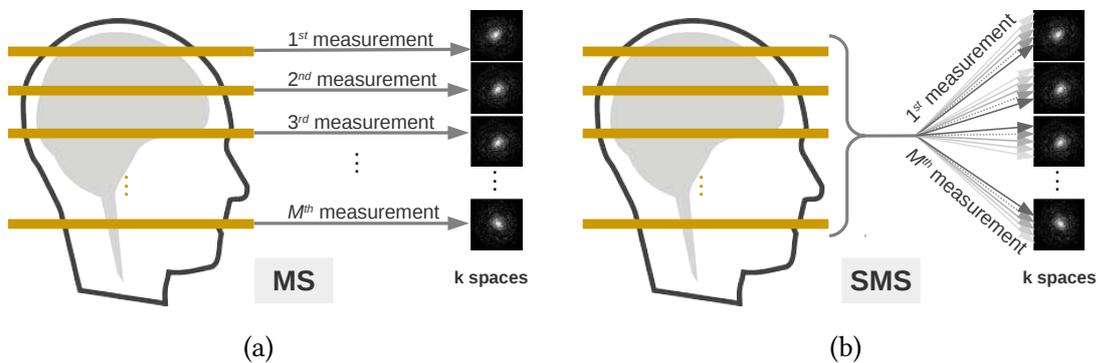

Fig. 3.1: Schematic of multi-slice measurement strategies. (a) Conventional multi-slice (MS): Sequential excitation of slices (represented by yellow bars). (b) Simultaneous multi-slice (SMS): Simultaneous excitation of slices.

## 3.2. Simultaneous multi-slice

With simultaneous multi-slice (SMS) imaging we obtain SNR improved results while maintaining the measurement time of conventional multi-slice MRI. Furthermore SMS provides advantages in terms of a more efficient elimination of aliasing. The number of slices $M$ considered in SMS experiments is referred to as multiband factor.

## 3.2.1. Basic principle

In SMS MRI we perform $M$ measurements to get information about $M$ parallel slices. Contrary to conventional multi-slice, in each measurement all $M$ slices are excited simultaneously (Fig. 3.1b), thus we acquire superposed data. In general, a specific encoding scheme such as Fourier-encoding, is used for superposition.[1] We therefore introduce the discrete Fourier transform matrix

$$\Xi_{pq} = e^{-2\pi i \frac{(p-1)\cdot(q-1)}{M}}, \quad p, q = 1, \dots, M. \tag{3.1}$$

Let $\boldsymbol{y}_q$ be the k-space of slice $q$ obtained by a single-slice experiment. The Fourier-encoded k-space $\tilde{\boldsymbol{y}}_p$ acquired in SMS measurement $p$ is then given by

$$\tilde{\boldsymbol{y}}_p := \sum_{q=1}^{M} \Xi_{pq} \boldsymbol{y}_q. \tag{3.2}$$

The normalized inverse of the encoding matrix is given by

$$\Xi_{qp}^{-1} = \frac{1}{M} \Xi_{qp}^{H} = \frac{1}{M} e^{2\pi i \frac{(p-1)\cdot(q-1)}{M}}. \tag{3.3}$$

If we deal with fully sampled k-spaces, we can use this inverse to recover the original k-space data,

$$\boldsymbol{y}_q^{\text{SMS}} = \sum_{p=1}^{M} \Xi_{qp}^{-1} \tilde{\boldsymbol{y}}_p. \tag{3.4}$$

Equation (3.4) implies that the SMS reconstruction of a slice is equivalent to an averaging process. This becomes more clear if we expand the equation for $\boldsymbol{y}_q^{\text{SMS}}$,

$$\boldsymbol{y}_q^{\text{SMS}} \overset{(3.4)}{=} \sum_{p=1}^{M} \Xi_{qp}^{-1} \tilde{\boldsymbol{y}}_p \overset{(3.2)}{=} \sum_{p=1}^{M} \sum_{q=1}^{M} \Xi_{qp}^{-1} \Xi_{pq} \boldsymbol{y}_q \overset{(3.1)}{=} \frac{\sum_{p=1}^{M} \boldsymbol{y}_q}{M}. \tag{3.5}$$

---

[1] For more information about the choice of the encoding matrix, see appendix A.6.

In Fig. 3.2 we depict the direct root sum of squares[2] (RSS) reconstruction $\tilde{m}_p^{\text{RSS}} = \sqrt{\sum_{j=1}^{N} |\mathcal{F}^{-1}(\tilde{y}_p^j)|^2}$ of a Nyquist sampled, Fourier-encoded k-space (multiband factor $M = 2$), together with its disentangled counterpart $m_q^{\text{RSS}} = \sqrt{\sum_{j=1}^{N} |\mathcal{F}^{-1}(y_q^{\text{SMS},j})|^2}$. Here $j = 1, \dots, N$ iterates the coils and $p = 1, 2$ or $q = 1, 2$ counts the Fourier-encoded measurements or disentangled slices. The images on the left show a superposition of two slices whereas on the right the overlay is disentangled.

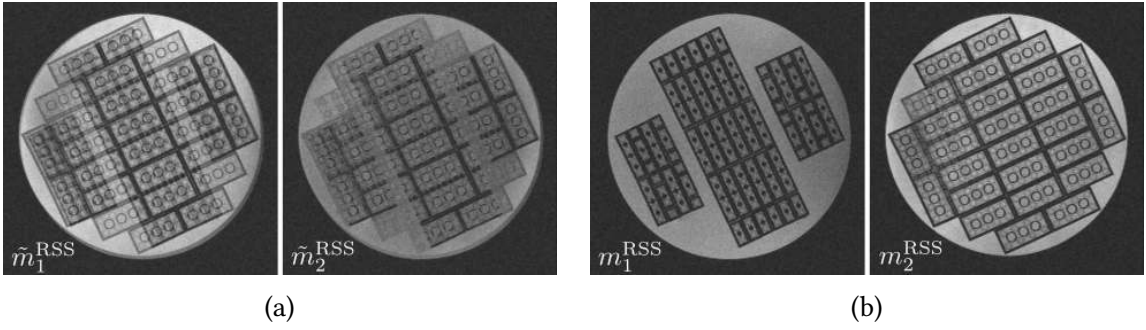

(a)             (b)

Fig. 3.2: Root sum of squares reconstructions of a SMS measurement with multiband factor $M = 2$. (a) Direct reconstruction of the Fourier-encoded k-spaces. (b) Reconstruction of the disentangled k-spaces.

### 3.2.2. Signal-to-noise benefit

The averaging process described in the previous section results in an improved SNR for SMS experiments compared to conventional multi-slice experiments. Let $\tilde{y}_p$ be a k-space acquired in a SMS experiment, which is contaminated by additive Gaussian white noise $\eta_p$ with standard deviation $\sigma_\eta$. Let $\tilde{\Upsilon}_p$ be the same k-space without noise. Then we can write

$$\tilde{y}_p := \tilde{\Upsilon}_p + \eta_p. \tag{3.6}$$

Similarly we define

$$y_q^{\text{SMS}} := \Upsilon_q^{\text{SMS}} + \eta_q^{\text{SMS}} \tag{3.7}$$

---

[2]The root sum of squares method is a common technique used to combine information from multiple receiver channels and to eliminate phase errors by only considering the magnitude of the reconstruction.

and rewrite (3.5),

$$
\begin{aligned}
\boldsymbol{y}_q^{\text{SMS}} \quad &:= \Upsilon_q^{\text{SMS}} + \eta_q^{\text{SMS}} \\
&\overset{(3.4)(3.6)}{=} \sum_{p=1}^{M} \Xi_{qp}^{-1}(\tilde{\Upsilon}_p + \eta_p) \\
&\overset{(3.3)}{\underset{(3.1)(3.2)}{=}} \frac{\sum_{p=1}^{M} \Upsilon_q}{M} + \frac{\sum_{p=1}^{M} e^{2\pi i \frac{(p-1)\cdot(q-1)}{M}} \eta_p}{M}.
\end{aligned} \tag{3.8}
$$

The summation of the ideal signal yields

$$
\sum_{p=1}^{M} \Upsilon_q = M \cdot \Upsilon_q, \tag{3.9}
$$

whereas the summation of Gaussian noise with zero mean results in

$$
\sum_{p=1}^{M} e^{2\pi i \frac{(p-1)\cdot(q-1)}{M}} \eta_p = \sqrt{M\sigma_{\text{k}}^2} = \sqrt{M}\sigma_{\text{k}}. \tag{3.10}
$$

Thus, the SMS acquisition leads to an improved noise behavior compared to conventional multi-slice experiments. The SNR benefit is given by

$$
\frac{\text{SNR}^{\text{SMS}}}{\text{SNR}^{\text{MS}}} = \sqrt{M}. \tag{3.11}
$$

### 3.2.3. Conventional reconstruction methods

To speed up the acquisition we usually deal with undersampled k-spaces. For each encoding a different undersampling pattern may be used (see section 7.3) and therefore, the reconstruction formula (3.4) can not always be applied. A variety of successful algorithms have been developed or adapted to solve this problem, e.g. SENSE [3, 26] and its extension CAIPIRINHA [31], GRAPPA [27], the hybrid SENSE-GRAPPA [32] or Slice-GRAPPA [33]. These methods are based on a sequential approach. As a first step a calibration from reference lines is performed. For SENSE-like algorithms, that operate in image space, this corresponds to the determination of the coil sensitivities. In GRAPPA-like algorithms, that act in k-space, convolution kernels are computed. As a second step, a linear reconstruction is performed to recover aliasing free images of each slice. A precise calibration is essential to obtain good image quality. Especially for SENSE-like algorithms an inadequate prediction of the coil profiles results in unwanted reconstruction artifacts.

In general, GRAPPA-like algorithms are more robust but limited to rather moderate reduction factors, due to an increased noise-amplification for high undersampling [6, 27, 34].

Particularly demanding are experiments where the object of interest is moving and where a high temporal resolution is necessary, e.g. in real-time cardiac imaging. For these kind of scans a high reduction factor is mandatory. At the same time, a precise prediction of coil sensitivities is difficult, since even small object movements alter its dielectric properties and therefore influence the receiver sensitivities. To overcome this problem in single-slice experiments, Uecker et al. proposed to perform image reconstruction by Regularized Nonlinear Inversion [6], where image content and coil sensitivities are jointly estimated. They demonstrated that this method leads to a more accurate estimation of coil sensitivities, since all k-space data, and not only a subset, is used for coil profile determination. Moreover, this method can easily be extended to non-Cartesian imaging [35]. In chapter 5 we extend the method of Regularized Nonlinear Inversion for the reconstruction of Fourier-encoded simultaneous multi-slice data. Before that, we introduce the fundamentals of simultaneous multi-slice excitation pulse design.

# 4. Simultaneous multi-slice excitation pulse design

In SMS MRI we want to excite spins of $M$ slices with thickness $\Delta z$ simultaneously. The adjacent slices are separated by the center-to-center slice distance $d$.[1] Without loss of generality we choose the slices to be perpendicular to the z-axis.

We have already discussed the technique of single-slice excitation in section 2.2.2. A magnetic field gradient $B_z = G_z z$ generates a spatially dependent Larmor frequency in the probe, thus a bandwidth limited transversal RF pulse $B_{\text{rf}}^{(1)} = B_{\text{env}}(\Delta\omega)e^{-i\omega_c t}$ excites only spins in a specific region. According to (2.54), $\omega_c = \gamma(B_0 + G_z z_c)$ is the carrier frequency which determines the slice center. The pulse envelope $B_{\text{env}}(\Delta\omega)$ with bandwidth $\Delta\omega$ specifies the slice shape and thickness. In the following section we will extend the description of slice selective excitation to multiple slices.

## 4.1. Pulse envelope design

To simultaneously excite $M$ equidistant slices of equal thickness, we use a superposition of single-slice excitation pulses with carrier frequencies $\omega_{cq}, q = 1, \ldots, M$, that are further specified in equation (4.2).

$$
\begin{aligned}
B_{\text{rf}}^{(M)}(\Delta\omega, \omega_{c1}, \ldots, \omega_{cM}) &= \sum_{q=1}^{M} B_{\text{rf}}^{(1)}(\Delta\omega, \omega_{cq}) \\
&= B_{\text{env}}(\Delta\omega) \sum_{q=1}^{M} e^{-i\omega_{cq}t}
\end{aligned}
\tag{4.1}
$$

The slice centers are located at $z_{cq} = (\omega_{cq} - \gamma B_0)/\gamma G_z$ and the slice distance $d$ is given by

$$
d = |z_{c(q+1)} - z_{cq}| = |(\omega_{c(q+1)} - \omega_{cq})/\gamma G_z|.
\tag{4.2}
$$

---

[1] To simplify the description, we assume a constant slice thickness and a constant slice distance between all adjacent slices. In general, arbitrary slice distances and thicknesses are possible.

We define

$$\Omega_c := \frac{\sum_{q=1}^M \omega_{cq}}{M}, \quad \Delta\Omega_q := \omega_{cq} - \Omega_c \tag{4.3}$$

$$B_{\text{env}q} := B_{\text{env}} e^{-i\Delta\Omega_q t}, \tag{4.4}$$

which yields[2]

$$
\begin{aligned}
B_{\text{rf}}^{(M)} &\overset{(4.1)(4.3)}{=} B_{\text{env}} e^{-i\Omega_c t} \sum_{q=1}^M e^{-i\Delta\Omega_q t} \\
&\overset{(4.4)}{=} \left( \sum_{q=1}^M B_{\text{env}q} \right) e^{-i\Omega_c t},
\end{aligned}
\tag{4.5}
$$

$$B_{\text{env}}^{(M)} := \left( \sum_{q=1}^M B_{\text{env}q} \right). \tag{4.6}$$

Equation (4.5) shows that an $M$-slice excitation pulse $B_{\text{rf}}^{(M)}$ can be written as a super-position of phase modulated envelop functions (4.4) with a single carrier frequency $\Omega_c$. We can use this fact to efficiently implement the SMS-sequence on a MRI scanner using the `SIEMENS IDEA`[3] software. For more information on the actual implementation see appendix A.5.

## 4.2. Fourier-encoding

In the previous section we have shown that multi-slice excitation can be accomplished by superposing envelope functions (see (4.5)). Now we introduce a method to obtain Fourier-encoded multi-slice k-spaces. Therefore, we multiply the envelopes (4.4) with phase factors $\Xi_{pq}$ from (3.1),

$$\tilde{B}_{\text{env}p}^{(M)} := \sum_{q=1}^M \Xi_{pq} B_{\text{env}q}, \tag{4.7}$$

$$B_{\text{rf}p}^{(M)} \overset{(4.5)}{:=} \tilde{B}_{\text{env}p}^{(M)} e^{-i\Omega_c t} \overset{(4.7)}{=} \left( \sum_{q=1}^M \Xi_{pq} B_{\text{env}q} \right) e^{-i\Omega_c t}. \tag{4.8}$$

---

[2]For convenience we do not explicitly denote the dependencies on $\Delta\omega$ and $\omega_{cq}$.
[3]`IDEA`: Integrated Development Environment for Applications.

The RF excitation pulse $B_{\text{rfp}}^{(M)}$ excites $M$ slices simultaneously and generates the Fourier-encoded k-space (3.2),

$$\tilde{y}_p := \sum_{q=1}^{M} \Xi_{pq} y_q.$$

The connection between (4.8) and (3.2) is motivated in appendix A.7.

As an example, in Fig. 4.1 we depict two Fourier-encoded multi-slice pulse envelopes and their corresponding frequency domain representations. The pulses excite four equidistant slices and are given by

$$\tilde{B}_{\text{env1}}^{(4)} = \sum_{q=1}^{4} \Xi_{1q} B_{\text{env}q} = B_{\text{env0}} + B_{\text{env1}} + B_{\text{env2}} + B_{\text{env3}}, \tag{4.9}$$

$$\tilde{B}_{\text{env2}}^{(4)} = \sum_{q=1}^{4} \Xi_{2q} B_{\text{env}q} = B_{\text{env0}} - iB_{\text{env1}} - B_{\text{env2}} + iB_{\text{env3}}, \tag{4.10}$$

where $B_{\text{env}q}$ are single-slice excitation envelopes multiplied with phase ramps according to (4.4) to shift their position. The frequency plots (Fig. 4.1c, 4.1d) clearly show bands that correspond to the four slices excited by the pulse. Considering the real and imaginary parts, we furthermore recognize the encoding schemes $\Xi_{1q} \overset{(3.1)}{=} (1, 1, 1, 1)$ and $\Xi_{2q} \overset{(3.1)}{=} (1, -i, -1, i)$.

## 4.3. Technical limitations

There are two major limitations for the construction of SMS RF excitation pulses. First, due to the superposition of single-slice excitation pulses, the peak amplitude in multi-slice pulses rises linearly with the number of simultaneously excited slices and consequently, the peak power rises quadratically. Therefore, the SMS pulse is prone to exceed the RF amplifier capabilities of the scanner. Second, the more slices we simultaneously excite, the more total power must be deposited, which might lead to a violation of the specific absorbtion rate (SAR)[4] limits. One possible method of overcoming these limitations is to increase the pulse duration, while keeping the flip angle and the bandwidth-time-product ($P_{\text{BWT}}$) constant. However, a prolonged excitation pulse might have adverse effects on sequence timing, or - due to a reduced bandwidth - the pulse becomes prone to off-resonance effects.

---

[4]The SAR limit depends on the patients mass and the investigated body region. Exceeding the SAR restrictions can lead to tissue heating and damage.

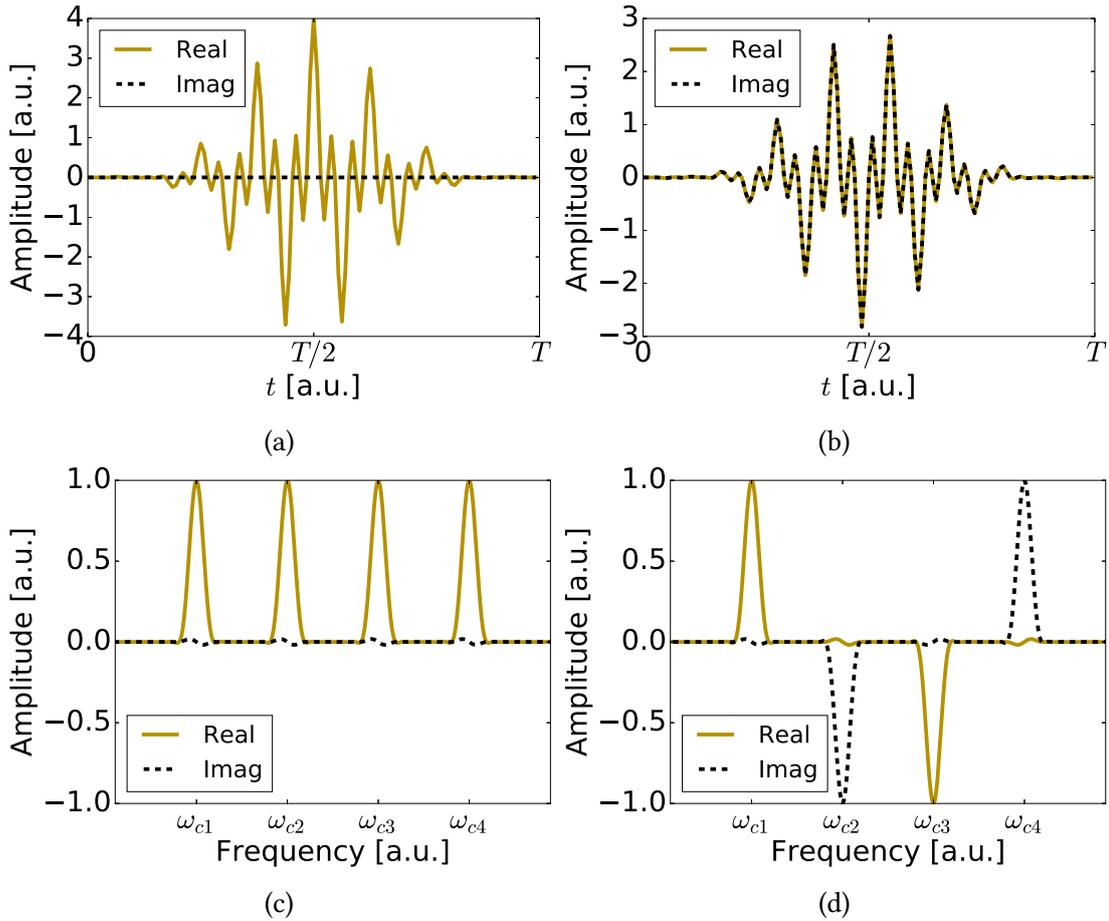

Fig. 4.1: Fourier-encoded multi-slice excitation pulse envelopes and their corresponding frequency domain representation. $T$ is the pulse duration and $\omega_{c1}, \ldots, \omega_{c4}$ are the center frequencies of the four bands. (a) Pulse envelope $\tilde{B}_{\text{env1}}^{(4)}$. (b) Pulse envelope $\tilde{B}_{\text{env2}}^{(4)}$. (c) Normalized Fourier transform of $\tilde{B}_{\text{env1}}^{(4)}$. (d) Normalized Fourier transform of $\tilde{B}_{\text{env2}}^{(4)}$.

A number of more advanced strategies were developed to tackle these limitations [36]: To prevent high peak amplitudes, a phase cycling scheme can be used. Instead of just superposing the single-slice excitation pulses of type (2.56), each pulse is multiplied with a specific phase factor $e^{i\varphi}$ before combination [37]. This is similar to what is done in Fourier-encoding with $\Xi_{pq}$ and $p > 1$. Alternatively, the single-slice excitation pulses can be shifted in time by $1 - 2\,\text{ms}$ to prevent a peak amplitude overlap [38, 39].

The total RF power of a slice selective pulse can be reduced by Variable Rate Excitation (VERSE) [40]. In this technique the amplitude of the slice selection gradient is temporarily reduced at times where the RF pulse deposits most energy. This corresponds to a decelerated k-space traversal and accordingly, to a diminished overall RF power. A second approach is called Power Independent of the Number of Slices (PINS) [41]. Here a series of non-selective RF pulses is interleaved with small slice-gradient blips. This has the same effect as sampling the actual slice selective pulse with a train of delta functions.

However, the sequence utilized in this thesis uses a small flip angle of 15° and therefore SAR and peak amplitude limits are not an issue.

# 5. Simultaneous multi-slice Regularized Nonlinear Inversion

In the following sections we introduce a multi-slice version of the Regularized Nonlinear Inversion algorithm (NLINV), developed by Uecker et al. [6], for the reconstruction of Fourier-encoded SMS data. We dub this extended algorithm simultaneous multi-slice Regularized Nonlinear Inversion (SMS-NLINV).

## 5.1. Signal equation in operator notation

We model the MRI signal equation for parallel imaging (2.81) as a nonlinear operator equation

$$F(X) = \tilde{Y}. \tag{5.1}$$

$X$ is the vector we want to reconstruct. It contains the image content $m(\boldsymbol{r})$ and the coil sensitivities $c^j(\boldsymbol{r})$, $j = 1, \ldots, N$, for each of the $M$ slices $q$.

$$X := \begin{pmatrix} \boldsymbol{x}_1 \\ \vdots \\ \boldsymbol{x}_M \end{pmatrix}, \quad \boldsymbol{x}_q := \begin{pmatrix} m_q \\ c_q^1 \\ \vdots \\ c_q^N \end{pmatrix} \tag{5.2}$$

Vector $\tilde{Y}$ contains the Fourier-encoded k-spaces for all $M$ encodings and all $N$ channels.

$$\tilde{Y} := \begin{pmatrix} \tilde{\boldsymbol{y}}_1 \\ \vdots \\ \tilde{\boldsymbol{y}}_M \end{pmatrix}, \quad \tilde{\boldsymbol{y}}_p := \begin{pmatrix} \tilde{y}_p^1 \\ \vdots \\ \tilde{y}_p^N \end{pmatrix} \tag{5.3}$$

$F$ is a nonlinear mapping function and will explicitly be introduced later.

We assume that $F(X) = \tilde{Y}$ is given in a discretized form and all functions are represented by vectors of point values on a rectangular grid. We solve this equation with the Iteratively Regularized Gauss-Newton Method (IRGNM) [42, 43] and closely follow the concept which Uecker et al. proposed in [6] for single-slice reconstruction.

## 5.2. Iteratively Regularized Gauss-Newton Method

The first step of IRGNM is to linearize (5.1) by choosing an adequate guess $X_n$.

$$\tilde{Y} \overset{!}{=} F(X_n + dX) \approx DF(X_n)dX + F(X_n), \tag{5.4}$$

with $DF(X_n)$ being the Jacobian of $F$ at point $X_n$. Then, we solve this equation for the update $dX$ which is used to calculate the subsequent guess

$$X_{n+1} = X_n + dX. \tag{5.5}$$

With suitable regularizations, this iteration scheme converges to a solution.

We use the Conjugate Gradient (CG) algorithm [44] to find an approximate solution for (5.4) . This algorithm requires a symmetric and positive-definite matrix which we can generate by multiplying (5.4) with the adjoint of $DF(X_n)$. A rearrangement yields

$$DF(X_n)^H DF(X_n)dX = DF(X_n)^H(\tilde{Y} - F(X_n)). \tag{5.6}$$

Due to the bad conditioning of the linearized system, Uecker et al. proposed the addition of a positive-definite regularization matrix $\beta_n I$. Here $I$ is the identity matrix and $\beta_n$ is the regularization parameter that is reduced in each iteration step $n$ according to $\beta_n = \beta_0 b^n$ with $b \in (0, 1)$. This results in the Levenberg-Marquardt algorithm

$$\left(DF(X_n)^H DF(X_n) + \beta_n I\right)dX = DF(X_n)^H(\tilde{Y} - F(X_n)). \tag{5.7}$$

The effect of the regularization becomes more clear if we use the Gauss normal equation to transform (5.7) into the unique minimizer of a functional, as we show in appendix A.8.

$$\min\left(\|DF(X_n)dX - (\tilde{Y} - F(X_n))\|^2 + \beta_n\|dX\|^2\right) \tag{5.8}$$

The term $\beta_n\|dX\|^2$ is known as Tikhonov regularization. Equations (5.7) and (5.8) are equivalent, i.e. they possess the same solution $dX$. Hence, the idea of Levenberg-Marquardt is to minimize (5.4) in the least-squares sense and at the same time prevent $dX$ from growing too big. For large $\beta_n$, the algorithm represents the gradient descent algorithm

$$(\beta_n I)dX = DF(X_n)^H(\tilde{Y} - F(X_n)), \tag{5.9}$$

which is very robust. In each iteration $\beta_n$ is reduced and for small values the algorithm essentially turns into the classical Gauss-Newton method

$$DF(X_n)^H DF(X_n)dX = DF(X_n)^H(\tilde{Y} - F(X_n)), \tag{5.10}$$

which is very fast. As we will justify further down, we get an even more stable algorithm by modifying the right term of the minimizer (5.8),

$$\min\left(\|DF(X_n)dX - (\tilde{Y} - F(X_n))\|^2 + \beta_n\|X_n + dX\|^2\right). \tag{5.11}$$

Thus, the regularization no longer applies solely to the update $dX$ but to the subsequent guess $X_{n+1} = X_n + dX$. Again, the Gauss normal equation provides us an equivalent equation to (5.11),

$$\left(DF(X_n)^H DF(X_n) + \beta_n I\right)dX = DF(X_n)^H(\tilde{Y} - F(X_n)) - \beta_n X_n. \tag{5.12}$$

For the actual implementation of the algorithm, we require the explicit representation of the forward operator $F$, its derivative and the adjoint of the derivative. For this means, we define the projection matrix

$$\boldsymbol{P} := \begin{pmatrix} P_1 & & 0 \\ & \ddots & \\ 0 & & P_M \end{pmatrix}, \tag{5.13}$$

where $P_p$ is the orthogonal projection onto the k-space trajectory used in encoding $p = 1, \ldots, M$. For Cartesian sampling $P_p$ is a diagonal matrix with ones at sample positions and zeros elsewhere. From the theory of Fourier-encoding (see section 4.2) we know that a valid model for $F$ is given by

$$F : X \mapsto \boldsymbol{P}\Xi \begin{pmatrix} \mathcal{F}(m_1\boldsymbol{c}_1) \\ \vdots \\ \mathcal{F}(m_M\boldsymbol{c}_M) \end{pmatrix}, \tag{5.14}$$

with

$$\mathcal{F}(m_q \boldsymbol{c}_q) := \begin{pmatrix} \mathcal{F}(m_q c_q^1) \\ \vdots \\ \mathcal{F}(m_q c_q^N) \end{pmatrix} \tag{5.15}$$

Here $\mathcal{F}$ is the (two-dimensional) Fourier transform and $\Xi$ is the $M \times M$ DFT matrix (3.1). The magnetization $m_q$ of each slice $q$ is weighted with the coil sensitivities $\boldsymbol{c}_q = (c_q^1, \ldots, c_q^M)^T$, transformed into k-space ($\mathcal{F}$), Fourier-encoded ($\Xi$) and sampled ($\boldsymbol{P}$).

Since the Fourier transform is a linear operation, we can calculate the derivative with the help of the product rule of derivatives,

$$DF(X) \begin{pmatrix} d\boldsymbol{x}_1 \\ \vdots \\ d\boldsymbol{x}_M \end{pmatrix} = \boldsymbol{P}\Xi \begin{pmatrix} \mathcal{F}(dm_1\boldsymbol{c}_1 + m_1 d\boldsymbol{c}_1) \\ \vdots \\ \mathcal{F}(dm_M\boldsymbol{c}_M + m_M d\boldsymbol{c}_M) \end{pmatrix}. \tag{5.16}$$

In appendix A.9 we deduce the adjoint of the derivative

$$DF^H(X) \begin{pmatrix} \tilde{\boldsymbol{y}}_1 \\ \vdots \\ \tilde{\boldsymbol{y}}_M \end{pmatrix} =$$

$$\begin{pmatrix} \begin{pmatrix} \boldsymbol{c}_1^H \\ m_1^H \end{pmatrix} & & 0 \\ & \ddots & \\ 0 & & \begin{pmatrix} \boldsymbol{c}_M^H \\ m_M^H \end{pmatrix} \end{pmatrix} \mathcal{F}^H \Xi^H \boldsymbol{P}^H \begin{pmatrix} \tilde{\boldsymbol{y}}_1 \\ \vdots \\ \tilde{\boldsymbol{y}}_M \end{pmatrix}, \tag{5.17}$$

with

$$\begin{pmatrix} \boldsymbol{c}_q^H \\ m_q^H \end{pmatrix} := \begin{pmatrix} c_q^{1*}, \ldots, c_q^{N*} \\ m_q^* \end{pmatrix}. \tag{5.18}$$

The asterisk $*$ denotes the pointwise complex conjugation.

The 2D Fourier transform $\mathcal{F}$ always appears in combination with the DFT matrix $\Xi$ in the operators $F$, $DF$ and $DF^H$. For a more elegant implementation we can define

$$\begin{aligned} \mathcal{F}_{3D} &:= \Xi\mathcal{F}, \\ \mathcal{F}_{3D}^H &:= \mathcal{F}^H\Xi^H. \end{aligned} \tag{5.19}$$

It becomes clear that we can implement $\mathcal{F}_{3D}$ and $\mathcal{F}_{3D}^H$ as a three-dimensional Fast Fourier transform and its adjoint.[1] Due to the small size of $\Xi$ we will not achieve great performance improvements by this substitution. Nevertheless, it shows how a given (single-slice) NLINV implementation can easily be extended to SMS-NLINV.

Fig. 5.1 shows a flow chart for the calculation of the operators $F$, $DF$ and $DF^H$.

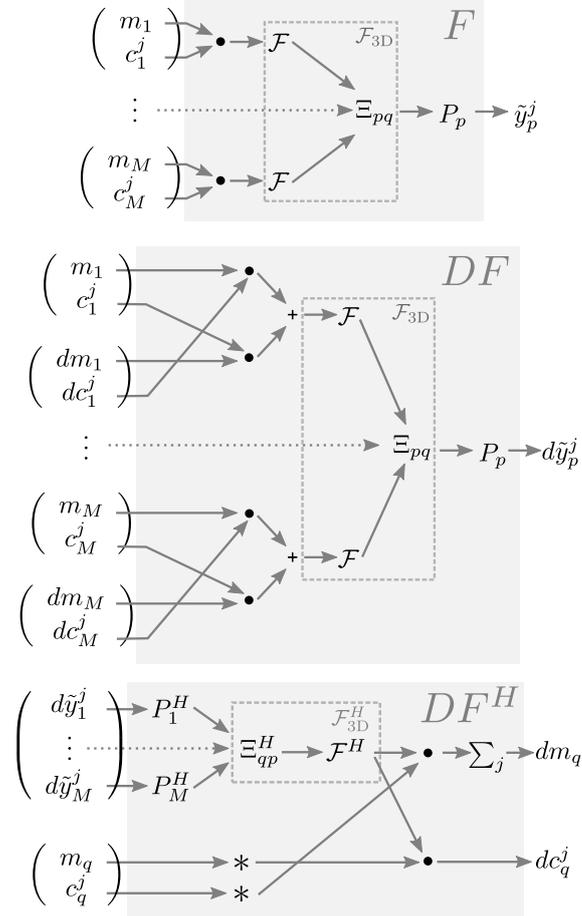

Fig. 5.1: Flow chart for the calculation of the forward operator ($F$), its derivative ($DF$) and the adjoint of the derivative ($DF^H$). $\tilde{y}$: Fourier-encoded k-space data. $m$: Magnetization. $c$: Coil sensitivity. $P$: Projection onto k-space trajectory. $\mathcal{F}$: 2D Fourier transform. $\Xi_{pq}$: DFT matrix. $\cdot$: Pointwise multiplication. $+$: Addition. $*$: Complex conjugation. More details about the notation can be found in Tab. A.1.

---

[1] Note, that the 2D Fourier transform $\mathcal{F}$ is a discretized version of a continuous Fourier transform, whereas the Fourier-encoding $\Xi$ is discrete by definition.

## 5.3. Regularization through prior knowledge

The application of the presented IRGNM algorithm without further modifications would not yield the correct solution for $X$. The coil sensitivities $c_q^j$ would contain part of the object information $m_q$ and vice versa. This is due to the fact that the equation system is highly underdetermined, which becomes obvious with the following consideration: Let

$$X := \begin{pmatrix} \boldsymbol{x}_1 \\ \vdots \\ \boldsymbol{x}_M \end{pmatrix}, \quad \boldsymbol{x}_q = (m_q, c_q^1, \ldots, c_q^N)^T, \tag{5.20}$$

be a solution for $F(X) = \tilde{Y}$. With (5.14) it is clear that for arbitrary complex functions $\xi_q \neq 0$ we can create a new solution simply by writing

$$\boldsymbol{x}_q' = (m_q \cdot \xi_q, c_q^1/\xi_q, \ldots, c_q^N/\xi_q)^T. \tag{5.21}$$

However, there is only one physical solution to the problem and we can find a reasonable estimate by adding prior knowledge about the object and the coil sensitivities. While image content can contain strong variations and edges, coil sensitivities in general are smooth. We can therefore apply a smoothness demanding norm for the coil profiles. Uecker et al. suggest a Sobolev norm [45][2]

$$\|f\|_{H^l} := \|2\pi(I - a\Delta)^{l/2} f\|, \tag{5.22}$$

with $a$ a scaling parameter and $\Delta = \partial_x^2 + \partial_y^2$ the 2D Laplacian. Hence, in Fourier space the standard $L^2$-norm has to weighted by the additional term $(1 + a\|k\|^2)^{l/2}$ (see appendix A.10), which penalizes high spatial frequencies. The actual implementation of this regularization is achieved by the transformation of $X = (\boldsymbol{x}_1, \ldots, \boldsymbol{x}_M)^T$ with a diagonal preconditioning matrix

$$\mathcal{W}^{-1} := \begin{pmatrix} W^{-1} & & 0 \\ & \ddots & \\ 0 & & W^{-1} \end{pmatrix}, \tag{5.23}$$

---

[2]In the scope of this thesis, the norm $\| \cdot \|$ without subscript stands for the $L^2$ norm.

$$W^{-1} := \begin{pmatrix} I & & & & 0 \\ & (1+a\|\vec{k}\|^2)^{l/2}\mathcal{F} & & & \\ & & \ddots & & \\ 0 & & & & (1+a\|\vec{k}\|^2)^{l/2}\mathcal{F} \end{pmatrix}. \tag{5.24}$$

We denote

$$X' := \mathcal{W}^{-1}X, \tag{5.25}$$

$$\boldsymbol{x}'_q := W^{-1}\boldsymbol{x}_q = \begin{pmatrix} m_q \\ c'^1_q \\ \vdots \\ c'^N_q \end{pmatrix}. \tag{5.26}$$

We then obtain a transformed but equivalent system of equations for the IRGNM algorithm

$$X' = \mathcal{W}^{-1}X, \tag{5.27a}$$

$$GX' := F\mathcal{W}X' = \tilde{Y}. \tag{5.27b}$$

Hence, instead of (5.11) the respective minimizer to be solved in every Newton step is

$$\min\left(\|DG(X'_n)dX' - (\tilde{Y} - G(X'_n))\|^2 + \beta_n\|X'_n + dX'\|^2\right). \tag{5.28}$$

The effect of the regularization term, i.e. the right summand of (5.28), becomes apparent by the following consideration:

$$\begin{aligned} \|X'_n + dX'\|^2 := \|X'\|^2 &= \|\mathcal{W}^{-1}X\|^2 \\ &= \sum_{q=1}^{M}\left(\|m_q\|^2 + \sum_{j=1}^{N}\left\|c'^j_q\right\|^2\right) \\ &= \sum_{q=1}^{M}\left(\|m_q\|^2 + \sum_{j=1}^{N}\left\|(1+a\|\vec{k}\|^2)^{l/2}\mathcal{F}\,c^j_q\right\|^2\right) \\ &= \sum_{q=1}^{M}\left(\|m_q\|^2 + \sum_{j=1}^{N}\left\|c^j_q\right\|^2_{H^l}\right) \end{aligned} \tag{5.29}$$

The applied regularization penalizes high spatial frequencies as it corresponds to the squared Sobolev norm $\|c^j_q\|^2_{H^l}$ of the coil sensitivities in the original space. Hence, smooth coil profiles are enforced.

## 5.4. Postprocessing

Although the mixing of image content and coil sensitivities is prevented by regularization, the reconstruction result may exhibit minor large scale intensity variations compared to a common RSS reconstruction. Again the reason is the underdetermination of the equation system. However, a simple post processing step can compensate for those differences. We just have to multiply the image content with the RSS of the coil profiles,

$$C_q^{\text{rss}} := \sqrt{\sum_{j=1}^{N} |c_q^j|^2}, \tag{5.30}$$

$$m_q^{\text{final}} = m_q \cdot C_q^{\text{rss}}. \tag{5.31}$$

This postprocessing procedure is not obligatory but allows for a better comparison of the presented SMS-NLINV method with other algorithms.

# 6. Hardware and materials

This section summarizes information about the used materials and hardware.

## 6.1. NMR scanner and coil array

All experiments were conducted on a `SIEMENS MAGNETOM Skyra 3T` scanner (Fig. 6.1a) at the University Medical Center Göttingen (UMG) . The system possesses an open bore with 70 cm diameter, XQ gradients (maximum gradient amplitude $45\,\mathrm{mT\,m^{-1}}$, maximum gradient slew rate $200\,\mathrm{T\,m^{-1}\,s^{-1}}$) and uses passive and active shimming. We utilize the `SIEMENS Head/Neck 20` coil (Fig. 6.1b), which consists of two rings of 8 elements and one ring with 4 elements, for parallel imaging. In the scope of this thesis, we do not use any other coil.

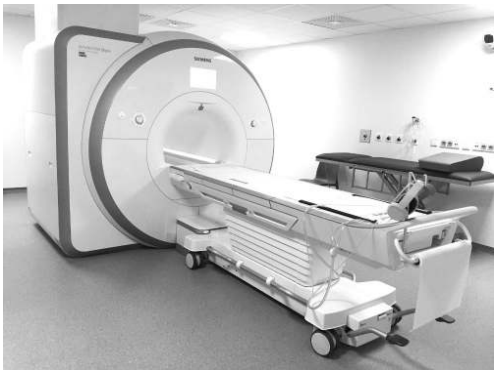

(a)

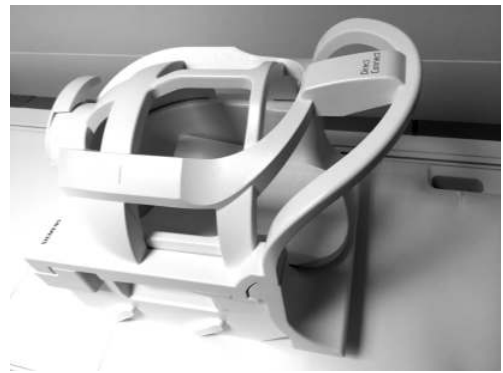

(b)

Fig. 6.1: Experimental hardware at the University Medical Center Göttingen. (a) Whole body MRI system `SIEMENS MAGNETOM Skyra 3T`. (b) 20-channel `SIEMENS Head/Neck 20` headcoil.

## 6.2. Phantoms

We utilize three different phantoms depicted in Fig. 6.2.

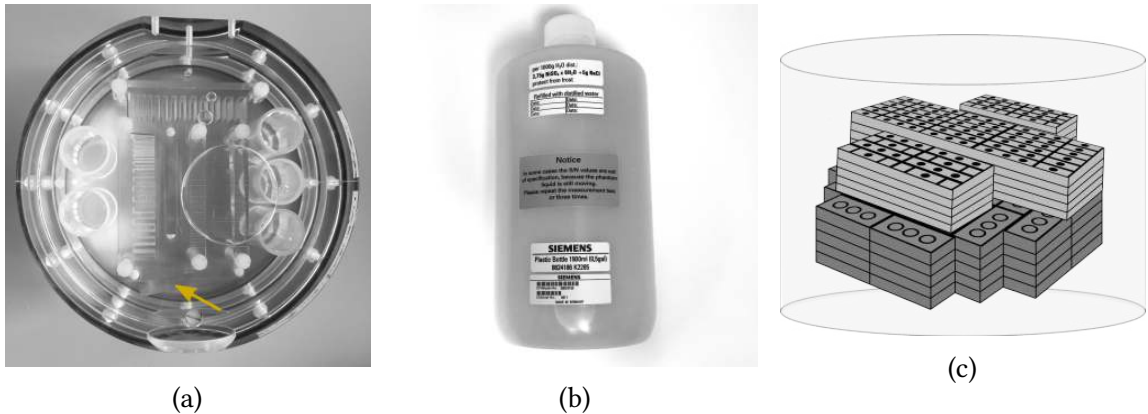

(a)                          (b)                          (c)

Fig. 6.2: Utilized phantoms. (a) Resolution phantom with isosceles triangle (high-lighted by the arrow). (b) Homogeneous phantom. (c) Brick phantom (schematic).

**Resolution phantom.** The resolution phantom is a 75 mM NaCl water phantom and contains several resolution objects including an isosceles triangle with base angle $\beta = 45°$. We use it in section 7.2.1 to measure and validate the slice distances of the multi-slice sequence.

**Homogeneous phantom.** The homogeneous phantom consists of a solution (3.75 g $NiSO_4 \times 6H_2O$ + 5 g NaCl per 1000 g $H_2O$) inside a plastic bottle (1900 mL). We use it for flip angle validation in section 7.2.2 and for signal-to-noise ratio calculations in section 7.2.3.

**Brick phantom.** The brick phantom is self-made using Acrylonitrile-Butadiene-Sty-rene (ABS) bricks known from Lego® and pure water. It is designed such that the proton density of the top and bottom part of the phantom clearly differ. Thus, it is easy to tell whether the reconstruction algorithm can or cannot disentangle the simultaneously excited slices.

# 7. Experiments

In this section we cover the experiments performed in our study. We first introduce a technique to check the accuracy of estimated coil profiles that we need in several subsequent experiments. We then validate the developed multi-slice sequence, i.e. we verify the fidelity of the slice distance, the flip angle and the SNR. Next we characterize the noise amplification of two multi-slice undersampling schemes using a linear SENSE reconstruction and finally investigate the characteristics of SMS-NLINV.

If not denoted differently, for all experiments we use the following settings:

Tab. 7.1: Default parameters for all experiments (if not denoted differently).

| | |
|---:|:---|
| Field of view | $FOV_{\text{read/phase}} = 170\,\text{mm}$ |
| Slice thickness | $\Delta z = 5\,\text{mm}$ |
| Repetition time | $TR = 9.1\,\text{ms}$ |
| Echo time | $TE = 4.8\,\text{ms}$ |
| Flip angle | $\alpha = 15°$ |
| Base resolution | $256 \times 256$ |
| RF pulse duration | $T_{\text{RF}} = 2560\,\mu\text{s}$ |
| Number of coils | $N = 20$ |

We use a 2D Cartesian FLASH sequence. An interleaved measurement scheme is used for the acquisition of the differently encoded k-spaces. By doing this, corresponding k-space lines of all measurements are acquired close in time, which allows for a more adequate Fourier decoding of the k-spaces. We always acquire full k-spaces and undersampling is performed retrospectively by multiplication with a pattern.[1] Hence, we can quickly test various undersampling schemes without having to conduct new measurements.

---

[1] This is a valid method since we measure static objects and thus, the acquisition is time-independent.

## 7.1. Coil sensitivity determination and validation

For some of the upcoming sections we require explicit knowledge about the coil sensitivities of a slice. Here, we demonstrate a method to determine and validate them.

**Methods.** To obtain the coil profiles of a slice, we acquire the full k-space and apply the ESPIRiT algorithm developed by Uecker et al. [46]. ESPIRiT combines the benefits of SENSE [26] and GRAPPA [27] and estimates the coil profiles by an eigenvalue decomposition. This algorithm is already implemented in the C/C++ program Berkeley Advanced Reconstruction Toolbox (BART) [47]. We furthermore examine the accuracy of the estimated $N$ coil sensitivities per slice with a projection test:

Let $m$ be the underlying magnetization image and $c^j$ be a diagonal matrix representing the coil profile of channel $j$. We find

$$m^j = c^j m, \tag{7.1}$$

where $m^j$ is the individual coil image, i.e. the magnetization image seen by coil $j$. We furthermore define the normalized sensitivity maps

$$\hat{c}^j := \left[ \sum_{l=1}^{N} c^{lH} c^l \right]^{-1/2} c^j. \tag{7.2}$$

It follows that for given coil profiles $\hat{c}^l$ and measured coil images $m^l_{\text{meas}}$ with $l = 1 \ldots, N$, we can calculate a projected coil image $m^j_{\text{proj}}$,

$$m^j_{\text{proj}} = \hat{c}^j \sum_{l=1}^{N} \hat{c}^{lH} m^l_{\text{meas}}. \tag{7.3}$$

The vector of coil sensitivities pointwise spans the vector of individual coil images. Thus, if the determined coil sensitivities are accurate, the difference matrices

$$\Delta m^j := m^j_{\text{proj}} - m^j_{\text{meas}}, \quad j = 1, \ldots, N, \tag{7.4}$$

solely consist of noise, i.e. do not possess any residuals from the magnetization image.

As an example, we acquire a slice of the brick phantom (Fig. 7.1a) using a conventional single-slice 2D FLASH sequence and 12 channels of the headcoil. We then determine the coil sensitivities with ESPIRiT and calculate the difference images using (7.4).

**Results.** All difference images solely contain noise, the underlying magnetization cannot be perceived. As an example we depict 6 of the 12 coil sensitivities (Fig. 7.1b) and difference images (Fig. 7.1c).

**Discussion.** Since no residual magnetization information can be found in the difference images, we assume that the profiles are correctly estimated by the ESPIRiT algorithm. Every time we explicitly need coil sensitivities we use the described procedure for determination and validation.

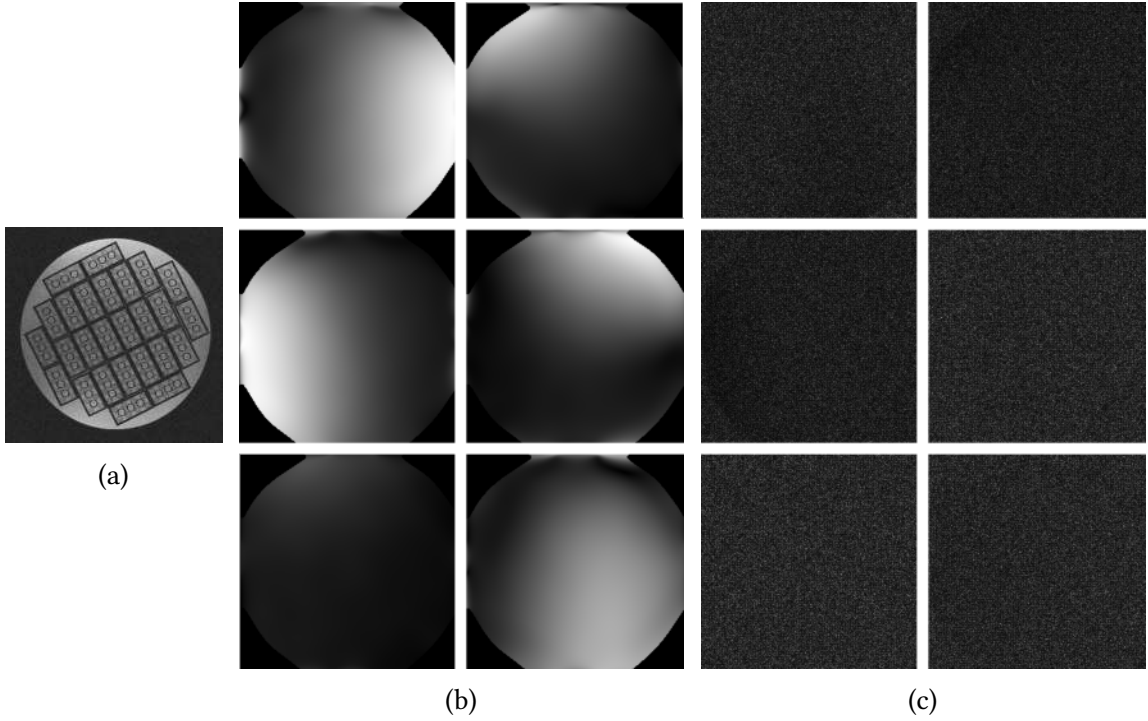

Fig. 7.1: Coil sensitivity validation using the projection test. (a) Actual RSS slice image $m_{\mathrm{meas}}^{\mathrm{RSS}}$. (b) 6 out of 12 coil sensitivities $\hat{c}^j$ estimated with ESPIRiT [46]. (c) 6 out of 12 difference images $\Delta m^j$.

## 7.2. Simultaneous multi-slice sequence validation

In the upcoming sections we investigate the fidelity of the developed SMS sequence.

### 7.2.1. Slice distance

For the SMS sequence we use phase-ramps to shift the slices in space and thus, adjust a certain slice distance. The aim of this section is to verify the accuracy of these shifts.

**Methods.** We use the resolution phantom (Fig. 6.2a) for validation. The phantom contains a triangle which itself does not generate any signal. Thus, in suitably positioned single-slice images we can observe black bars of zero intensity, as illustrated in Fig. 7.2 for two slices. With the SMS-sequence we can acquire a superposed k-space using the Fourier-encoding coefficients $\Xi_{1i} \equiv 1$ from (3.1), which corresponds to a coherent superposition of the single-slice signals. Hence, after performing an inverse Fourier transform on the encoded k-space, we expect to find two bars with two different gray scales (Fig. 7.2). As the base angle is given by $\beta = 45°$, the actual slice distance is equal to the length of the light-gray bar and can easily be measured using the online measurement tool of the SIEMENS SYNGO software installed on the scanner.

We perform this experiment for slice distances $d = 20$ mm and 60 mm (multiband factor $M = 2$), $d = 20$ mm ($M = 3$) and $d = 10$ mm ($M = 4$). We choose a small slice thickness ($\Delta z = 2$ mm) to get a sharp transition between the gray bars.

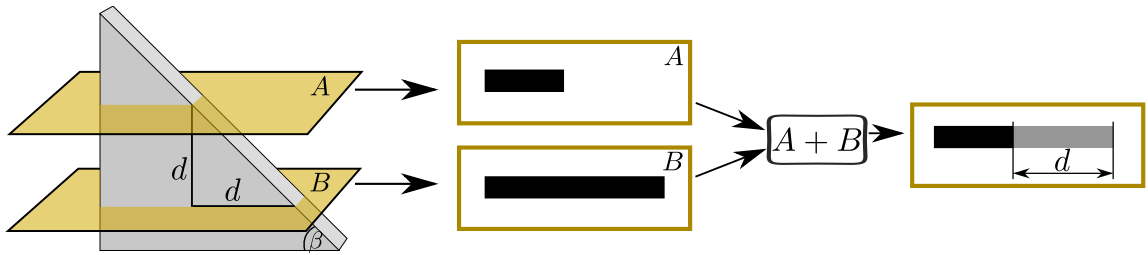

Fig. 7.2: Schematic for the validation of the slice distance $d$ of the SMS sequence. Isosceles triangle with $\beta = 45°$ (*left*), corresponding single-slice images (*center*) and measured superposition of single-slice images (*right*). *White:* full signal, *Black:* no signal.

**Results.** All investigated slice distances match the expectations. As an example, in Fig. 7.3 we show the reconstructed images for a dual-band acquisition with preset slice dis-tances 20 mm and 60 mm. The measured lengths are $d = 20.0(2)$ mm and $d = 60.0(2)$ mm respectively.

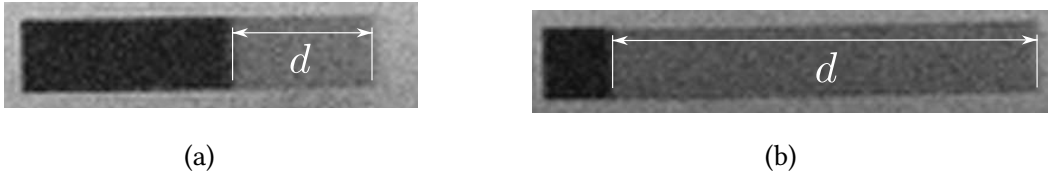

(a)                                              (b)

Fig. 7.3: Indirect measurement of the slice distance using the resolution phantom. (a) Set slice distance $d_{\text{set}} = 20$ mm, measured slice distance $d = 20$ mm. (b) $d_{\text{set}} = 60$ mm, $d = 60.0(2)$ mm.

**Discussion.** We can confirm the accuracy of the slice distances for multiband factors $M = 2 - 4$. We are confident that our implementation of the SMS-sequence also provides slice excitation with correct slice distances for higher multiband factors, since the shifts are implemented using a fixed formula which is the same for all multiband factors.

## 7.2.2. Flip angle

In the following we assure that the developed SMS-sequence flips the spins by the predefined angle.

**Methods.** From equation (2.91) we know that the signal intensity $S$ of a MRI measurement depends on the flip angle $\alpha$. For $TR \gg T_1$ and $TE \ll T_2^*$ the signal intensity is approximately given by $S = m_0 \sin \alpha$ (see (2.92)). In section 3.2.2 we showed that with reconstructions of type (3.4) we expect to find equal signal intensities for SMS and single-slice acquisitions, i.e. $S^{\text{SMS}} = S$. Hence, we can validate the flip angle of the developed SMS sequence by comparing its signal intensity to an equivalent single-slice sequence with verified flip angle.

To calculate the signal intensities, we use (3.4) and reconstruct the k-space $\boldsymbol{y}_q^{\text{SMS}}$ of slice $q$ from a SMS measurement. $\boldsymbol{y}_q$ denotes the k-space obtained by a conventional single-slice experiment. We do the signal intensity analysis in image space,

$$\begin{aligned} \boldsymbol{m}_q^{\text{SMS}} &= \mathcal{F}^{-1}(\boldsymbol{y}_q^{\text{SMS}}), \\ \boldsymbol{m}_q &= \mathcal{F}^{-1}(\boldsymbol{y}_q). \end{aligned} \tag{7.5}$$

Since we use multiple receiver coils we have information about the magnetization in each of the $N$ channels $\boldsymbol{m}_q = (m_q^1, \ldots, m_q^N)^T$. We combine the information of the coil array using the minimum-variance unbiased estimator[2],

$$m_q := \sum_{j=1}^{N} \frac{c_q^{j\,H} m_q^j}{\sum_{l=1}^{N} c_q^{l\,H} c_q^l}. \tag{7.6}$$

For this procedure the coil sensitivities $c_q^j$ for the $j$th coil and the $q$th slice are determined using the ESPIRiT algorithm [46] and verified as described in section 7.1. We then calculate the signal intensities,

$$S_q^{\mathrm{SMS}} := \sum_{\boldsymbol{r} \in \mathrm{ROI}} |m_q^{\mathrm{SMS}}(\boldsymbol{r})|,$$
$$S_q := \sum_{\boldsymbol{r} \in \mathrm{ROI}} |m_q(\boldsymbol{r})|, \tag{7.7}$$

where the ROI lies entirely inside the phantom.

We perform single-slice and SMS measurements with multiband factor $M = 2$ on the homogeneous phantom with flip angles from 5° to 30° and an increment of $\Delta = 5°$. The slice distance is $d = 20\,\mathrm{mm}$ and $N = 12$ channels of the headcoil are used. To fulfill the necessary conditions for approximation (2.92), we choose a long $TR = 5000\,\mathrm{ms}$ and a short $TE = 4.8\,\mathrm{ms}$. This prevents $T_1$ and $T_2$ weighting. To avoid long measurement times we reduce the base resolution to $64 \times 64$. We restrict ourselves to the analysis of slice $q = 1$ only.

**Results.**  Fig. 7.4 shows the signal intensity as a function of flip angle for the single- and the simultaneous multi-slice sequence as well as as a reference plot which implies the expected $\sin \alpha$ behavior. Both measurements show the trend $S \sim \sin \alpha$ predicted by (2.92). Furthermore, the values for the single-slice signal intensities are equal to the (normalized) SMS signal intensities, i.e. $S^{\mathrm{SMS}} \approx S$.

**Discussion.**  The equality of both intensities is a proof for the accuracy of the flip angle induced by the SMS sequence. In Fig. 7.4 we find small deviations from the theory with increasing $\alpha$. A possible explanation could be the approach of the limit for the small flip angle approximation, which we use for the pulse envelope design. In the scope of this thesis the flip angle $\alpha = 15°$ is utilized as it provides reasonable signal intensities while not exceeding the regime for which the small flip angle approximation holds.

---

[2]The proof for equation (7.6) is given by the Gauss-Markov theorem.

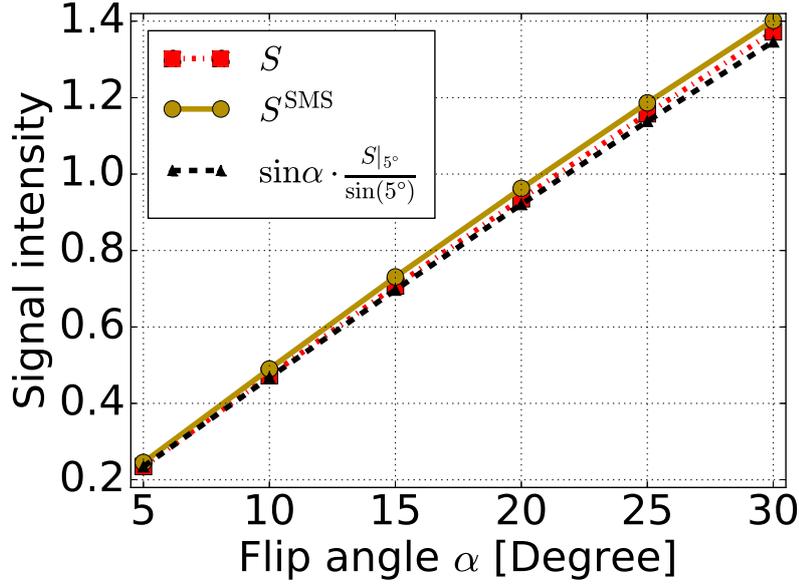

Fig. 7.4: Signal intensity as a function of the flip angle $\alpha$ for a single-slice sequence ($S$) and a SMS sequence ($S^{\text{SMS}}$, $M = 2$). The expected $S \sim \sin \alpha$ relation is plotted as a reference and is scaled to the value of $S$ at $\alpha = 5°$.

### 7.2.3. Signal-to-noise ratio

For a simultaneous multi-slice experiment with $M$ simultaneously excited slices, we expect a SNR increase of $\sqrt{M}$ compared to a conventional single-slice experiment (see section 3.2.2). This is one of the main advantages of SMS compared to conventional multi-slice. Here we verify this behavior for the developed sequence.

**Methods.** We utilize SMS sequences with multiband factor $M \geqslant 2$ as well as a single-slice sequence ($M = 1$) as a reference and perform measurements on the homogeneous phantom. In each receiver coil $j = 1, \ldots, N$ a fully sampled, Fourier-encoded k-space $\tilde{y}_p^j$ is acquire, where $p = 1, \ldots, M$ iterates the $M$ encodings. We decode this k-space using the normalized inverse DFT matrix (3.3). Hence, for each receiver coil $j$ and each slice $p$ we obtain the SNR-improved k-space $y_p^{\text{SMS},j}$. We determine the corresponding coil sensitivity $c_p^j$ using the ESPIRiT algorithm and apply the inverse Fourier transform $m_p^j = \mathcal{F}^{-1}(y_p^{\text{SMS},j})$. To compare the SNR for different measurements, we combine all $N$ coil images $m_p^j$ in each slice $p$ using the minimum variance unbiased estimator (7.6), which yields $m_p$. Since this operation is linear, it does not alter the SNR.

We use the differences method [48–50] to determine the actual SNR: Two measurements are performed for each investigated sequence. The SNR of slice $p = p_0$ is then given by

$$\text{SNR}_{p_0} = \frac{\frac{1}{2}\underset{\boldsymbol{r} \in \text{ROI}}{\text{mean}}\left(|m_{p_0}^I(\boldsymbol{r})| + |m_{p_0}^{II}(\boldsymbol{r})|\right)}{\frac{1}{\sqrt{2}}\underset{\boldsymbol{r} \in \text{ROI}}{\text{stdv}}(|m_{p_0}^I(\boldsymbol{r})| - |m_{p_0}^{II}(\boldsymbol{r})|)}. \tag{7.8}$$

Here $|m_{p_0}^{I/II}|$ are the magnitude values in image space of slice $p_0$ of measurement one/two, mean() returns the mean value and stdv() returns the standard deviation of their arguments with respect to the variables below them. The vector $\boldsymbol{r} \in \text{ROI}$ stands for a spatial position inside the ROI. We choose the ROI to lie entirely in the area of the phantom. In this region we can assume to have Gaussian noise although magnitude images are considered (see section 2.2.9). We make sure that we investigate the same slice $p_0$ when we compare the single-slice with the SMS measurements.

**Results.** We find very good agreement between the measured and the predicted SNR for all considered multiband factors. Fig. 7.5 depicts the SNR for multiband factors $M = 2, \ldots, 6$ compared to a single-slice measurement. On the left, the reconstructed slice is shown for $M = 1$ and $M = 6$. The ROI which is used for (7.8) is highlighted by the white dashed line. It can clearly be observed that the single-slice reconstruction contains more noise than the SMS one. On the right the SNR is plotted against the multiband factor. It increases from $\text{SNR}_{M=1} = 36$ with $\sqrt{M}$ to $\text{SNR}_{M=6} = 89$.

**Discussion.** Our sequence shows the expected SNR benefit of $\sqrt{M}$. We find both visual and quantitative improvements of the image quality. With the successful validation of the slice distance (section 7.2.1), the flip angle (section 7.2.2) and the SNR, we are confident that the developed SMS-sequence is accurate.

## 7.3. Undersampling scheme analysis

For fast imaging k-space undersampling is mandatory. Therefore, we need to find Cartesian undersampling patterns that provide both a high reduction factor and a low noise amplification (g-factor). For this pattern study we use a conventional linear SENSE reconstruction implemented in BART [47].

**Methods.** We choose a fixed multiband factor of $M = 2$, i.e. we have two k-spaces (one for each encoding) which can individually be undersampled. We introduce a Full/Ref and a CAIPIRINHA-like SMS undersampling scheme that both possess the same reduction factor. All utilized k-space patterns have a full center with $L_{ref} = 30$ reference lines. For the Full/Ref scheme we acquire the full k-space of one encoding, but only reference lines in the center of the other one. The second scheme resembles CAIPIRINHA-like [31] undersampling, where k-space line acquisition alternates between measurements. For this experiment the periphery of k-space is undersampled by a factor of $R = 2$, i.e. $R_{eff} = 1.79$. However, in general the scheme can be extended to higher reduction factors $R > 2$ and be applied to higher multi-band factors $M > 2$. If $R \geqslant M$, line acquisition alternates between all measurements such that in the periphery no line is recorded twice. If $R < M$, strict alternation is not possible and the CAIPIRINHA scheme is repeated cyclically to cover all measurements. Fig. 7.6 shows a schematic of different undersampling schemes.

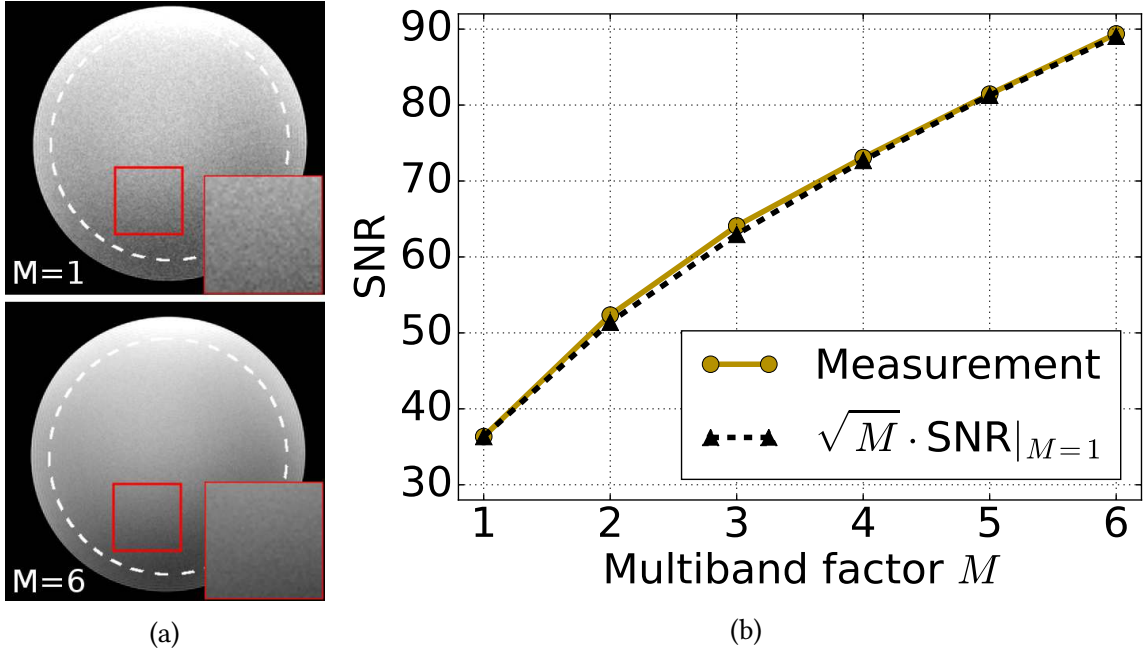

                  (b)

Fig. 7.5: SNR for different multiband factors. (a) Reconstructed slice image (minimum variance unbiased estimator) for multiband factors $M = 1$ and $M = 6$. The white dashed line indicates the ROI used for the SNR calculation. On the bottom right, a zoomed version of the red square is shown. (b) SNR as a function of the multiband factor $M$. The dashed line is used as a reference to show the expected SNR $\sim \sqrt{M}$ behavior. We have scaled the reference to the SNR value for $M = 1$.

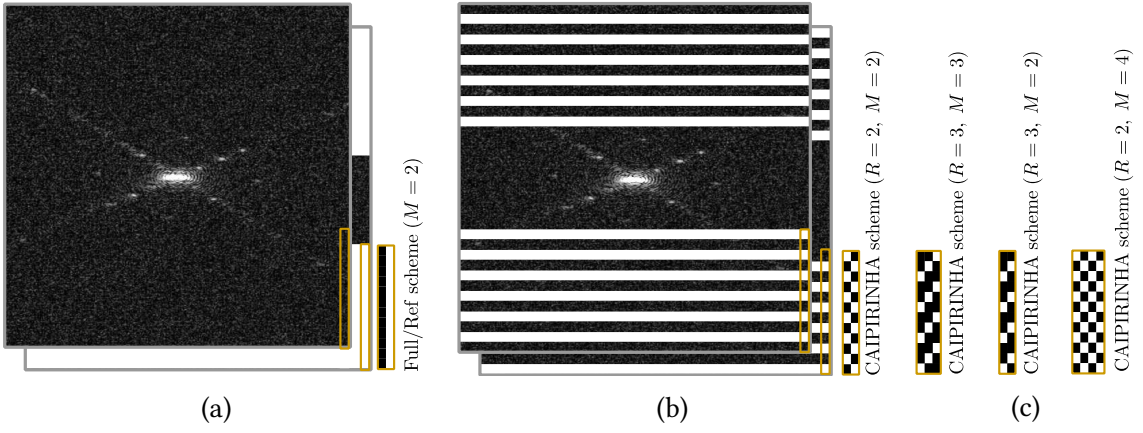

Fig. 7.6: Undersampling schemes for SMS acquisitions. Each subfigure (a) and (b) depicts a schematic of k-spaces with two undersampling patters - one for each measurement in a $M = 2$ acquisition. The yellow rectangle highlights the actual undersampling pattern in the periphery of k-space. (a) Full/Ref scheme: The k-space of one encoding is fully sampled, for the other encoding only reference lines in the center are acquired. (b) CAIPIRINHA scheme: The acquired k-space lines alternate between the two measurements. The centers are fully sampled. (c) CAIPIRINHA patterns for different values of $R$ and $M$.

The g-factor (2.89) for a reconstructed slice $q$ is given by,

$$g_q(\boldsymbol{r}) := \frac{\sqrt{(\sigma_q^{\text{red}}(\boldsymbol{r}))^2}}{\sqrt{R_{\text{eff}}} \cdot \sqrt{(\sigma_q^{\text{full}}(\boldsymbol{r}))^2}} = \frac{\sigma_q^{\text{red}}(\boldsymbol{r})}{\sqrt{R_{\text{eff}}} \cdot \sigma_q^{\text{full}}(\boldsymbol{r})}. \tag{7.9}$$

We use Monte-Carlo simulations to determine the pixelwise standard deviations $\sigma_q^{\text{full}}(\boldsymbol{r})$ and $\sigma_q^{\text{red}}(\boldsymbol{r})$ in image space. A schematic of the procedure is presented in Fig. 7.7. We start with artificial k-spaces for two encodings and $N$ coils that solely consist of Gaussian white noise with zero mean,

$$\begin{aligned} \tilde{Y}^{\text{full}} &:= (\tilde{\boldsymbol{y}}_1^{\text{full}}, \tilde{\boldsymbol{y}}_2^{\text{full}}), \\ \tilde{\boldsymbol{y}}_p^{\text{full}} &:= (\tilde{y}_p^{1,\text{full}}, \dots, \tilde{y}_p^{N,\text{full}}), \quad p = 1, 2. \end{aligned} \tag{7.10}$$

The corresponding undersampled k-spaces are denoted $\tilde{Y}^{\text{red}}$. We then apply a 3D SENSE[3] algorithm to $\tilde{Y}^{\text{full}}$ and $\tilde{Y}^{\text{red}}$ respectively and get noise images $\eta_q^{\text{full}}$ and $\eta_q^{\text{red}}$ for each slice $q$.

---

[3]The 3D SENSE algorithm takes the Fourier-encoding of the slice-dimension into account.

We repeat this procedure 400 times and calculate the pixelwise standard deviations

$$\sigma_q^{\text{full}}(\boldsymbol{r}) = \operatorname*{stdv}_{i=1,\ldots,400}(\eta_{q,i}^{\text{full}}(\boldsymbol{r})),$$
$$\sigma_q^{\text{red}}(\boldsymbol{r}) = \operatorname*{stdv}_{i=1,\ldots,400}(\eta_{q,i}^{\text{red}}(\boldsymbol{r})),$$

(7.11)

where $i$ counts the repetitions. We get g-factor maps for different slice distances by using appropriate coil sensitivities in the SENSE reconstruction. We obtain them by applying the ESPIRiT [46] algorithm to additional scans for different slice positions on the homogeneous phantom. The sensitivities are verified as described in section 7.1.

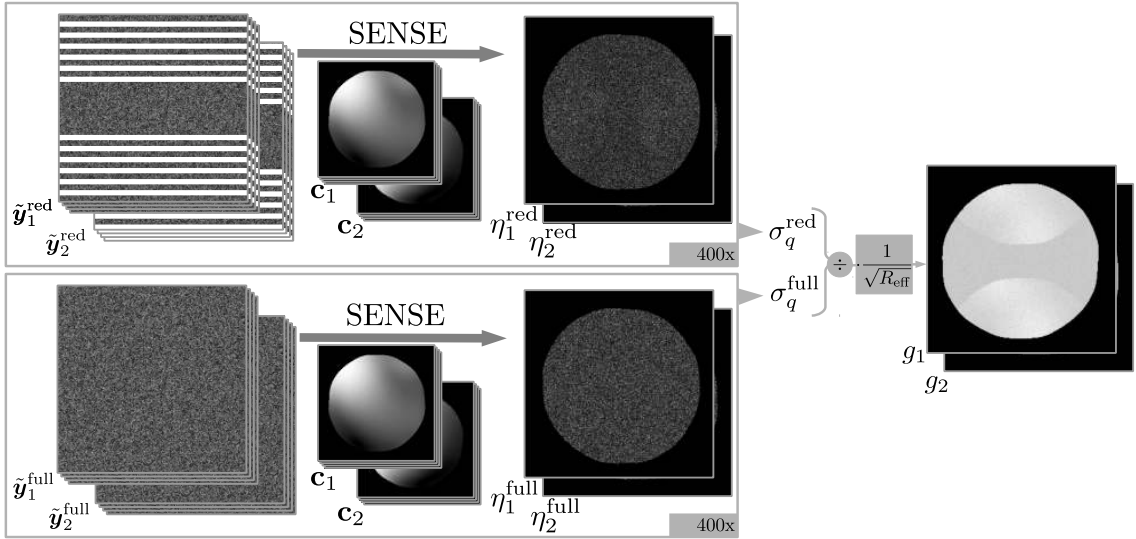

Fig. 7.7: Determination of multi-slice (multiband factor $M = 2$) g-factor maps for a SENSE reconstruction using Monte-Carlo simulations. Artificially created full/reduced k-spaces $\tilde{y}_1^{\text{full/red}}$ and $\tilde{y}_2^{\text{full/red}}$, which solely consist of Gaussian white noise, are reconstructed using the SENSE algorithm and explicit knowledge of the $N$ coil profiles $\boldsymbol{c}_p = (c_p^1, \ldots, c_p^N)^T$ of slice $p = 1, 2$. The pixelwise standard deviations $\sigma_q^{\text{full}}$ and $\sigma_q^{\text{red}}$ of 400 reconstructed image sets are calculated, divided and multiplied with $1/\sqrt{R_{\text{eff}}}$ to obtain the g-factor maps for both slices.

We compare the g-factor maps of the Full/Ref and the CAIPIRINHA scheme for slice distances $d = 10 - 90\,\text{mm}$ with increment $\Delta d = 10\,\text{mm}$. Of particular interest is furthermore the quantity $g_{\text{max}}$ from (2.90), but as we use Monte-Carlo simulations the maximum g-factor is unstable. Instead we introduce the definition

$$g_{\text{max}} := \max\left(\max_{99\%}(g_1(\boldsymbol{r})), \max_{99\%}(g_2(\boldsymbol{r}))\right),$$

(7.12)

where we determine the g-factor that is greater than 99 % of all the other values for both slices and define the bigger one as $g_{\text{max}}$.

**Results.** Fig. 7.8 shows the maximum g-factor $g_{\text{max}}$ of the Full/Ref and CAIPIRINHA scheme as a function of the slice distance $d$. The amplification of noise for both schemes is very different, although the effective reduction factor $R_{\text{eff}} = 1.79$ is the same. For all investigated slice distances the CAIPIRINHA scheme shows a lower $g_{\text{max}}$ than the Full/Ref one. For $d = 90\,\text{mm}$ we find the moderate g-factors $g_{\text{max}}^{\text{CAIPI}}|_{d=90\,\text{mm}} = 1.17$ and $g_{\text{max}}^{\text{Full/Ref}}|_{d=90\,\text{mm}} = 1.97$. Going to smaller slice distances, the noise amplification of both schemes rise. However, the g-factor of the CAIPIRINHA scheme increases only about 7 % and converges to $g_{\text{max}}^{\text{CAIPI}}|_{d=10\,\text{mm}} = 1.25$, whereas the Full/Ref scheme diverges to $g_{\text{max}}^{\text{Full/Ref}}|_{d=10\,\text{mm}} = 18.5$.

Fig. 7.9 shows the g-factor maps of the Full/Ref scheme (*top*) and CAIPIRINHA scheme (*bottom*) for slice distances $d = 10\,\text{mm}$ and $d = 90\,\text{mm}$. We only depict the maps $g_1(\boldsymbol{r})$ of slice one, since the maps for the second slice are similar. The Full/Ref maps show a relatively homogeneous noise amplification, which is notably bigger than in the CAIPIRINHA maps - especially for $d = 10\,\text{mm}$. The CAIPIRINHA maps show an enhanced g-factor where aliasing occurs due to the chosen undersampling pattern. In general we find higher g-factors for smaller slice distances.

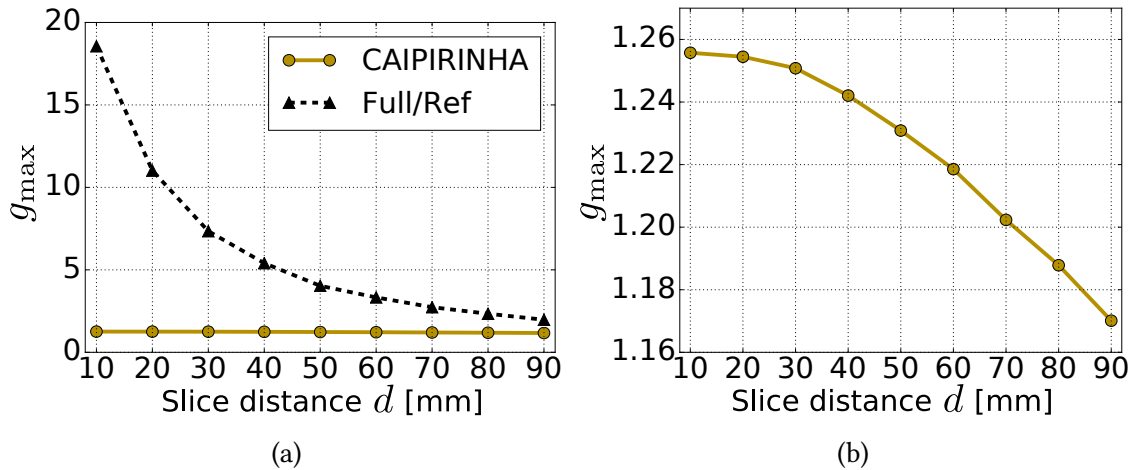

(a)  (b)

Fig. 7.8: Maximum g-factor $g_{\text{max}}$ as a function of the slice distance $d$ for a SENSE reconstruction with multiband factor $M = 2$ and reduction factor $R = 2$ ($L_{\text{ref}} = 30$ and $R_{\text{eff}} = 1.79$). Results of Monte-Carlo simulations with 400 realizations for each depicted slice distance. (a) $g_{\text{max}}$ for the CAIPIRINHA and Full/Ref scheme. (b) $g_{\text{max}}$ of the CAIPIRINHA scheme. (Same data as in (a) but windowed for more details.)

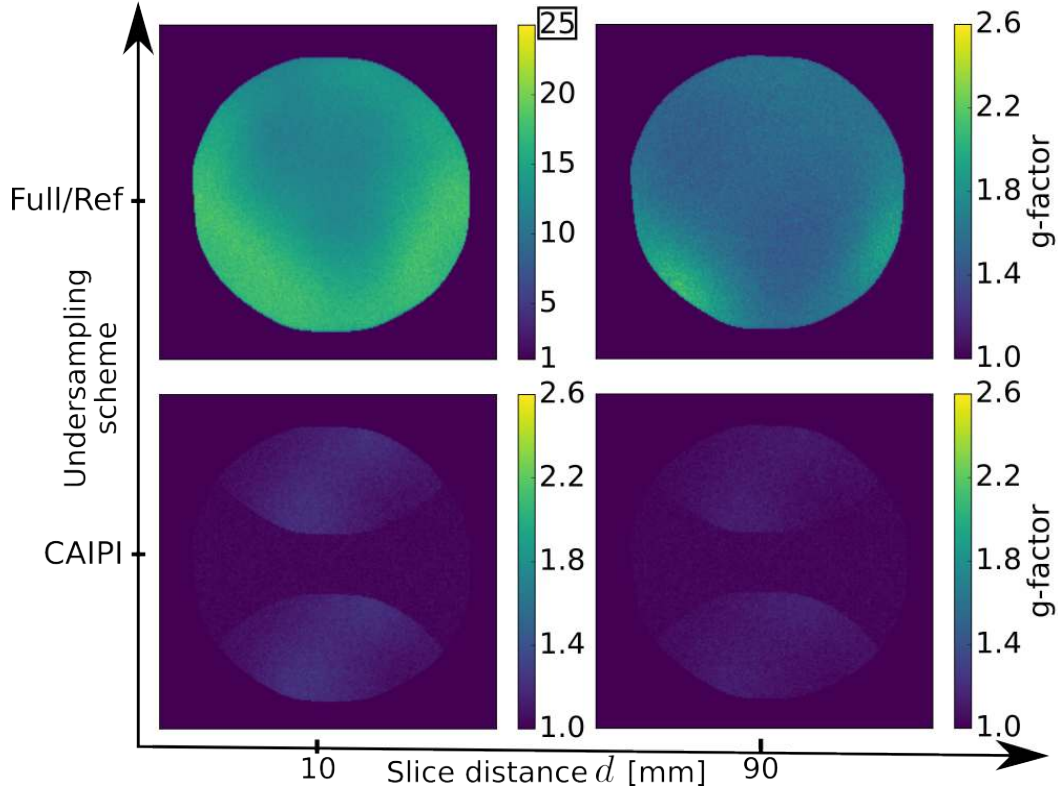

Fig. 7.9: g-factor maps of a SENSE reconstruction for different slice distances and undersampling schemes. Results of Monte-Carlo simulations with 400 realizations for each map, multiband factor $M = 2$ and reduction factor $R = 2$ ($L_{ref} = 30$ and $R_{eff} = 1.79$). Note the different scale of the top-left image.

**Discussion.** The SENSE algorithm utilizes coil sensitivities to disentangle the slices. For small slice distances, these sensitivities do not differ much in axial direction, which makes accurate disentangling difficult and causes an increased noise amplification. However, this negative effect is much less pronounced in the CAIPIRINHA scheme. This observation is well known and has been reported before [31, 36]. In the Full/Ref scheme we cannot take advantage of the Fourier-encoding of the slices, since only the periphery of one k-space is available. Therefore, the disentanglement of the slices solely relies on the coil sensitivities. By contrast, using the CAIPIRINHA scheme we can also exploit the information provided by the Fourier-encoding. There is an illustrative explanation why this leads to a beneficial noise amplification behavior of the CAIPIRINHA scheme: This scheme enforces an alternating sign in the periphery of the k-space of one slice, which corresponds to a FOV/2 shift of the respective slice in the image domain. Hence, the coil sensitivities of even very close slices will vary significantly due to in-plane sensitivity variations. This leads to a notably reduced g-factor. For large slice distances the coil profiles differ naturally, for what reason the g-factor of the Full/Ref scheme approaches the CAIPIRINHA scheme. For all future experiments we utilize the CAIPIRINHA scheme because of its favorable noise amplification characteristics.

## 7.4. Nonlinear reconstruction - SMS-NLINV

In this section we present the experiments and results for the newly developed simultaneous multi-slice Regularized Nonlinear Inversion (SMS-NLINV) algorithm (see section 5). We have implemented SMS-NLINV in BART [47]. For the initial guess $X_0$ we follow the suggestions given in the original NLINV paper [6] and set the magnetization of each slice $q$ to $m_q = 1$ and the coil sensitivities to $c_q^j = 0$. We choose the default values defined in BART for all other parameters of the algorithm, i.e. the Sobolev[4] index $l = 32$ and scaling factor $a = 220$, the initial regularization parameter[5] $\beta_0 = 1$ and its reduction factor $b = 1/2$.

We investigate the influence of the number of iteration steps, identify the benefit of SMS-NLINV over conventional multi-slice NLINV and analyze the noise amplification of SMS-NLINV. Finally, we test the reconstruction on SMS in-vivo data of a human brain.

### 7.4.1. Number of iteration steps

The SMS-NLINV algorithm applied to an accelerated SMS measurement has to disentangle the Fourier-encoded slices and eliminate aliasing, which occurs due to undersampling. To achieve these demands, a certain number of iteration steps (Newton steps) is necessary. In this section we inspect the image quality subject to the number of iterations.

**Methods.**   We perform SMS-NLINV reconstructions on the brick phantom for reduced k-space data ($R = 2, L_{ref} = 12, R_{eff} = 1.91$, CAIPIRINHA scheme), multiband factors $M = 2-5$ and a slice-to-slice distance of $d = 20\,\text{mm}$. We visually examine the resulting images to obtain the minimum number of iteration steps that provides clean reconstructions.

**Results.**   For the investigated multiband factors we can eliminate all occurring artifacts with 8 iteration steps. As an example, in Fig. 7.10 we show the reconstructed images of a dual-slice acquisition ($M = 2$) after $it = 5 - 8$ iteration steps. The two different slices can already be distinguished after $it = 5$ steps. However, we still find strong wrap-around artifacts and observe the other slice shining through in both images. Aliasing and incomplete disentangling are reduced by increasing the number of Newton steps and completely disappear for $it = 8$.

---

[4]See equation (5.22).
[5]See equation (5.7).

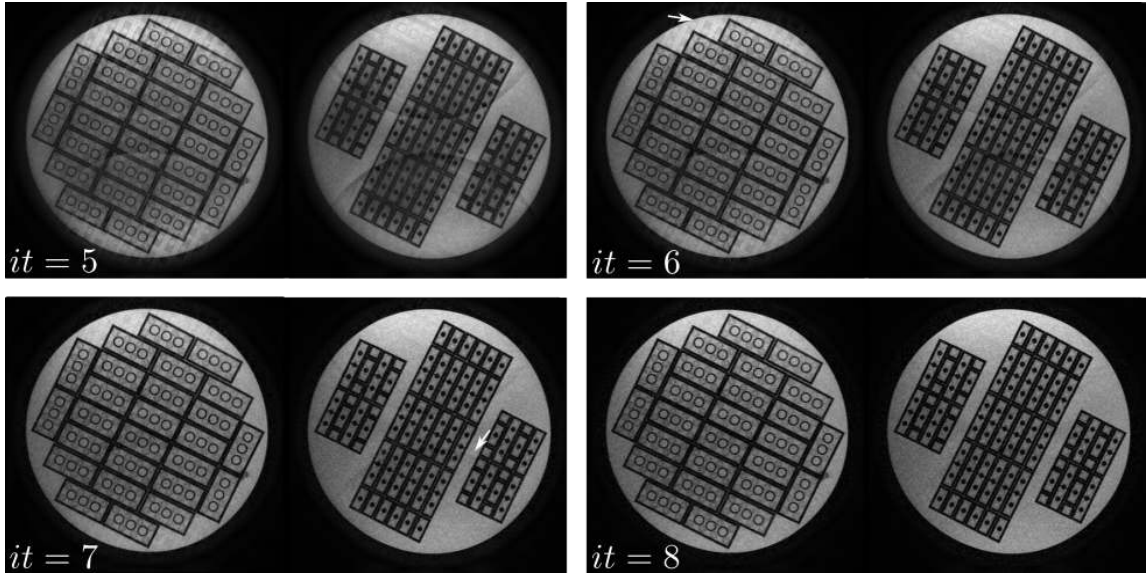

Fig. 7.10: Influence of the number of interation steps (*it*) on the SMS-NLINV reconstructions (reduction factor $R = 2$, $L_{ref} = 12$, $R_{eff} = 1.91$, CAIPIRINHA scheme) of two slices (multiband factor $M = 2$) and slice distance $d = 20$ mm. The white arrows highlight artifacts.

**Discussion**  For the presented experiment we obtain clear results with 8 Newton steps. However, the number of required iteration steps is influenced by the undersampling pattern, the slice distance, the number of used coils, the general quality of the measurement, the nature of the phantom and in particular the reduction factor (see section 7.4.2). Nevertheless, the proposed $it = 8$ Newton steps together with the presented initial values and choice of parameters should give reasonably good results for most reconstruction problems. There are several publications that suggest methods to find the optimal stopping parameter for inverse problems [51, 52], but to the best of our knowledge no satisfactory solution has yet been found. Note that we always have to deal with a trade-off between reduction of aliasing and noise amplification. We therefore recommend to choose the number of Newton steps to be as small as possible, since otherwise the reduction time and the g-factor will rise. Alternatively, a method called aNLINV [53] could be used to limit the noise amplification in image estimates even for a large number of Newton steps, by applying additional regularizations and incorporating the noise covariance of the coil array in the minimization function.

If not stated differently, we use $it = 8$ iteration steps for all further SMS-NLINV reconstructions.

## 7.4.2. Advantage of SMS over conventional multi-slice

In this section we emphasize the advantage of SMS imaging over a conventional multi-slice experiment, where each slice is recorded and reconstructed individually.

**Methods.** We compare a SMS measurement with a series of single-slice acquisitions. We utilize the brick phantom and measure $M = 6$ slices with slice-to-slice distance $d = 15\,\mathrm{mm}$. For the SMS sequence we use a CAIPIRINHA undersampling scheme with reduction factor $R = 4$ ($L_{\mathrm{ref}} = 12$, $R_{\mathrm{eff}} = 3.51$). For all single-slice measurements we utilize a conventional (single-slice) pattern of the same kind. We perform the reconstruction with SMS-NLINV or NLINV respectively and use equivalent parameters for both algorithms.

**Results.** The SMS-NLINV algorithm needs fewer iteration steps to eliminate aliasing and achieves visually better results than what we get from single-slice reconstructions with NLINV. In Fig. 7.11 we show two of the six images that we get from single-slice NLINV reconstructions after $it = 10 - 12$ newton steps, together with the corresponding SMS-NLINV images after $it = 10$ iterations. In the NLINV images we observe aliasing for $it = 10$ and $it = 11$. After $it = 12$ iterations the wrap-around artifacts are almost completely eliminated, but due to the high noise the images are practically useless. By contrast, the SMS-NLINV images do not possess any noticeable artifacts after $it = 10$ iterations while the noise is still reasonable. The accumulated duration of six NLINV reconstructions with $it = 12$ iterations is approximately the same as the $it = 10$ SMS-NLINV reconstruction.

**Discussion.** The single-slice NLINV reconstructions cannot eliminate aliasing without causing unacceptable noise amplification, whereas SMS-NLINV provides useful results. An explanation can be given by considering the actual reconstruction strategies of both techniques. NLINV performs solely in-plane reconstruction, i.e. the coil profile information of a particular slice is exploited to fill the undersampled k-space of the same slice. By contrast, with SMS-NLINV sensitivity encoding is not only performed in phase- but also in axial direction. Therefore, we can exploit the additional information provided by the coil array to accelerate in the third dimension, too. However, the advantage of SMS-NLINV over NLINV becomes less pronounced for small slice distances. The difference between the coil profiles of adjacent slices vanishes, thus sensitivity information in axial direction is no longer given. Hence, we can solely exploit in-plane sensitivity variations and only the SNR benefit remains.

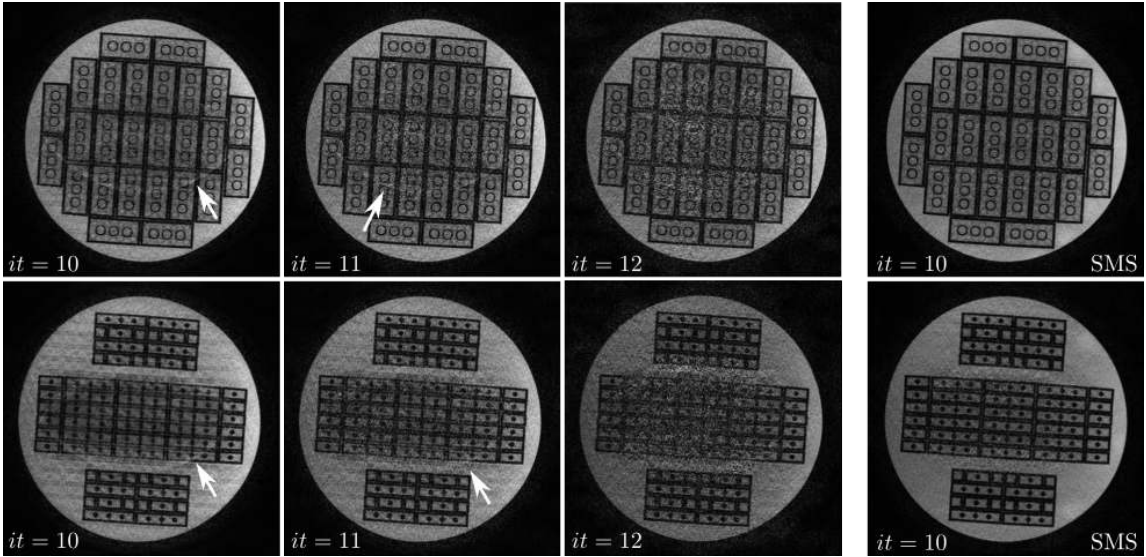

Fig. 7.11: Comparison of conventional multi-slice and simultaneous multi-slice reconstructions. Acquisition of $M = 6$ slices (two depicted), slice-to-slice distance $d = 15\,\text{mm}$ and reduction factor $R = 4$ ($L_{\text{ref}} = 12$, $R_{\text{eff}} = 3.51$, CAIPIRINHA scheme for SMS). *Left*: NLINV reconstruction of individual slices after $it = 10 - 12$ Newton steps. *Right*: SMS-NLINV reconstruction after $it = 10$ Newton steps. The white arrows indicate aliasing artifacts.

### 7.4.3. Noise amplification

In this section we evaluate the noise amplification of SMS-NLINV reconstructions.

**Methods.** We choose the multiband factor $M = 2$, the CAIPIRINHA undersampling scheme with reduction factor $R = 2$ ($L_{\text{ref}} = 12$, $R_{\text{eff}} = 1.91$) and use Monte-Carlo simulations to determine the g-factor maps. We follow the idea described in section 7.3 but adapt the simulations to make it work for SMS-NLINV. This algorithm needs actual object information in k-space and does not work properly with noise only. We therefore start with a dummy phantom (homogeneous disk) $m_{\text{disk}}$ and mimic a measurement by multiplying it with coil sensitivities $c_q = (c_q^j, \ldots, c_q^N)^T$ determined in previous measurements,

$$m_{\text{disk}} \cdot c_q = m_q. \qquad (7.13)$$

Here $q = 1, 2$ indexes the slices. We transform the resulting magnetization images $m_q = (m_q^1, \ldots, m_q^N)^T$ into k-space using a Fourier transform and Fourier-encode the k-spaces using the DFT matrix $\Xi_{pq}$ (3.1). This yields the noiseless k-spaces $\tilde{\mathbf{Y}}_p$. We add Gaussian white noise with zero mean and standard deviation $\sigma_\eta$ to these k-spaces and get $\tilde{y}_{1/2}^{\text{full}}$ or - application of the CAIPIRINHA scheme - $\tilde{y}_{1/2}^{\text{red}}$. For reconstruction we use SMS-NLINV and obtain the g-factor maps as described in section 7.3. For analysis we choose a ROI determined by the dummy phantom. A flow chart of this process is depicted in Fig. 7.12.

We use coil sensitivities $c_{1/2}$ that correspond to the slice distance $d = 60\,\text{mm}$ and investigate the maximum g-factor for different ratios $S_{\text{max}}/\sigma_\eta$, where $S_{\text{max}}$ is the maximum k-space signal (i.e. the absolute value of the DC component) and $\sigma_\eta$ the noise standard deviation in k-space. We determine values $g_{\text{max}}$ for ratios from $S_{\text{max}}/\sigma_\eta \approx 25$ to $S_{\text{max}}/\sigma_\eta \approx 3000$.[6] We furthermore calculate $g_{\text{max}}$ for slice distances $d = 10 - 90\,\text{mm}$ and fixed $S_{\text{max}}/\sigma_\eta \approx 3000$.

**Results.** We find an explicit dependence of the g-factor on $S_{\text{max}}/\sigma_\eta$. Fig. 7.13a shows that the maximum g-factor is approximately constant ($g_{\text{max}} \approx 1.15$) for $S_{\text{max}}/\sigma_\eta > 10^2$. For lower ratios the g-factor increases up to $g_{\text{max}} = 1.27$ at $S_{\text{max}}/\sigma_\eta \approx 25$. In Fig. 7.13b we observe a behavior similar to the linear case (Fig. 7.8b). The maximum g-factor is low for large slice distances ($g_{\text{max}}|_{d=90\,\text{mm}} = 1.112$) and for smaller distances converges to a 6 % higher value ($g_{\text{max}}|_{d=10\,\text{mm}} = 1.175$).

Fig. 7.14 shows the g-factor maps for slice distances $d = 10\,\text{mm}$ and $d = 90\,\text{mm}$ and for $S_{\text{max}}/\sigma_\eta \approx 25$ and $S_{\text{max}}/\sigma_\eta \approx 3000$. The maps are windowed to the size of the dummy phantom. Similar to the g-factor maps of the linear reconstruction (Fig. 7.9) we find a pronounced noise amplification where wrap-around artifacts have to be eliminated. Furthermore, the g-factor in the aliasing region is noticeably bigger for smaller slice distances than for large ones, whereas for the non-aliased area there is not much difference. We find a uniform increase of the g-factor in the entire phantom when we go from the high value $S_{\text{max}}/\sigma_\eta \approx 3000$ to the significantly lower $S_{\text{max}}/\sigma_\eta \approx 25$.

---

[6]The actual values of $S_{\text{max}}/\sigma_\eta$ are determined by considering the k-spaces with coherent superposition (i.e. $\Xi_{1q}$) and choosing the minimum of all channels. However, the order of the values is the same for all coils and encodings.

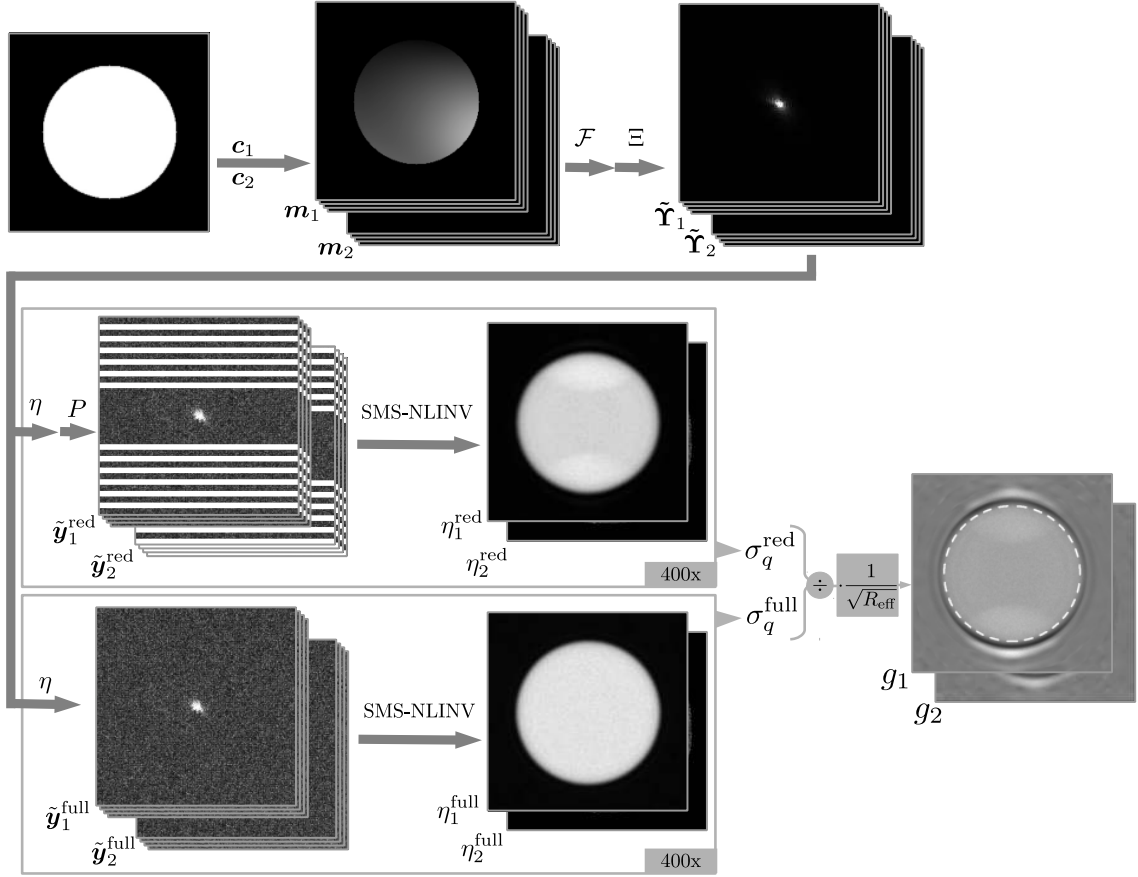

Fig. 7.12: Determination of multi-slice (multiband factor $M = 2$) g-factor maps for SMS-NLINV using Monte-Carlo simulations. A dummy phantom (homogeneous plate) is weighted with the coil sensitivities $c_q$ of slice $q = 1$ and $q = 2$. The resulting coil images $m_q$ are transformed into k-space ($\mathcal{F}$) and Fourier-encoded ($\Xi$) which results in the noiseless k-spaces $\tilde{\Upsilon}_p$. We add Gaussian white noise ($\eta$) to these k-spaces and reconstruct the full/reduced k-spaces $\tilde{y}_{1/2}^{\text{full/red}}$ to get the slice images $\eta_{1/2}^{\text{full/red}}$. The undersampling is achieved using the sampling operator $P$. The pixelwise standard deviations $\sigma_q^{\text{full}}$ and $\sigma_q^{\text{red}}$ of the 400 reconstructed image sets $\eta_q^{\text{full/red}}$ are calculated, divided by one another and multiplied with $1/\sqrt{R_{\text{eff}}}$ to obtain the g-factor maps $g_1$ and $g_2$. We analyze only the region inside the dummy phantom (dashed circle).

**Discussion.** Whereas for linear reconstructions the g-factor is independent from the k-space noise standard deviation, the situation is different for nonlinear methods. With SMS-NLINV we do not have (and do not require) any prior information - such as coil profiles - about the system, but rely on a signal which stands out from the noise. If signal contamination is too severe, we get non-linear effects and the reconstruction amplifies

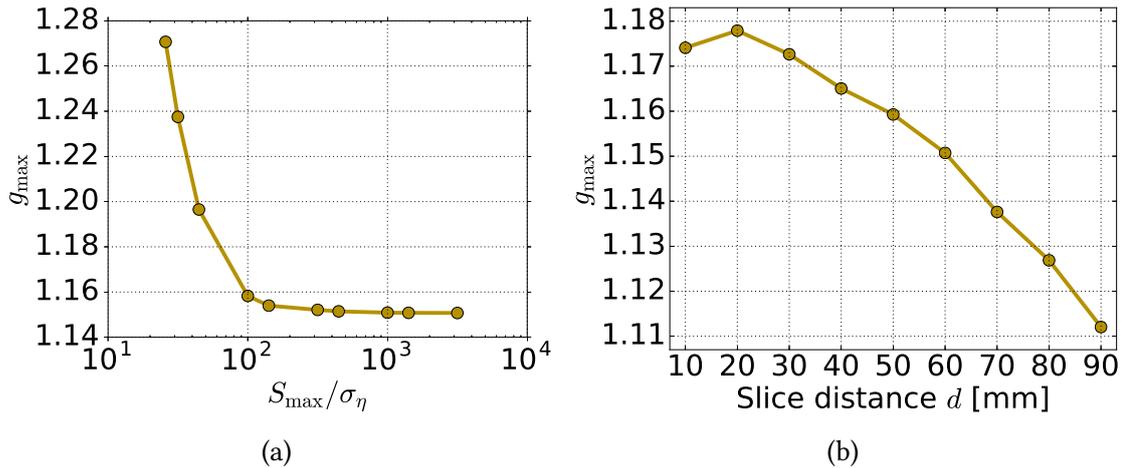

(a)    (b)

Fig. 7.13: Maximum g-factor $g_{max}$ of SMS-NLINV reconstructions with multiband
factor $M = 2$ and reduction factor $R = 2$ ($L_{ref} = 12$, $R_{eff} = 1.91$, CAIPIR-
INHA scheme). Results of Monte-Carlo simulations with 400 realizations
for each value of the maximum-k-space-signal to noise ratio $S_{max}/\sigma_\eta$ and
slice distance $d$. (a) $g_{max}$ as a function of $S_{max}/\sigma_\eta$ for $d = 60$ mm (logarithmic
scale). (b) $g_{max}$ as a function of $d$ for $S_{max}/\sigma_\eta \approx 3000$.

noise more than linear algorithms do, whereas for higher values of $S_{max}/\sigma_\eta$ the g-factor
stays constant. SMS measurements ($M = 2$) on the homogeneous phantom reveal that for
our scanner and coil system we have $S_{max}/\sigma_\eta > 10^2$ for any coil. Thus, we do not expect
any nonlinear noise amplification effects for reconstructions of actual measurements.
Moreover, Sénégas and Uecker found that noise amplification for single-slice NLINV and
SENSE is comparable [53], which should also hold for SMS experiments.

In analogy to section 7.3, the differing coil profiles of faraway slices and the correspond-
ing exploitation of sensitivity encoding in axial direction improves the g-factor compared
to spatially close slices. Therefore, we would expect to find the maximum value of $g_{max}$ at
the smallest slice distance $d = 10$ mm. However, we find it at $d = 20$ mm which is probably
an averaging error due to the finite number of simulations. One possibility to correct
it could be to perform more than the used 400 simulations for g-factor determination.
Alternatively one could investigate whether formula (2.90) for $g_{max}$ should be adapted to
get more accurate results. Nevertheless, the overall behavior of Fig. 7.13b matches the
expectations well.

Note that even if we approach regions where $S_{max}/\sigma_\eta < 10^2$, i.e. nonlinear reconstruc-
tion effects come into play, the actual g-factor maps show a homogeneously increased
noise amplification distributed over the entire ROI and do not exhibit unexpected artifacts
such as singularities.

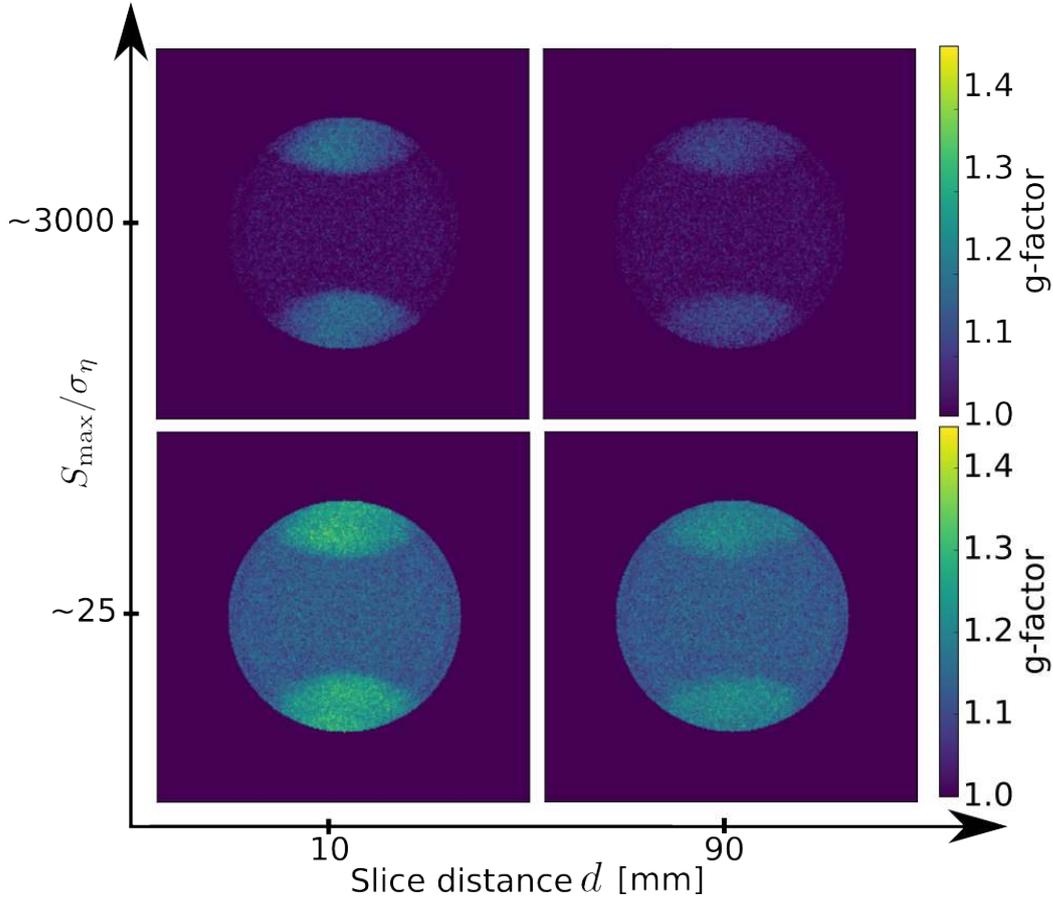

Fig. 7.14: g-factor maps of a SMS-NLINV reconstruction for different slice distances $d$ and maximum-k-space-signal to noise ratios $S_{max}/\sigma_\eta$. Results of Monte-Carlo simulations with 400 realizations for each map, multiband factor $M = 2$ and reduction factor $R = 2$ ($L_{ref} = 12$, $R_{eff} = 1.91$, CAIPIRINHA scheme). The maps are windowed to the size of the dummy phantom.

### 7.4.4. In-vivo SMS-NLINV reconstructions

In this section we test the developed SMS sequence and SMS-NLINV reconstruction in-vivo for different slice distances and reduction factors.

**Methods.** We investigate a human brain and utilize the multiband factor $M = 5$ (FOV$_{read/phase}$ = 210 mm). We perform two experiments: In the first one we set the slice distance to $d = 10$ mm and the reduction factor to $R = 2$ ($L_{ref} = 12$, $R_{eff} = 1.91$, CAIPIRINHA scheme). In the second experiment we use $d = 20$ mm and $R = 4$ ($L_{ref} = 12$, $R_{eff} = 3.51$, CAIPIRINHA scheme). To recover aliasing-free slices, we perform $it = 8$ iterations for the first experiment and $it = 11$ for the second. To get a reference, we disentangle the Fourier-encoded, full k-spaces according to (3.4) and reconstruct the slice images by a conventional inverse Fast Fourier transform followed by a RSS combination.

**Results.**   Fig. 7.15 shows all 5 slices reconstructed with SMS-NLINV as well as the corresponding reference images for both experiments. All slices could be recovered successfully without noticeable wrap-around artifacts. In the first experiment ($d = 10\,\mathrm{mm}$, $R = 2$) the SMS-NLINV reconstruction exhibits only marginally more noise than the reference. In the second experiment ($d = 20\,\mathrm{mm}$, $R = 4$) the quality in the central areas of the SMS-NLINV images begins to deteriorate, but we still get competent results. In Fig. 7.16 we also depict an excerpt of coil sensitivities estimated with SMS-NLINV for the second experiment. All profiles are smooth as expected.

**Discussion.**   The results show that the SMS sequence together with the SMS-NLINV reconstruction gives promising results in-vivo. For moderate reduction factors such as $R = 2$ we get good results that are comparable to fully sampled reconstructions. By increasing the reduction factor we also increase the number of wrapped replicates. It follows that particularly in the image center we have to disentangle many superposed pixel which leads to increased noise corruption. The reduction factor $R = 4$ seems rather moderate compared to earlier publications [6, 54]. However, the reported higher reduction factors could only be achieved for slices extracted from 3D scans, where undersampling can be distributed among both slice dimensions. This is known to improve the image quality [55, 56]. For 1D sensitivity encoding, the critical reduction factor is approximately $R = 4$ [57] and higher values of $R$ are only possible by exploiting coil profile variations in more dimensions [56]. In our experiments we do not achieve reduction factors higher than $R = 4$, although we make use of sensitivity encoding in phase and axial direction, because the utilized 20 channel headcoil seems not to provide sufficient sensitivity variation - in particular not in axial direction. The profiles of adjacent slices of any receive channel are mostly quite similar, which limits the exploitation of through-plane sensitivity encoding. More coil profile variations and thus higher reduction factors should be feasible when using coils with more channels.

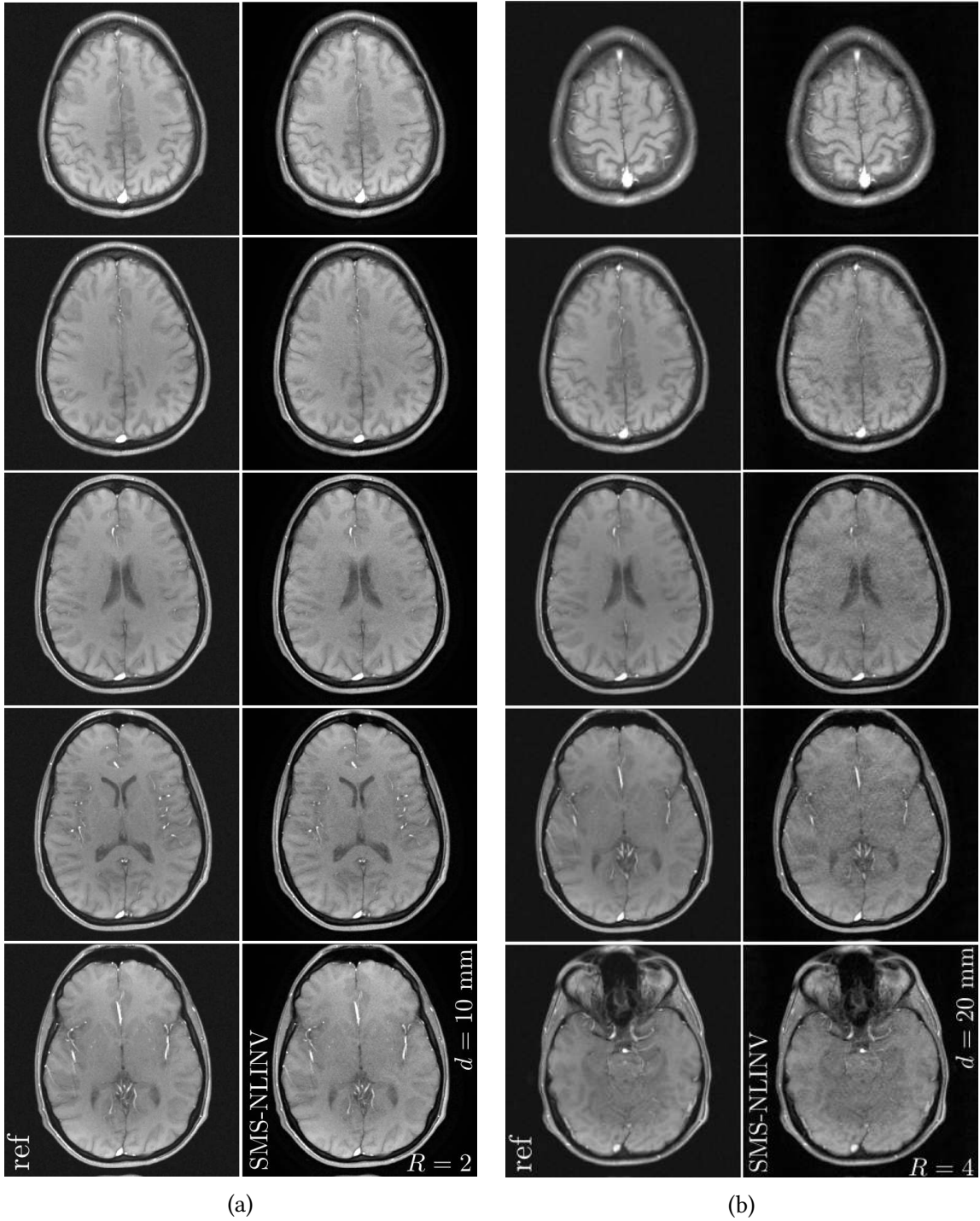

(a)                                    (b)

Fig. 7.15: Reconstructed slices of an in-vivo SMS measurement of a human brain with multiband factor $M = 5$. RSS reference reconstructions from Nyquist sampled single-slice k-spaces and SMS-NLINV reconstructions from undersampled Fourier-encoded k-spaces are depicted. (a) Slice distance $d = 10\,\text{mm}$, reduction factor $R = 2$ ($L_{\text{ref}} = 12$, $R_{\text{eff}} = 1.91$, CAIPIRINHA scheme), $it = 8$ Newton steps. (b) Slice distance $d = 20\,\text{mm}$, reduction factor $R = 4$ ($L_{\text{ref}} = 12$, $R_{\text{eff}} = 3.51$, CAIPIRINHA scheme), $it = 11$ Newton steps.

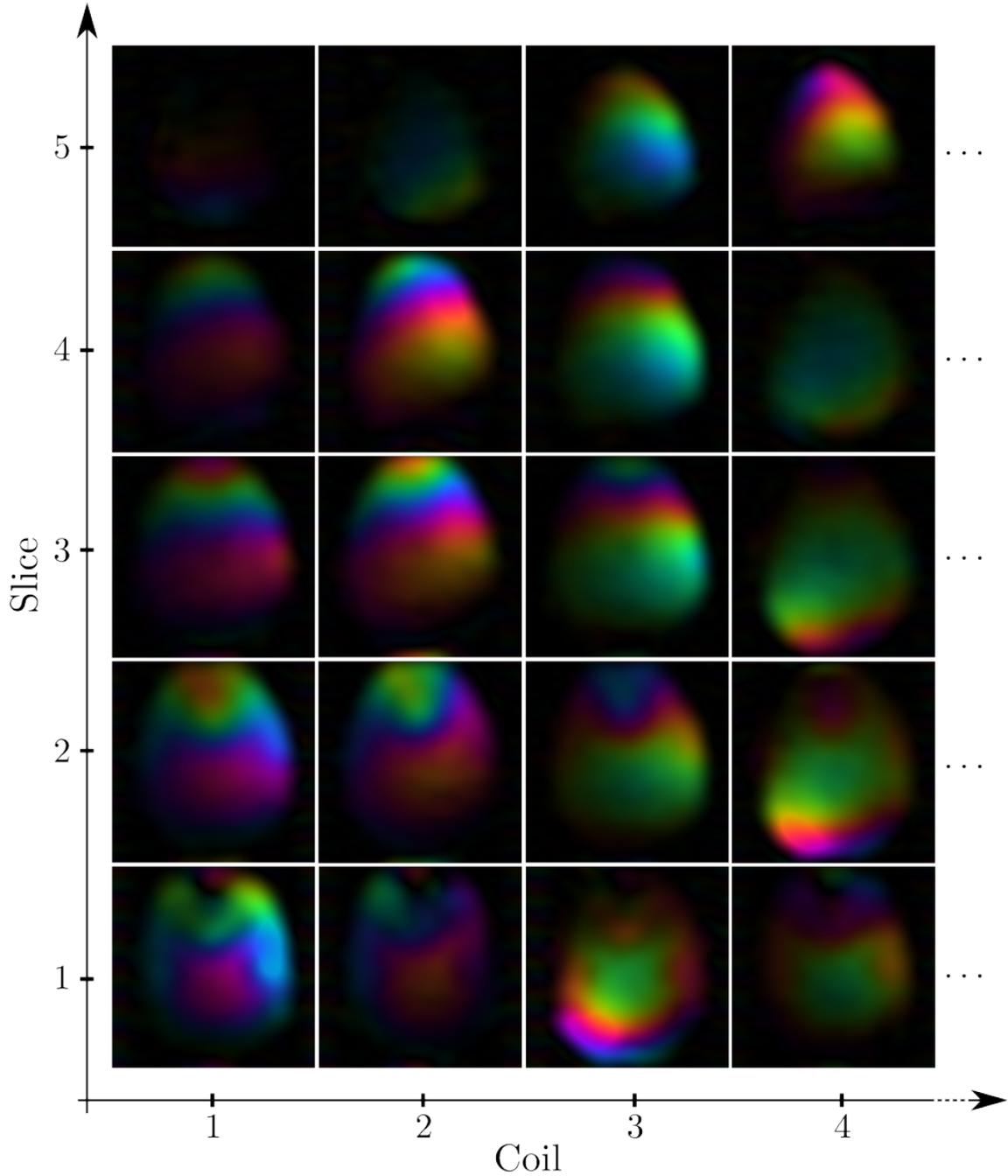

Fig. 7.16: Complex coil sensitivities obtained by a SMS-NLINV reconstruction of an in-vivo measurement of a human brain with multiband factor $M = 5$, slice distance $d = 20\,\text{mm}$, reduction factor $R = 4$ ($L_{\text{ref}} = 12$, $R_{\text{eff}} = 3.51$, CAIPIRINHA scheme), $it = 11$ Newton steps. The phase is given by the color, the amplitude is given by the brightness. Depicted are 4 out of 20 profiles for each slice. The sensitivities correspond to the reconstruction in Fig. 7.15b.

# 8. Conclusion and outlook

In this thesis we have developed a Cartesian simultaneous multi-slice (SMS) sequence and a nonlinear reconstruction approach (SMS-NLINV) which allows for a joint estimation of coil sensitivities and image content for all slices.

We have performed several tests to verify the accuracy of the sequence. In particular we have checked the fidelity of the slice distance as well as the flip angle. Moreover we have confirmed the square-root-like signal-to-noise ratio benefit of SMS compared to conventional multi-slice experiments. A g-factor analysis of SMS undersampling schemes revealed the beneficial characteristics of CAIPIRINHA-like patterns, where k-space line acquisition is alternated between different measurements. Especially for small slice distances this undersampling strategy results in a significantly reduced noise amplification.

We have extended the Regularized Nonlinear Inversion (NLINV) [6] algorithm for the reconstruction of Fourier-encoded simultaneous multi-slice data (SMS-NLINV). NLINV is known to provide an improved estimation of coil sensitivities and a considerable reduction of artifacts for high acceleration factors compared to linear autocalibrating parallel MRI [6]. We can now exploit these benefits for simultaneous multi-slice experiments. What is more, we have demonstrated that SMS-NLINV can deal with even higher reduction factors and achieves yet better slice images than NLINV due to an enhanced SNR and the utilization of through-plane sensitivity encoding. The reconstruction time for all slices is not notably affected. From g-factor studies we could conclude that non-linear noise amplification effects do not have a notable impact on the reconstructed images. Last but not least, the performed experiments on a human brain with reduction factors up to $R = 4$ have revealed the capability of SMS-NLINV for in-vivo studies. We expect significant improvements in terms of $R$ and a even more pronounced benefit of SMS-NLINV over single-slice NLINV, when experiments are performed with better coils such as the SIEMENS 32-Channel Head Coil or the SIEMENS Head/Neck 64.

Nevertheless, it was not the intention of this thesis to push the reduction factor for Cartesian sampling to its limit. It is generally acknowledged that non-Cartesian sampling schemes provide significantly improved results for high values of $R$ [58, 59]. For time-critical imaging such as real-time cardiac MRI, radial k-space trajectories are the means of choice [35, 60, 61]. This sampling scheme consists of a set of rotated lines (spokes) that all cross the k-space center, which entails several benefits compared to conventional Cartesian sampling. (i) Radially undersampled data suffer from streaking artifacts which deteriorate the image muss less than aliasing does in Cartesian sampling. (ii) Rectilinear sampling lines are not equivalent as they encode either low or high spatial frequencies (in PE-direction). By contrast, all spokes cover both high and low frequencies, which results in a more consistent sampling, especially for moving objects. (iii) Radial trajectories provide an inherently oversampled k-space center. First, a dense sampling of low spatial frequencies is a requirement for SMS-NLINV. Second, local undersampling increases the more we approach the k-space periphery which contains information that is less important for the image quality. Fortunately, (SMS-)NLINV can be extended to non-Cartesian sampling schemes simply by introducing a gridding operator [35]. In Fig. 8.1 we depict the first result for the SMS-NLINV reconstructions of one measurement on the brick phantom with radial k-space trajectory (192 spokes with base resolution 192, slice distance $d = 60$ mm).

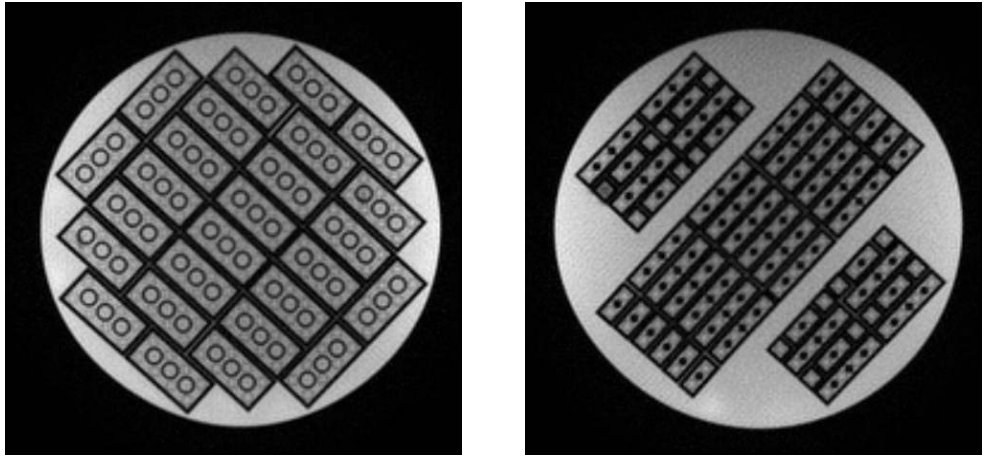

Fig. 8.1: SMS-NLINV reconstruction of a measurement on the brick phantom with radial k-space trajectory (192 spokes with base resolution 192, slice distance $d = 60$ mm).

A logical next step is the implementation of simultaneous multi-slice into the conventional multi-slice real-time MRI sequence that is based on NLINV and which is already extensively being used in clinical and pre-clinical practice by several groups in Göttingen [61–64]. The adaption of the sequence and the offline reconstruction should not pose any major problems. Nevertheless, to accomplish actual live movies, much effort has to be put into a fast and efficient implementation of SMS-NLINV on graphical processor units (GPUs).

Another promising idea would be the use of a balanced Steady State Free Precession (bSSFP) sequence, which provides a higher SNR and a better contrast, especially for cardiac imaging[1], than FLASH. However, bSSFP is more prone to off-resonance artifacts and magnetic field inhomogeneities. Furthermore, strategies to reduce the SAR and the RF peak amplitude in SMS bSSFP sequences have to be considered, which we could avoid by choosing FLASH. What is more, because of the alternating phase of the RF pulse in the bSSFP sequence, the implementation of the Fourier-encoding becomes non-trivial.

To reduce noise in the reconstructions, Knoll et al. described an approach to include variational penalties in parallel imaging with nonlinear inversion [54]. They showed how the advantageous properties of e.g. Total Variation (TV) based regularizations can be exploited to improve the image quality especially for high reduction factors. This would be a promising extension to the described L2-regularized SMS-NLINV algorithm.

After all, SMS-NLINV is a very general reconstruction approach and can therefore be applied to all kinds of experiments that make use of simultaneous multi-slice, such as perfusion, Diffusion Tensor Imaging, functional MRI or $T_1/T_2$ quantification.

---

[1]In particular, a better contrast between blood and the myocardium is found when using bSSFP instead of FLASH [65].

# A. Appendix

## A.1. Symbols, notation and abbreviations

The terms coil profiles/sensitivities are used interchangeably.

Tab. A.1: Symbols, notation and abbreviations.

| | | | |
|---|---|---|---|
| NMR | Nuclear magnetic resonance | $d$ | Slice distance |
| MRI | Magnetic resonance imaging | $T_{\mathrm{RF}}$ | RF excitation pulse duration |
| SMS | Simultaneous multi-slice | $\boldsymbol{y}$ | k-spaces for all coils |
| SNR | Signal-to-noise ratio | $y^j$ | k-space of coil $j$ |
| FOV | Field of view | $y_q$ | k-space of slice $q$ |
| RSS | Root sum of squares | $\boldsymbol{m}$ | Magnetizations for all coils |
| ROI | Region of interest | $m^j$ | Magnetization of coil $j$ |
| SAR | Specific absorption rate | $m_q$ | Magnetization of slice $q$ |
| RF | Radiofrequency | $\boldsymbol{c}$ | Coil sensitivities for all coils |
| PE | Phase-encoding | $c^j$ | Coil sensitivity of coil $j$ |
| RO | Read-out | $c_q$ | Coil sensitivity of slice $q$ |
| BART | Berkeley Advanced Reco. Toolbox [47] | $\hat{c}$ | Normalized coil sensitivity |
| NLINV | Regularized Nonlinear Inversion [6] | $\tilde{\phantom{x}}$ | Fourier-encoded quantity |
| SENSE | Sensitivity Encoding [26] | $^H$ | Adjoint |
| $TE$ | Echo time | $^T$ | Transposed |
| $TR$ | Repetition time | $^*$ | Complex conjugate |
| $\Delta z$ | Slice thickness | | |

## A.2. Precession in the classical vector model

The classical vector model provides an intuitive insight into NMR. Precession and the Larmor frequency can by derived by using quantities and equations known from classical mechanics. In this model spins are treated just like angular momenta $S$. Its relationship to the magnetic moment $\boldsymbol{\mu}$ is

$$\boldsymbol{\mu} = \gamma S. \tag{A.1}$$

In MRI, we measure all spins in a voxel. Thus, we are not interested in the behavior of a single spin but of a spin-ensemble. We therefore define the net spin $S_{\text{net}}$ and the magnetization vector $\boldsymbol{m}$ of a macroscopic volume $\mathcal{V}$,

$$S_{\text{net}} = \sum_{n=1}^{N_s} S_n, \tag{A.2}$$

$$\boldsymbol{m} = \frac{\sum_{n=1}^{N_s} \boldsymbol{\mu}_n}{\mathcal{V}}. \tag{A.3}$$

Here $N_s$ represents the total number of spins in $\mathcal{V}$, $\boldsymbol{\mu}_n$ is the magnetic moment and $S_n$ the spin of the $n$th nucleus. Classical mechanics provides us the relation between the angular momentum $S_{\text{net}}$ and the torque $\boldsymbol{\tau}$,

$$\boldsymbol{\tau} = \frac{d}{dt} S_{\text{net}}. \tag{A.4}$$

Furthermore, $\boldsymbol{m}$ experiences a torque from the external magnetic field $\boldsymbol{B}_0$.

$$\boldsymbol{\tau} = \boldsymbol{m} \times \boldsymbol{B}_0 \tag{A.5}$$

Using equations (A.1), (A.2) and (A.3) and combining equations (A.4) and (A.5), we get

$$\frac{d\boldsymbol{m}}{dt} = \gamma \boldsymbol{m} \times \boldsymbol{B}_0, \tag{A.6}$$

which is the equation of motion of the magnetization vector $\boldsymbol{m}$ in the external field $\boldsymbol{B}_0$. Its solution is the precession of $\boldsymbol{m}$ around the axis of $\boldsymbol{B}_0$ with the so called Larmor frequency [17]

$$\omega_0 = \gamma B_0. \tag{A.7}$$

## A.3. Spinor representation and Pauli matrices

The spinor representation, a vector representation of the spin eigenstates, is given by

$$|\uparrow\rangle \rightarrow \begin{pmatrix} 1 \\ 0 \end{pmatrix}, \quad |\downarrow\rangle \rightarrow \begin{pmatrix} 0 \\ 1 \end{pmatrix}. \tag{A.8}$$

Eq. (2.9) can then be written as

$$\Psi = \Phi \begin{pmatrix} c_\uparrow \\ c_\downarrow \end{pmatrix}. \tag{A.9}$$

In this notation many calculations can be performed using the matrix formalism. With (2.5a) and (2.5b), the matrix representation of $\hat{S}_z$ is given by

$$\hat{S}_z = \frac{\hbar}{2} \begin{pmatrix} 1 & 0 \\ 0 & -1 \end{pmatrix}. \tag{A.10}$$

Further considerations lead us to matrix representations of $\hat{S}_x$ and $\hat{S}_y$. We introduce the Pauli matrices

$$\sigma_x := \begin{pmatrix} 0 & 1 \\ 1 & 0 \end{pmatrix}, \quad \sigma_y := \begin{pmatrix} 0 & -i \\ i & 0 \end{pmatrix}, \quad \sigma_z := \begin{pmatrix} 1 & 0 \\ 0 & -1 \end{pmatrix} \tag{A.11}$$

and $\boldsymbol{\sigma} := (\sigma_x, \sigma_y, \sigma_z)^T$. Then, the spin operator (2.1) can be expressed by

$$\hat{S} = \frac{\hbar}{2} \boldsymbol{\sigma}. \tag{A.12}$$

We can now easily calculate the effect of $\hat{S}_x$ and $\hat{S}_y$ on the eigenstates $|\uparrow\rangle$ and $|\downarrow\rangle$ [15].

$$\hat{S}_x |\uparrow\rangle = \frac{\hbar}{2} \begin{pmatrix} 0 & 1 \\ 1 & 0 \end{pmatrix} \begin{pmatrix} 1 \\ 0 \end{pmatrix} = \frac{\hbar}{2} |\downarrow\rangle, \quad \hat{S}_x |\downarrow\rangle = \frac{\hbar}{2} \begin{pmatrix} 0 & 1 \\ 1 & 0 \end{pmatrix} \begin{pmatrix} 0 \\ 1 \end{pmatrix} = \frac{\hbar}{2} |\uparrow\rangle \tag{A.13a}$$

$$\hat{S}_y |\uparrow\rangle = \frac{\hbar}{2} \begin{pmatrix} 0 & -i \\ i & 0 \end{pmatrix} \begin{pmatrix} 1 \\ 0 \end{pmatrix} = i\frac{\hbar}{2} |\downarrow\rangle, \quad \hat{S}_y |\downarrow\rangle = \frac{\hbar}{2} \begin{pmatrix} 0 & -i \\ i & 0 \end{pmatrix} \begin{pmatrix} 0 \\ 1 \end{pmatrix} = -i\frac{\hbar}{2} |\uparrow\rangle \tag{A.13b}$$

## A.4. Rotating reference frame

We transform the Bloch equation without relaxation terms,

$$\frac{d}{dt}\langle\hat{\boldsymbol{m}}\rangle = \gamma\langle\hat{\boldsymbol{m}}\rangle \times \boldsymbol{B}, \tag{A.14}$$

into a rotating reference frame with angular frequency $\omega_{\text{rf}}$. In general, the time derivative of some vector $\boldsymbol{u}$ that rotates about $\boldsymbol{\omega}_{\text{rf}}$ is given by

$$\frac{d}{dt}\boldsymbol{u} = \omega_{\text{rf}} \times \boldsymbol{u}. \tag{A.15}$$

Now let $\boldsymbol{u} = \boldsymbol{u}(t)$ be time-dependent and $\boldsymbol{e}_x$, $\boldsymbol{e}_y$ and $\boldsymbol{e}_z$ the axis vectors of the rotating coordinate system described in the static laboratory-frame. Then, the time derivative of $\boldsymbol{u}$ with respect to the lab frame coordinates is

$$\frac{d}{dt}\boldsymbol{u} = \dot{u}_x\boldsymbol{e}_x + u_x\dot{\boldsymbol{e}}_x + \dot{u}_y\boldsymbol{e}_y + u_y\dot{\boldsymbol{e}}_y + \dot{u}_z\boldsymbol{e}_z + u_z\dot{\boldsymbol{e}}_z$$
$$\overset{(A.15)}{=} \dot{u}_x\boldsymbol{e}_x + \dot{u}_y\boldsymbol{e}_y + \dot{u}_z\boldsymbol{e}_z + \boldsymbol{\omega}_{\text{rf}} \times \boldsymbol{u}. \tag{A.16}$$

The term

$$\dot{u}_x\boldsymbol{e}_x + \dot{u}_y\boldsymbol{e}_y + \dot{u}_z\boldsymbol{e}_z := \left(\frac{d}{dt}\boldsymbol{u}\right)_r \tag{A.17}$$

can be identified as the time derivative of $\boldsymbol{u}$ with respect to the rotating coordinates. We substitute $\boldsymbol{u}$ by $\langle\hat{\boldsymbol{m}}\rangle$ which yields the relationship

$$\frac{d}{dt}\langle\hat{\boldsymbol{m}}\rangle \overset{(A.16)(A.17)}{=} \left(\frac{d}{dt}\langle\hat{\boldsymbol{m}}\rangle\right)_r + \boldsymbol{\omega}_{\text{rf}} \times \langle\hat{\boldsymbol{m}}\rangle. \tag{A.18}$$

Equation (A.14) is then given by

$$\left(\frac{d}{dt}\langle\hat{\boldsymbol{m}}\rangle\right)_r = \gamma\langle\hat{\boldsymbol{m}}\rangle \times \left(\boldsymbol{B} - \boldsymbol{\omega}_{\text{rf}}/\gamma\right). \tag{A.19}$$

Hence, in the rotating frame the Bloch equation has the same shape as the static one except that a reduced magnetic field $\boldsymbol{B} - \boldsymbol{\omega}_{\text{rf}}/\gamma$ prevails [9].

## A.5. Implementation of the SMS sequence in IDEA

For sequence programming we use the proprietary SIEMENS IDEA software. The sinc excitation pulse is defined as

$$B_{\text{sinc}}(t_i) := \text{sinc}(\pi \, \gamma \, G_z \, E \, \Delta z \, t_i). \tag{A.20}$$

Here $E$ is an empirical broadening factor, $\Delta z$ is the slice thickness and $t_i$ the discrete time. The gradient strength $G_z$ is implemented as

$$G_z := \frac{P_{\text{BWT}}}{\Delta z \, T \, \gamma}. \tag{A.21}$$

$P_{\text{BWT}}$ is the product of pulse bandwidth and pulse duration and named bandwidth-time product.[1]

SIEMENS provides a carefully revised single-slice FLASH sequence. During scan preparation, the user can set the position and orientation of a single-slice via the GUI. Internally the program calculates the necessary slice selection gradient and direction as well as the RF pulse carrier frequency. Equation (4.5) states that by modifying the envelope function we can turn the single-slice FLASH sequence into a multi-slice sequence while the carrier frequency remains unchanged. In IDEA this is relatively easy to accomplish and no major changes in the GUI have to be implemented.

## A.6. Choice of the encoding matrix

In general, an encoding matrix should be unitary. A unitary matrix solely possess eigenvalues of modulus 1. So, the inherent condition number of the matrix is also 1, which makes decoding robust. One possible choice often used in MRI are $M \times M$ Hadamard matrices $H_M$ [66]. However, a necessary (but not sufficient) condition for the existence of $H_M$ is, that $M$ must be divisible by 4 for $M > 2$. This imposes undesired restrictions on the choice of the multiband factor. Since Fourier-encoding of the third dimension is very natrual to MRI, the use of a discrete Fourier transform matrix of type (3.1) is a logical choice for slice encoding. Fourier-encoding also makes the implementation of the SMS-NLINV algorithm more elegant (see section 5.2).

---

[1] Thus, we can write $G_z = \Delta \omega T / \Delta z T \gamma$ and recover (2.55). Its default value is $P_{\text{BWT}} = 2.7$ and fixed. A certain slice thickness is therefore achieved by variation of the gradient strength $G_z$.

## A.7. Fourier-encoding and radiofrequency excitation pulse

We deduce how the excitation pulse envelope (4.8)

$$B_{\mathrm{rf}p}^{(M)} := \tilde{B}_{\mathrm{env}p} e^{-i\Omega_c t} = \left( \sum_{q=1}^{M} \Xi_{pq} B_{\mathrm{env}q} \right) e^{-i\Omega_c t}$$

leads to the Fourier-encoded k-space (3.2)

$$\tilde{y}_p := \sum_{q=1}^{M} \Xi_{pq} y_q.$$

Starting point is the Bloch equation (2.57). We use the small flip-angle approximation $\langle \hat{m}_z(t) \rangle \approx m_0$ and set $\omega_{\mathrm{rf}} := \Omega_c$ and $\Delta\Omega(z) := \omega(z) - \Omega_c$, i.e. we transform the equation into the rotating frame with angular frequency $\Omega_c$. With the definition $\zeta := i\gamma e^{-i\Delta\Omega t} m_0$ the solution (2.62) of the Bloch equation for Fourier-encoded envelopes $\tilde{B}_{\mathrm{env}p}$ is given by

$$
\begin{aligned}
\langle \hat{m}_{\perp p}(t) \rangle &\stackrel{(2.62)}{=} \zeta \int_0^t \tilde{B}_{\mathrm{env}p}(\tau) e^{i\Delta\Omega\tau} d\tau \\
&\stackrel{(4.8)}{=} \zeta \int_0^t \left( \sum_{q=1}^{M} \Xi_{pq} B_{\mathrm{env}q}(\tau) \right) e^{i\Delta\Omega\tau} d\tau \\
&= \sum_{q=1}^{M} \Xi_{pq} \left( \zeta \int_0^t B_{\mathrm{env}q}(\tau) e^{i\Delta\Omega\tau} d\tau \right) \\
&\stackrel{(2.62)}{=} \sum_{q=1}^{M} \Xi_{pq} \langle \hat{m}_{\perp q}(t) \rangle
\end{aligned}
\tag{A.22}
$$

The expectation value of the transversal magnetization $\langle \hat{m}_{\perp}(t) \rangle$ is closely connected to the measured k-space signal as described in section 2.2.4. Hence (A.22) shows the Fourier-encoding that we demand in (3.2).

## A.8. Gauss normal equation

We derive the equality of (5.7) and (5.8) with the help of the Gauss normal equation. This equation states: $u$ is a least-squares solution of $Au = v$ if and only if the normal equation $A^H A u = A^H v$ holds.

We start with equation (5.8)

$$\min \left( \| DF(X_n)dX - (\tilde{Y} - F(X_n)) \|^2 + \beta_n \| dX \|^2 \right), \tag{A.23}$$

which can be written as

$$\min \left( \left\| \begin{pmatrix} DF(X_n) \\ \sqrt{\beta_n} I \end{pmatrix} dX - \begin{pmatrix} (\tilde{Y} - F(X_n)) \\ 0 \end{pmatrix} \right\|^2 \right). \tag{A.24}$$

We identify

$$A := \begin{pmatrix} DF(X_n) \\ \sqrt{\beta_n} I \end{pmatrix}, \quad u := dX, \quad v := \begin{pmatrix} (\tilde{Y} - F(X_n)) \\ 0 \end{pmatrix} \tag{A.25}$$

and write down the equivalent normal equation $A^H A u = A^H v$ as

$$\begin{aligned}
\left( DF^H(X_n), \ \sqrt{\beta_n} I \right) & \begin{pmatrix} DF(X_n) \\ \sqrt{\beta_n} I \end{pmatrix} dX \\
&= \left( DF^H(X_n), \ \sqrt{\beta_n} I \right) \begin{pmatrix} (\tilde{Y} - F(X_n)) \\ 0 \end{pmatrix}.
\end{aligned} \tag{A.26}$$

Explicitly carrying out the vector products yields

$$\left( DF(X_n)^H DF(X_n) + \beta_n I \right) dX = DF(X_n)^H (\tilde{Y} - F(X_n)), \tag{A.27}$$

which is the desired equation (5.7).

## A.9. Adjoint of the derivative

We derive $DF^H(X)$, which is the adjoint of the derivative of the mapping function $F$. All operators and vectors are given in discretized form.

With the definition

$$d\boldsymbol{x}_q := \begin{pmatrix} (dm_q) \\ \begin{pmatrix} dc_q^1 \\ \vdots \\ dc_q^N \end{pmatrix} \end{pmatrix} \tag{A.28}$$

we can rewrite eq. (5.16),

$$DF(X) \begin{pmatrix} d\boldsymbol{x}_1 \\ \vdots \\ d\boldsymbol{x}_M \end{pmatrix} =$$

$$\boldsymbol{P}\Xi\mathcal{F} \begin{pmatrix} (\boldsymbol{c}_1, m_1) & & 0 \\ & \ddots & \\ 0 & & (\boldsymbol{c}_M, m_M) \end{pmatrix} \begin{pmatrix} d\boldsymbol{x}_1 \\ \vdots \\ d\boldsymbol{x}_M \end{pmatrix}. \tag{A.29}$$

Now the construction of the adjoint is trivial,

$$DF^H(X) \begin{pmatrix} \tilde{\boldsymbol{y}}_1 \\ \vdots \\ \tilde{\boldsymbol{y}}_M \end{pmatrix} =$$

$$\begin{pmatrix} (\boldsymbol{c}_1, m_1)^H & & 0 \\ & \ddots & \\ 0 & & (\boldsymbol{c}_M, m_M)^H \end{pmatrix} \mathcal{F}^H \Xi^H \boldsymbol{P}^H \begin{pmatrix} \tilde{\boldsymbol{y}}_1 \\ \vdots \\ \tilde{\boldsymbol{y}}_M \end{pmatrix},$$

Eq. (A.30) equals (5.17).

# A.10. Sobolev norm in Fourier space

We derive the Sobolev norm in Fourier space. Therefore, we need the well known Fourier relations

$$f(\vec{x}) = \frac{1}{(2\pi)^2} \int_{\mathbb{R}_2} F(\vec{k}) e^{i\vec{k}\cdot\vec{x}} dk^2, \tag{A.30a}$$

$$\Delta_{\vec{x}} \int_{\mathbb{R}_2} F(\vec{k}) e^{i\vec{k}\cdot\vec{x}} dk^2 = -\int_{\mathbb{R}_2} \|\vec{k}\|^2 F(\vec{k}) e^{i\vec{k}\cdot\vec{x}} dk^2, \tag{A.30b}$$

$$\delta(\vec{k}) = \frac{1}{(2\pi)^2} \int_{\mathbb{R}_2} e^{i\vec{k}\cdot\vec{x}} dx^2. \tag{A.30c}$$

We assume $l$ to be even and rearrange the (squared) Sobolev norm,

$$\|f(\vec{x})\|_{H^l}^2$$
$$= \|2\pi(I - a\Delta_{\vec{x}})^{l/2} f(\vec{x})\|^2$$
$$= (2\pi)^2 \int_{\mathbb{R}_2} \left( (I - a\Delta_{\vec{x}})^{l/2} f(\vec{x}) \right) \left( (I - a\Delta_{\vec{x}})^{l/2} f(\vec{x}) \right)^* dx^2$$
$$\overset{(A.30a)}{=} \frac{1}{(2\pi)^2} \int_{\mathbb{R}_2} \left( (I - a\Delta_{\vec{x}})^{l/2} \int_{\mathbb{R}_2} F(\vec{k}_1) e^{i\vec{k}_1\cdot\vec{x}} dk_1^2 \right) \cdot$$
$$\left( (I - a\Delta_{\vec{x}})^{l/2} \int_{\mathbb{R}_2} F(\vec{k}_2) e^{i\vec{k}_2\cdot\vec{x}} dk_2^2 \right)^* dx^2$$
$$\overset{(A.30b)}{=} \frac{1}{(2\pi)^2} \int_{\mathbb{R}_2} \left( \int_{\mathbb{R}_2} \left( I + a\|\vec{k}_1\|^2 \right)^{l/2} F(\vec{k}_1) e^{i\vec{k}_1\cdot\vec{x}} dk_1^2 \right) \cdot$$
$$\left( \int_{\mathbb{R}_2} \left( I + a\|\vec{k}_2\|^2 \right)^{l/2} F(\vec{k}_2) e^{i\vec{k}_2\cdot\vec{x}} dk_2^2 \right)^* dx^2$$
$$= \iint_{\mathbb{R}_2} \left( I + a\|\vec{k}_1\|^2 \right)^{l/2} \left( I + a\|\vec{k}_2\|^2 \right)^{l/2^*} \frac{1}{(2\pi)^2} \int_{\mathbb{R}_2} e^{i(\vec{k}_1 - \vec{k}_2)\vec{x}} dx^2 \cdot$$
$$F(\vec{k}_1) F^*(\vec{k}_2) dk_1^2 dk_2^2$$
$$\overset{(A.30c)}{=} \iint_{\mathbb{R}_2} \left( I + a\|\vec{k}_1\|^2 \right)^{l/2} \left( I + a\|\vec{k}_2\|^2 \right)^{l/2^*} \cdot$$
$$F(\vec{k}_1) F^*(\vec{k}_2) \delta(\vec{k}_1 - \vec{k}_2) dk_1^2 dk_2^2$$
$$= \int_{\mathbb{R}_2} (I + a\|\vec{k}\|^2)^{l/2} (I + a\|\vec{k}\|^2)^{l/2^*} F(\vec{k}) F^*(\vec{k}) dk^2$$
$$= \|(I + a\|\vec{k}\|^2)^{l/2} F(\vec{k})\|^2$$

# Bibliography

[1] A. Maudsley. "Multiple-line-scanning spin density imaging". *J. Magn. Reson.* 41 (1980), pp. 112–126.

[2] S. Müller. "Multifrequency selective rf pulses for multislice MR imaging". *Magn. Reson. Med.* 6 (1988), pp. 364–371.

[3] D. J. Larkman et al. "Use of multicoil arrays for separation of signal from multiple slices simultaneously excited". *J. Magn. Reson. Imaging* 13 (2001), pp. 313–317.

[4] S. Moeller et al. "Multiband multislice GE-EPI at 7 tesla, with 16-fold acceleration using partial parallel imaging with application to high spatial and temporal whole-brain fMRI". *Magn. Reson. Med.* 63 (2010), pp. 1144–1153.

[5] D. A. Feinberg et al. "Multiplexed echo planar imaging for sub-second whole brain FMRI and fast diffusion imaging". *PLoS One* 5 (2010), e15710.

[6] M. Uecker et al. "Image reconstruction by regularized nonlinear inversion—joint estimation of coil sensitivities and image content". *Magn. Reson. Med.* 60 (2008), pp. 674–682.

[7] E. Haacke et al. *Magnetic Resonance Imaging: Physical Principles and Sequence Design.* Wiley, 1999. ISBN: 9780471351283.

[8] Z. Liang and P. Lauterbur. *Principles of Magnetic Resonance Imaging: A Signal Processing Perspective.* IEEE Press Series on Biomedical Engineering. Wiley, 1999. ISBN: 9780780347236.

[9] M. Bernstein, K. King, and X. Zhou. *Handbook of MRI Pulse Sequences.* Elsevier Science, 2004. ISBN: 9780080533124.

[10] B. Dale, M. Brown, and R. Semelka. *MRI: Basic Principles and Applications.* Wiley, 2015. ISBN: 9781119013037.

[11] R. Dreizler and C. Lüdde. *Theoretische Physik 3: Quantenmechanik 1.* Springer-Lehrbuch. Springer Berlin Heidelberg, 2007. ISBN: 9783540488026.

[12]  F. Schwabl. *Quantenmechanik (QM I): Eine Einführung*. Springer-Lehrbuch. Springer Berlin Heidelberg, 2013. ISBN: 9783662096291.

[13]  J. S. Rigden. "Quantum states and precession: The two discoveries of NMR". *Rev. Mod. Phys.* 58 (1986), p. 433.

[14]  A. Karaus. "Aufbau und Anwendung von Verfahren der Magnetresonanztomografie mit stimulierten Echos". PhD thesis. Würzburg, Bayrische Julius-Maximilians-Universität Würzburg, 2010.

[15]  C. Cohen-Tannoudji, B. Diu, and F. Laloë. *Quantenmechanik*. Quantenmechanik Bd. 1. de Gruyter, 2007. ISBN: 9783110193244.

[16]  D. Griffiths. *Introduction to Quantum Mechanics*. Always learning. Pearson, 2013. ISBN: 9781292024080.

[17]  M. Duer. *Solid State NMR Spectroscopy: Principles and Applications*. Wiley, 2008. ISBN: 9780470999387.

[18]  G. Findenegg and T. Hellweg. *Statistische Thermodynamik*. Springer Berlin Heidelberg, 2015. ISBN: 9783642378720.

[19]  K. T. Block. "Spiralförmige Abtastung des k-Raumes bei der Magnetresonanz-Tomographie". MA thesis. Georg August Universität Göttingen, 2004.

[20]  F. Bloch. "Nuclear Induction". *Phys. Rev.* 70 (1946), pp. 460–474.

[21]  M. Bryon. *MRI transmitter, receiver, and coils*. Accessed 2016/09/12. URL: http://www.mouser.de/applications/medical-imaging-overview/.

[22]  D. A. Yablonskiy, A. L. Sukstanskii, and J. J. Ackerman. "Image artifacts in very low magnetic field MRI: the role of concomitant gradients". *J. Magn. Reson.* 174 (2005), pp. 279–286.

[23]  P. M. Joseph, L. Axel, and M. O'Donnell. "Potential problems with selective pulses in NMR imaging systems". *Med. Phys.* 11 (1984), pp. 772–777.

[24]  J. Smith et al. *Spectral Audio Signal Processing*. W3K, 2011. ISBN: 9780974560731.

[25]  A. Haase et al. "FLASH imaging. Rapid NMR imaging using low flip-angle pulses". *J. Magn. Reson.* 67 (1986), pp. 258–266.

[26]  K. P. Pruessmann et al. "SENSE: sensitivity encoding for fast MRI". *Magn. Reson. Med.* 42 (1999), pp. 952–962.

[27]  M. A. Griswold et al. "Generalized autocalibrating partially parallel acquisitions (GRAPPA)". *Magn. Reson. Med.* 47 (2002), pp. 1202–1210.

[28]   K. P. Pruessmann et al. "Advances in sensitivity encoding with arbitrary k-space trajectories". *Magn. Reson. Med.* 46 (2001), pp. 638–651.

[29]   H. Gudbjartsson and S. Patz. "The Rician distribution of noisy MRI data". *Magn. Reson. Med.* 34 (1995), pp. 910–914.

[30]   A. Haase. "Snapshot flash mri. applications to t1, t2, and chemical-shift imaging". *Magn. Reson. Med.* 13 (1990), pp. 77–89.

[31]   F. A. Breuer et al. "Controlled aliasing in parallel imaging results in higher acceleration (CAIPIRINHA) for multi-slice imaging". *Magn. Reson. Med.* 53 (2005), pp. 684–691.

[32]   M. Blaimer et al. "Accelerated volumetric MRI with a SENSE/GRAPPA combination". *J. Magn. Reson. Imaging* 24 (2006), pp. 444–450.

[33]   K. Setsompop et al. "Blipped-controlled aliasing in parallel imaging for simultaneous multislice echo planar imaging with reduced g-factor penalty". *Magn. Reson. Med.* 67 (2012), pp. 1210–1224.

[34]   M. Uecker. "Nonlinear Reconstruction Methods for Parallel Magnetic Resonance Imaging". PhD thesis. Georg-August-Universität Göttingen, 2009.

[35]   M. Uecker, S. Zhang, and J. Frahm. "Nonlinear inverse reconstruction for real-time MRI of the human heart using undersampled radial FLASH". *Magn. Reson. Med.* 63 (2010-06), pp. 1456–1462.

[36]   M. Barth et al. "Simultaneous multislice (SMS) imaging techniques". *Magn. Reson. Med.* 75 (2016), pp. 63–81.

[37]   J. Hennig. "Chemical shift imaging with phase-encoding RF pulses". *Magn. Reson. Med.* 25 (1992), pp. 289–298.

[38]   G. Goelman. "Two methods for peak RF power minimization of multiple inversion-band pulses". *Magn. Reson. Med.* 37 (1997), pp. 658–665.

[39]   E. J. Auerbach et al. "Multiband accelerated spin-echo echo planar imaging with reduced peak RF power using time-shifted RF pulses". *Magn. Reson. Med.* 69 (2013), pp. 1261–1267.

[40]   S. Conolly et al. "Variable-rate selective excitation". *J. Magn. Reson.* 78 (1988), pp. 440–458.

[41]   D. G. Norris et al. "Power independent of number of slices (PINS) radiofrequency pulses for low-power simultaneous multislice excitation". *Magn. Reson. Med.* 66 (2011), pp. 1234–1240.

[42]   H. W. Engl, M. Hanke, and A. Neubauer. *Regularization of inverse problems*. Vol. 375. Springer Science & Business Media, 1996.

[43]   A. Bakushinsky and M. Kokurin. *Iterative Methods for Approximate Solution of Inverse Problems*. Mathematics and Its Applications. Springer Netherlands, 2005. ɪꜱʙɴ: 9781402031212.

[44]   M. R. Hestenes and E. Stiefel. "Methods of conjugate gradients for solving linear systems". *J. Res. Nat. Bur. Stand.* 49 (1952).

[45]   G. Folland. *Introduction to Partial Differential Equations*. Princeton University Press, 1995. ɪꜱʙɴ: 9780691043616.

[46]   M. Uecker et al. "ESPIRiT—an eigenvalue approach to autocalibrating parallel MRI: where SENSE meets GRAPPA". *Magn. Reson. Med.* 71 (2014), pp. 990–1001.

[47]   M. Uecker et al. "Berkeley Advanced Reconstruction Toolbox". In: *Proc. Intl. Soc. Mag. Reson. Med. 23, Toronto; 2486.* 2015.

[48]   B. Murphy et al. "Signal-to-noise measures for magnetic resonance imagers". *Magn. Reson. Imaging* 11 (1993), pp. 425–428.

[49]   M. J. Firbank et al. "A comparison of two methods for measuring the signal to noise ratio on MR images". *Phys. Med. Biol.* 44 (1999), N261.

[50]   O. Dietrich et al. "Measurement of signal-to-noise ratios in MR images: Influence of multichannel coils, parallel imaging, and reconstruction filters". *J. Magn. Reson. Imaging* 26 (2007), pp. 375–385.

[51]   H. Engl, M. Hanke, and A. Neubauer. *Regularization of Inverse Problems*. Mathematics and Its Applications. Springer Netherlands, 1996. ɪꜱʙɴ: 9780792341574.

[52]   F. Bauer and T. Hohage. "A Lepskij-type stopping rule for regularized Newton methods". *Inverse Prob.* 21 (2005), p. 1975.

[53]   J. Sénegas and M. Uecker. "Non-linear Inversion in Parallel MRI: Co nsiderations on Noise Amplification in the Joint Estimation of Image and Coil Sensitivities". In: *Proc. Intl. Soc. Mag. Reson. Med. 18, Stockholm: 2873.* 2010.

[54]   F. Knoll et al. "Parallel imaging with nonlinear reconstruction using variational penalties". *Magn. Reson. Med.* 67 (2011-06), pp. 34–41.

[55] V. Athalye, M. Lustig, and M. Uecker. "Parallel magnetic resonance imaging as approximation in a reproducing kernel Hilbert space". *Inverse Prob.* 31 (2015), p. 045008.

[56] M. Weiger, K. P. Pruessmann, and P. Boesiger. "2D sense for faster 3D MRI". *Magn. Reson. Mater. Phys., Biol. Med.* 14 (2002), pp. 10–19.

[57] F. Wiesinger, P. Boesiger, and K. P. Pruessmann. "Electrodynamics and ultimate SNR in parallel MR imaging". *Magn. Reson. Med.* 52 (2004), pp. 376–390.

[58] K. Scheffler and J. Hennig. "Reduced circular field-of-view imaging". *Magn. Reson. Med.* 40 (1998), pp. 474–480.

[59] D. C. Peters et al. "Undersampled projection reconstruction applied to MR angiography". *Magn. Reson. Med.* 43 (2000), pp. 91–101.

[60] S. Zhang, K. T. Block, and J. Frahm. "Magnetic resonance imaging in real time: advances using radial FLASH". *J. Magn. Reson. Imaging* 31 (2010), pp. 101–109.

[61] M. Uecker et al. "Real-time MRI at a resolution of 20 ms". *NMR Biomed.* 23 (2010), pp. 986–994.

[62] D. Voit et al. "Real-time cardiovascular magnetic resonance at 1.5 T using balanced SSFP and 40 ms resolution". *J. Cardiov. Magn. Reson.* 15 (2013), p. 1.

[63] A. Joseph et al. "Real-time flow MRI of the aorta at a resolution of 40 msec". *J. Magn. Reson. Imaging* 40 (2014), pp. 206–213.

[64] A. Joseph et al. "Real-time phase-contrast MRI of cardiovascular blood flow using undersampled radial fast low-angle shot and nonlinear inverse reconstruction". *NMR Biomed.* 25 (2012), pp. 917–924.

[65] J. Bogaert et al. *Clinical Cardiac MRI*. Medical Radiology / Diagnostic Imaging. Springer Berlin Heidelberg, 2012. ISBN: 9783642230356.

[66] S. Souza et al. "SIMA: simultaneous multislice acquisition of MR images by Hadamard-encoded excitation." *J. Comput. Assist. Tomogr.* 12 (1988), pp. 1026–1030.

# Acknowledgements

First of all, I want to thank Prof. Martin Uecker for the possibility to do research and write my master thesis in the amazing field of magnetic resonance imaging. I am very grateful for his outstanding support and supervision. Furthermore, I want to thank Prof. Christoph Schmidt for reviewing this thesis as second referee.

A big thanks goes to my colleagues Robin Wilke and Christian Holme. It is a pleasure and so much fun to work with them. It really feels like being part of a team and I appreciate their support a lot. I also want to thank all other colleagues, technical assistants and M.D candidates from our group - especially my roommate Tom Stümpfig. They have all contributed to the extremely positive atmosphere and spirit that I felt throughout the whole year. Special thanks also goes to Volkert Roeloffs and Zhengguo Tan for valuable discussions and support.

Finally, I want to thank Andreas Fischer, Nicolas Triltsch and Jan Scholüke. Without them the study of physics would have been a lot more difficult and a lot less fun.

My studies would have not been possible without the great and unconditional support of my family. Thank you for all your assistance and aid – especially in the hard times!

After all... what I have and what I can accomplish, I owe everything to our creator - *S. D. G.*